\RequirePackage{fix-cm}
\documentclass[12pt]{article}
\usepackage{epsf,rotating,latexsym,amssymb} 
\usepackage{epsfig}   
\usepackage{epstopdf} 
\usepackage{rotating,latexsym,amssymb}           
\usepackage{mathptmx} 
\usepackage{graphicx}
\usepackage{times}
\usepackage{natbib}
\usepackage{lineno}       
\usepackage{color}
\newcommand{\kms}{$\mathrm {km\,s^{-1}}$}

\newcommand{\dpl}{DESTINY$^+$}

\begin{document}
\bibliographystyle{apalike}


\bigskip
\noindent
{\bf \large Modelling \dpl\ interplanetary and interstellar dust measurements en route to the active asteroid (3200) Phaethon} 

\bigskip

\noindent
{\normalsize
Harald Kr\"uger$^1$,  
Peter Strub$^1$, 
Ralf Srama$^2$,  
Masanori Kobayashi$^3$, 
Tomoko Arai$^3$,  
Hiroshi Kimura$^3$,
Takayuki Hirai$^3$,  
Georg Moragas-Klostermeyer$^2$,  
Nicolas Altobelli$^4$,  
Veerle J. Sterken$^5$,  
Jessica Agarwal$^1$, 
Maximilian Sommer$^2$,
Eberhard Gr\"un$^6$ 
}

\bigskip

\noindent
{\small
$^1$MPI f\"ur Sonnensystemforschung, G\"ottingen, Germany \\
$^2$Institut f\"ur Raumfahrtsysteme, Universit\"at Stuttgart, Germany \\
$^3$Planetary Exploration Research Center, Chiba Institute of Technology, Narashino, Japan \\
$^4$European Space Agency, ESAC, Madrid, Spain \\
$^5$Institute of Applied Physics, University of Bern, Switzerland \\
$^6$MPI f\"ur Kernphysik, Heidelberg, Germany
}




\begin{abstract}
The JAXA/ISAS spacecraft DESTINY$^+$  will be launched to the active 
asteroid (3200) Phaethon in  2022. Among the proposed core payload is the \dpl\ Dust Analyzer 
(DDA) which is an upgrade of the Cosmic Dust Analyzer flown on the Cassini spacecraft to 
Saturn \citep{srama2011}.
We use two up-to-date computer models, the ESA Interplanetary Meteoroid Engineering Model
\citep[IMEM,][]{dikarev2005a,dikarev2005b}, and the interstellar dust module of the Interplanetary 
Meteoroid environment for EXploration model
\citep[IMEX;][]{sterken2013a,strub2019} to study the detection conditions and 
fluences of interplanetary and interstellar 
dust with DDA. Our results show that a statistically 
significant number of 
interplanetary and interstellar dust particles 
will be  detectable with DDA during the 4-years interplanetary cruise of \dpl. 
The particle impact direction and speed can be used to descriminate
between interstellar and interplanetary particles and likely also to distinguish between
cometary and asteroidal particles. 

\end{abstract}

\section{Introduction}

The \dpl\ (Demonstration and Experiment of Space Technology for INterplanetary voYage Phaethon fLyby 
and dUst Science)  mission has been selected by the Japanese 
space agency JAXA/ISAS \citep{kawakatsu2013,arai2018}. The mission target is the active near-Earth asteroid (3200) 
Phaethon \citep{jewitt2010,jewitt2013}. 
\dpl\ will be launched in  2022 initially into 
a low elliptical Earth orbit. Driven by its ion engine, the spacecraft will raise its altitude until 
reaching the Moon for a series of gravity assists \citep{sarli2018}
and will ultimately  be injected into a heliocentric trajectory. A flyby at Phaethon 
is presently planned for
August 2026 at a heliocentric distance of 0.87~AU. A later gravity assist at Earth may redirect 
the spacecraft to another target. The \dpl\ mission schedule is given in Table~\ref{tab:schedule}.
\dpl\ is a three-axis stabilised spacecraft.
  
The proposed science instruments on board \dpl\ are two cameras \citep[the Telescopic CAmera for Phae\-thon, 
TCAP, and  the Multiband CAmera for Phaethon, MCAP;][]{ishibashi2018}, and the \dpl\ 
Dust Analyzer  \citep[DDA;][]{kobayashi2018b}. DDA is an 
upgrade of the Cassini Cosmic Dust Analyzer (CDA)
which  very successfully investigated dust throughout the Saturnian system \citep{srama2009,srama2004,srama2011}. 
DDA will be an impact ionization time-of-flight mass spectrometer capable of 
analyzing sub-micron and micron sized dust particles with a mass resolution of $m/\Delta m \approx 100 - 150$,
 and a trajectory sensor. 
In addition to the elemental and isotopic composition of impacting 
dust particles the instrument will measure the particle's mass, velocity vector,  electrical charge and impact direction.
DDA will measure the particle impact speed with an accuracy of approximately 10\,\% and the
 impact direction with an accuracy of about $10^{\circ}$. This will allow us to constrain 
the  trajectory, and thus, the source body of each  detected particle individually \citep{hillier2007}. 
As such, the in-situ particle analysis will  provide compositional information on the source object
where the particles originated. DDA will 
be equipped with two sensor heads, allowing for the measurement of positively and negatively charged ions released from  
impacting dust particles.
   
Phaethon is an extraordinary near-Earth asteroid with a diameter of 5.8~km \citep{taylor2018}. Its perihelion 
distance is presently 0.14~AU 
with an orbital period of 1.433~yr. Around perihelion its surface temperature reaches more than 1000~K.
Phaethon is the source of the Geminids, one of the most active meteor showers visible in the Earth's night sky. 
While parent bodies of meteor showers are mostly comets, Phaethon is an Apollo-type asteroid with a
carbonaceous B-type reflectance spectrum, similar to aqueously altered CI/CM meteorites, and of 
hydrated minerals \citep{licandro2007}. 
Recurrent dust ejection and a dust tail were reported at perihelion \citep{li2013,jewitt2013},
while no coma was observed around $1.0 - 1.5$~AU \citep{hsieh2005,jewitt2013,ye2018,kimura2019}. 
Phaethon's dust ejection mechanism remains unknown. 
Its physical properties were recently summarized by \citet{hanus2016}. Ground-based observations during 
a close encounter with  Earth in December 2017 showed that Phaethon 
has a non-hydrated surface \citep{takir2018}, and the light reflected from 
its surface shows an  unusually strong polarization 
which may be due to relatively large ($\mathrm{\sim 300\, \mu m}$) particles and/or high porosity of the 
surface material \citep{ito2018}. Finally, Phaethon shows indications for compositional variations 
across its surface \citep{kareta2018}.

 \begin{table}[b]
  \centering
      \caption{\dpl\ mission schedule based on the trajectory EAEXX01$^{\dagger}$ provided by JAXA/ISAS.}
      \begin{tabular}{lc} 
      & \\[-2ex]
\hline
            Launch      & 07 October 2022 \\
            Escape from Earth orbit & 24 September 2024 \\
            (3200) Phaethon flyby & 01 August 2026 \\
            Earth swing-by & 07 October 2028 \\
            \hline
   \end{tabular}
   
  {$\dagger$ We use an updated trajectory as compared to the one studied by \citet{sarli2018}}
   \label{tab:schedule}
\end{table}

\dpl\ will fly by Phaethon at a distance of 500~km or less, and the DDA instrument will directly 
analyse dust released from Phaethon's surface
\citep{kimura2019,szalay2019}. 
DDA will investigate the dust trail, the spatial distribution and composition of meteoroids 
in Phaethon's vicinity, as well as its surface composition and geology. 

\begin{figure}[tb]
	\centering
	\vspace{-7.4cm}
		\includegraphics[width=1.1\textwidth]{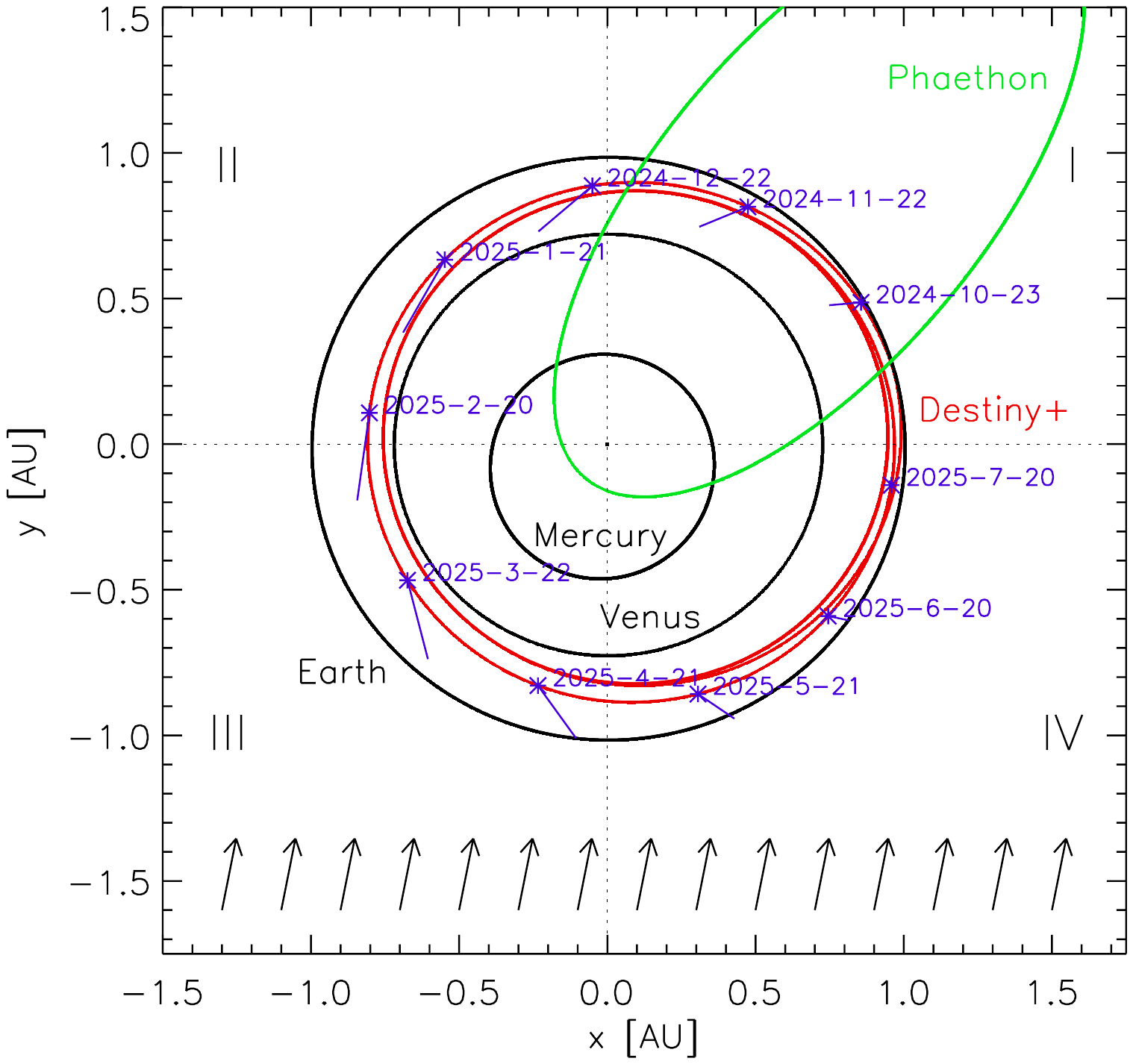}
		\vspace{-2.7cm}
	\caption{Trajectories of \dpl\ 
	 (red) and (3200) Phaethon (green) projected onto the ecliptic plane. Black arrows at the bottom indicate the flow of interstellar 
	 dust particles with a ratio of solar radiation pressure over gravity  $\beta=1$, which is assumed to be co-aligned 
	 with the flow of interstellar neutral helium \citep{witte2004b} 
	 \citep{wood2015}. 
The approach direction of these 
	 particles in the spacecraft-fixed coordinate system  
	is indicated by blue lines for selected times during the first year of the mission, the line 
	length is proportional to the particle impact speed (see also Figure~\ref{fig:imem_speed}). Vernal 
	equinox is to the right. \dpl\ trajectory from JAXA/ISAS (EAEXX01), updated from the earlier one 
	studied by \citet{sarli2016}.}
	\label{fig:dest_traj}
\end{figure}

In addition to the dust measurements at the Phaethon flyby, DDA will be able to measure dust in 
interplanetary space  during its four years of interplanetary 
voyage between the orbits of Venus and Earth (Figure~\ref{fig:dest_traj}). Interstellar 
particles with a radius $\mathrm{r_d \gtrsim 0.1\,\mu m}$ pass the heliospheric bow shock and enter 
the heliosphere \citep[][and references therein]{linde2000,slavin2012,sterken2015,mann2010}. 
As a consequence, interstellar dust constitutes the dominant 
known particulate component in the outer solar system (in number flux, not in mass flux).

Interstellar dust particles condense in the extended atmospheres of evolved stars and in stellar 
explosions and are injected into the interstellar medium. The particles originally carry the 
elemental and isotopic signatures from the environment 
where they were formed. Ultraviolet irradiation, interstellar shock waves, and mutual collisions 
subsequently modify the signatures and deplete particles in the interstellar medium. 
In dense molecular clouds particles can grow by agglomeration and accretion \citep[see, e.g.][]{tielens2005}. 
In our local
environment, the Sun and the heliosphere are surrounded by the diffuse Local Interstellar Cloud (LIC) 
of warm gas and dust where dust contributes about 1\% of the cloud mass. The Sun's motion with 
respect to this cloud causes an inflow of interstellar matter into the heliosphere
\citep{frisch1999a}.

Interstellar dust  in the solar system was  undoubtedly detected by the Ulysses spacecraft far 
from the ecliptic plane 
and outside the inner solar system \citep{gruen1993a}. Although the interstellar dust flow is 
modulated by the Lorentz force, solar radiation pressure and solar gravity 
\citep{landgraf2000b,sterken2012a,sterken2013a}, interstellar dust of some sizes can reach the 
region inside Earth's orbit \citep{altobelli2003,altobelli2005a,altobelli2006,strub2019,krueger2019b}. 
A small number of interstellar particles was successfully collected and returned to Earth by the 
Stardust spacecraft \citep{westphal2014b}, and a few dozen particles were analysed with  
CDA when Cassini was in orbit about Saturn \citep{altobelli2016}. See \citet{mann2010} and \citet{sterken2019} 
for comprehensive reviews of interstellar 
dust in the solar system,  the latter 
with a focus on the last approximately 10 years of research.

These observations open the possibility for spacecraft in 
the inner solar system like \dpl\  to detect and analyse interstellar dust in-situ near 
Earth's orbit \citep{gruen2005,strub2019}. The measurements with DDA will address several
major open 
questions in the study of interstellar dust, including:   
 A) Search for complex organic compounds in big -- 
approximately  sub-micron and micron-sized -- particles \citep{kimura2015};
B) Study the variability of  particle compositions, 
also in the context  of 
elemental depletions  observed in the interstellar medium \citep{slavin2008,frisch2011};
and C) Investigate the particle dynamics and trajectory modulation in  the heliosphere
\citep{sterken2015}. 

In addition to  the analysis of interstellar dust, the examination of interplanetary dust forming the 
zodiacal cloud will be another  major objective of the DDA measurements. 
The dominant sources for interplanetary dust in the inner solar system are comets and 
asteroids \citep{nesvorny2010}, however, the contributions of each of these sources to the zodiacal
dust cloud are still not well known. 
DDA will be able to 
measure the dynamical properties of each individual detected particle \citep{hillier2007} from an accurate 
measurement of the particle trajectories. Based on
the investigation of the physical properties and the composition of 
a large number of interplanetary  particles  DDA will  determine the abundances of 
asteroidal and cometary particles in the zodiacal cloud. This will 
lead to a better characterisation of the
dust sources feeding the zodiacal dust complex and it will ultimately help to improve existing 
 interplanetary dust models \citep{dikarev2005a}. DDA will also search for dust 
particles from  other sources like the 
Kuiper belt \citep{landgraf2002} or the Oort cloud and may provide information on particle alteration 
during their voyage to the Earth. A recent review on interplanetary dust was published  
by \citet{engrand2019}.

In this paper we study the detection conditions for interplanetary and interstellar dust 
particles with the DDA instrument on board \dpl. Dust detections during the Phaethon flyby   
 are discussed by \citet{kimura2019} and \citet{szalay2019}. In Section~\ref{sec_idp}
we present simulation results for interplanetary and interstellar dust during 
the four years of interplanetary voyage of \dpl. In  
Section~\ref{sec_discussion} we discuss  the detection conditions for these dust populations,
and in Section~\ref{sec_conclusions} we summarize our conclusions.


\section{Interplanetary and Interstellar Dust Simulations}

\label{sec_idp}

In this Section we study the detection conditions for interplanetary and interstellar dust particles 
with the DDA instrument during four years of interplanetary voyage of \dpl. To this end, we use
two dynamical models developed during the
recent years. For modelling 
interplanetary dust, we use the Interplanetary Meteoroid Engineering Model 
\citep[IMEM;][]{dikarev2005a,dikarev2005b}. We simulate interstellar dust with the
interstellar dust module of the Interplanetary Meteoroid environment for EXploration model
\citep[IMEX;][]{sterken2012a,sterken2013a,strub2019}.
Both computer models simulate dust densities in interplanetary space, and  they are
the  most up to date models presently available for the  dynamics of micrometer and sub-micrometer sized dust 
in the inner solar system.

In order to calculate dust densities and fluxes along the heliocentric orbit of \dpl\ 
we use  trajectory data
provided by JAXA/ISAS \citep[EAEXX01, cf.~Figure~\ref{fig:dest_traj};][]{sarli2016}. The trajectory 
covers a time period of 1474~days, from 24 September 2024 to 07 October 2028, beginning with 
the spacecraft's escape from Earth orbit (Table~\ref{tab:schedule}). We study
fluxes, impact speeds and impact directions on to the DDA sensor, i.e. in the spacecraft-centered
reference frame.

We calculate dust  fluences from our simulations by assuming a  total DDA sensitive area for two sensor heads
of $\mathrm{0.035\,m^2}$, and a 
detection threshold of $10^{-19}\,\mathrm{kg}$.
 For an assumed spherical particle with density 
typical of astrophysical silicates ($\mathrm{\rho=3300\,kg\,m^{-3}}$) the detection threshold 
corresponds to a radius $r_d=\mathrm{0.02\,\mu m}$.  Particle fluxes and fluences  
are given in the spacecraft frame of reference throughout the paper.

\subsection{Interplanetary Dust}

\label{sec_imem}

\begin{figure}[tb]
	\centering
	\vspace{-2.3cm}
		\includegraphics[width=0.88\textwidth]{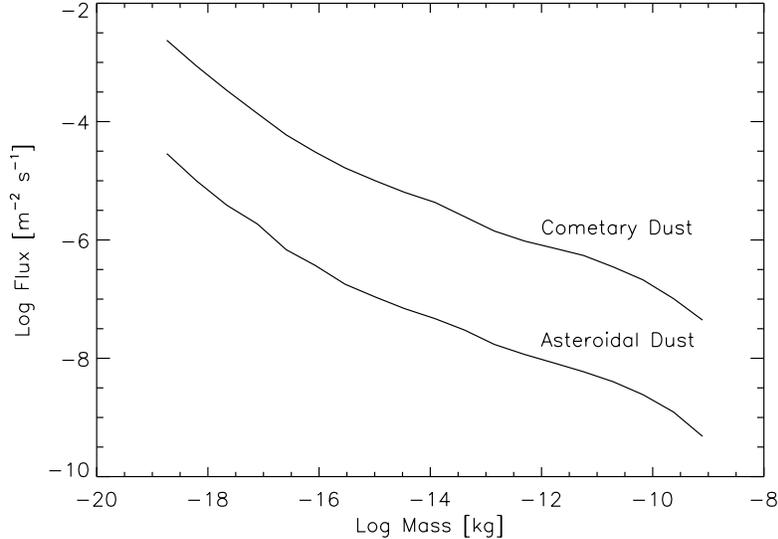}
		\vspace{-7.8cm}
	\caption{Mass distributions in the IMEM model \citep{dikarev2005a,dikarev2005b} 
	for asteroidal dust (population 2) and for cometary dust 
	(population 4) 
	 in the heliocentric distance range traversed by \dpl, 
	i.e. between 0.75~AU and 1.0~AU.  
	}
	\label{fig:imem_mass}
\end{figure}

The Interplanetary Meteoroid Engineering Model (IMEM), developed by \citet{dikarev2005a,dikarev2005b}, 
is the most sophisticated model to predict the dynamics and fluxes of interplanetary dust particles 
for  space missions 
presently available. It was developed under ESA contract for use by space engineers to study potential dust
hazards to interplanetary spacecraft. IMEM simulates the dynamics of  
five populations of cometary and asteroidal dust as well as interstellar dust in various mass ranges
\citep[we use the labelling employed by][]{dikarev2005a}:
(1) Asteroid Collisions ($m \geq 10^{-8}\,\mathrm{kg}$);
(2) Asteroid Poynting-Robertson ($m < 10^{-8}\,\mathrm{kg}$);
(3) Comet Collisions ($m \geq 10^{-8}\,\mathrm{kg}$);
(4) Comet Poynting-Robertson ($m < 10^{-8}\,\mathrm{kg}$);
(5) Interstellar Dust ($10^{-18}\,\mathrm{kg} < m \leq 10^{-12}\,\mathrm{kg}$). 

IMEM  was calibrated with infrared observations of the zodiacal cloud by the Cosmic Background 
Explorer (COBE) DIRBE instrument, in-situ flux
measurements by the dust detectors on board the Galileo and Ulysses spacecraft, and the crater size distributions 
on lunar rock samples retrieved by the Apollo missions. Within the model, the orbital distributions are 
expanded into a sum of contributions from a number of known sources, including the asteroid belt, with 
the emphasis on the prominent families Themis, Koronis, Eos and Veritas, as well as comets on 
Jupiter-encountering orbits \citep{dikarev2004,dikarev2005a,dikarev2005b}. In order to calculate
the particle dynamics,
solar gravity and the velocity-dependent tangential component of radiation pressure 
(Poynting-Robertson effect) are taken into account, while  the radial component of solar 
radiation pressure and the electromagnetic interaction of electrically charged dust particles are neglected. 

We do not consider populations 1 (asteroid collisions) and 3 (comet collisions) here
because the fluxes of these relatively big particles are very low and, therefore, not relevant
in our context. Here we concentrate on  
particles with masses  $m \lesssim 10^{-11}\,\mathrm{kg}$ which corresponds to a particle
radius of approximately $r_{d} \lesssim 10\,\mathrm{\mu m}$. 
The dust mass distributions
used by IMEM are shown in Figure~\ref{fig:imem_mass}. IMEM is a time-independent model, i.e. 
 temporal variations are only introduced by the spacecraft motion around the Sun. 

For interstellar dust (population 5) IMEM uses a very simple assumption, i.e. 
a mono-directional stream of  particles with an assumed ratio of gravitational force to 
solar radiation pressure  $\beta = 1$. Because of this simplification we do not use the IMEM interstellar 
dust module. Instead we 
use a more realistic  model that includes temporal variations in the dust dynamics due to the time-varying 
interplanetary magnetic field and realistic $\beta$ values for astronomical silicates 
\citep[Interplanetary Meteoroid environment for EXploration, IMEX; cf. Section~\ref{sec_imex}, see also][]{strub2019}. 

\begin{figure}[tb]
	\centering
	\vspace{-1.3cm}
		\includegraphics[width=0.84\textwidth]{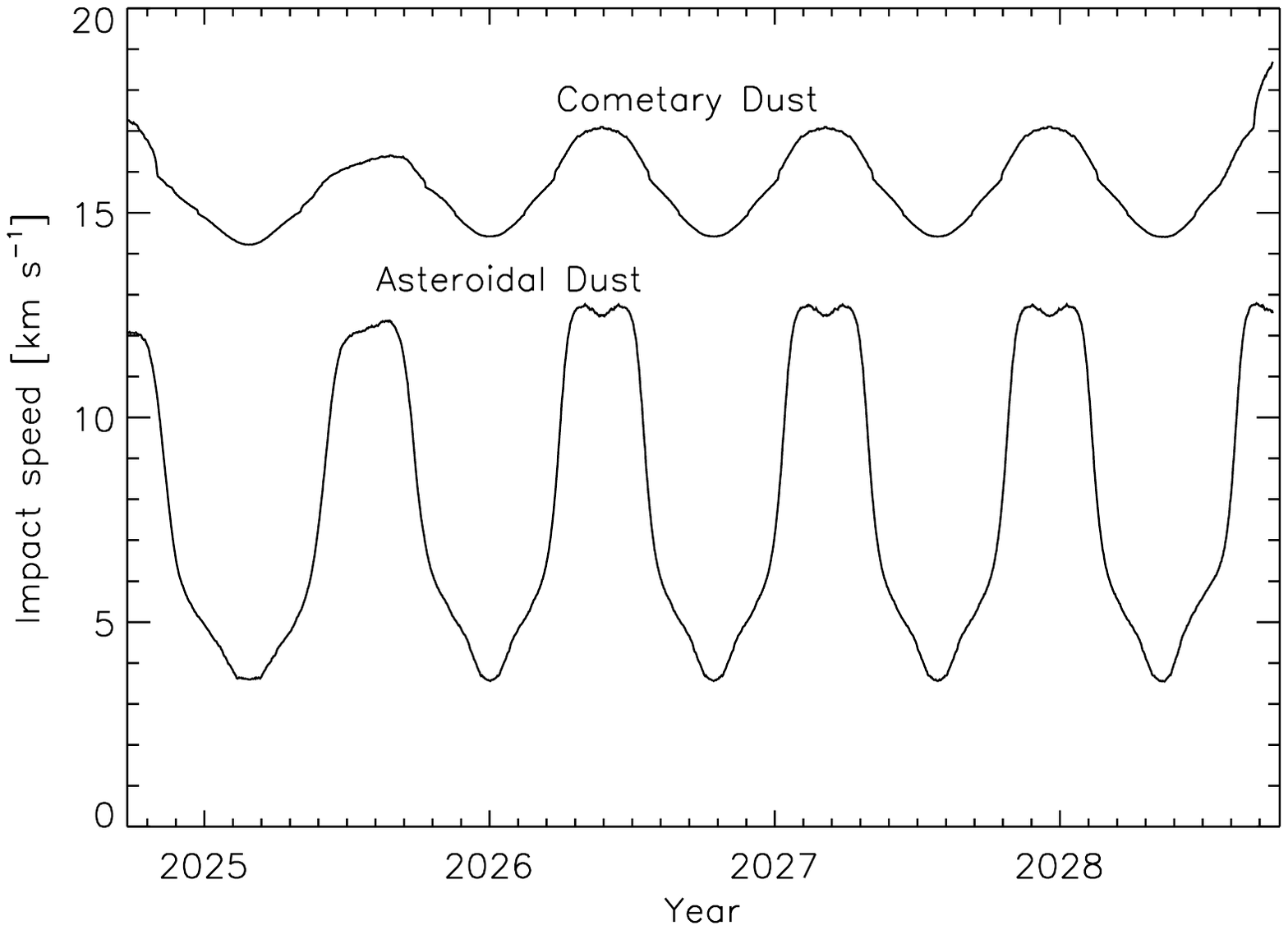}
		\vspace{-9.cm}
		
		\includegraphics[width=0.84\textwidth]{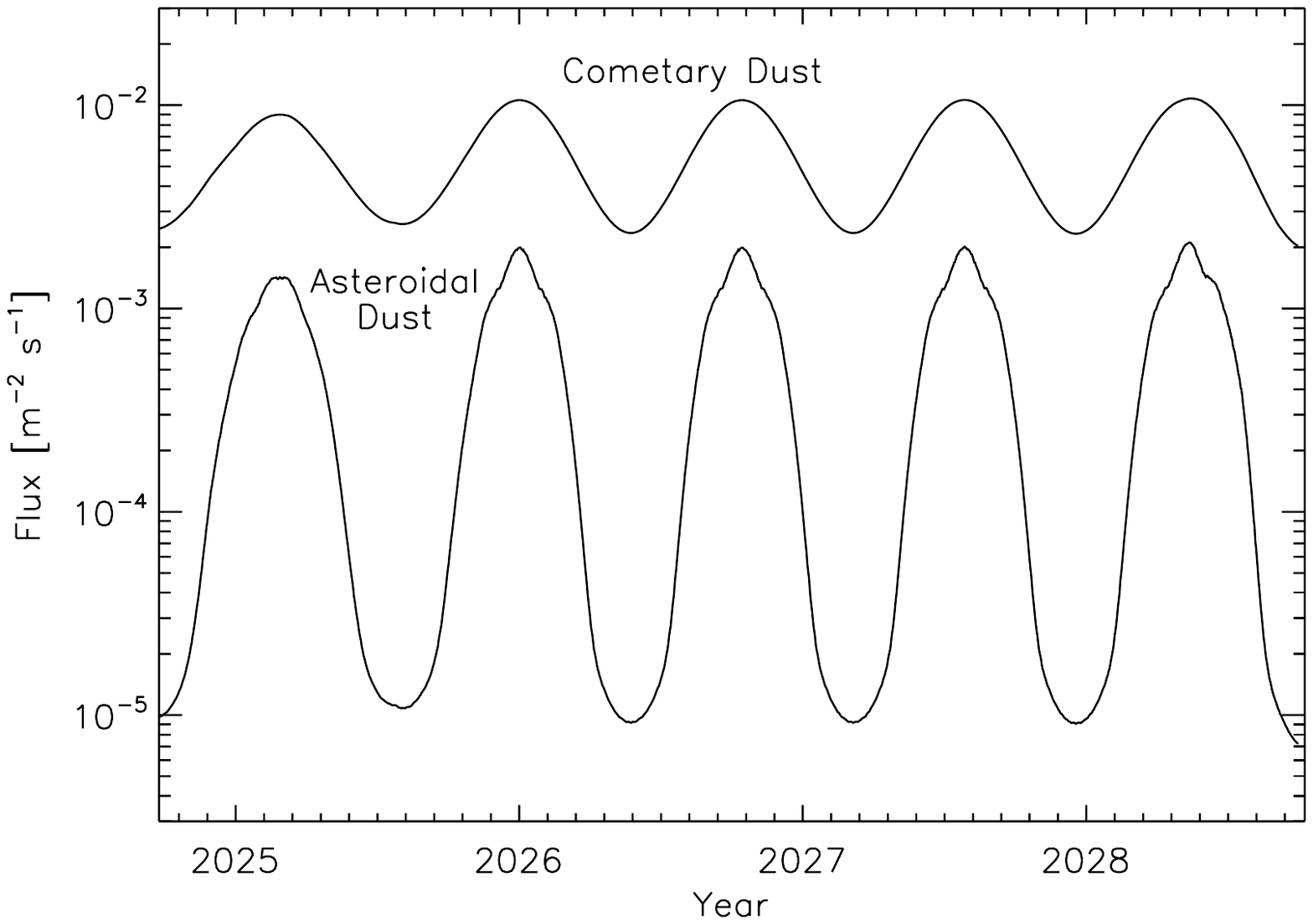}		
		\vspace{-7.5cm}
	\caption{Average particle impact speed ({\em top panel}) and flux ({\em bottom panel})  
	of asteroidal dust (population 2) and cometary dust (population 4) on to 
	\dpl\  in the spacecraft reference frame, for a flat plate sensor whose normal vector points
	parallel to the spacecraft velocity vector (continuous apex pointing). 
	  }
	\label{fig:imem_speed}
\end{figure}

When the Ulysses/Galileo in-situ data were incorporated in the IMEM model 
simultaneously with the COBE/DIRBE infrared sky maps, these two datasets turned out to be incompatible  
\citep{dikarev2005c}. 
Therefore, the ion charge calibration of the in-situ detectors was revised by a factor of 16, that is
to require an impactor 16 times more massive than derived from the standard instrument calibration 
derived by \citet{gruen1995a}.
This leads to an increase in particle sizes by a factor of 2.5. 

\subsubsection{Dust Flux and Impact Speed}

Up to now there is no detailed information on the spacecraft orientation available for the 
\dpl\ mission. For the IMEM simulations we therefore 
assume that the normal vector of the DDA sensor surface is oriented parallel to 
the direction of the spacecraft velocity vector all the time  (continuous apex pointing). 
Other sensor orientations may provide 
higher dust fluxes.
As a first step, we assume the sensor to be a flat plate. IMEM allows the inclusion 
of a detailed model for the sensor field-of-view (FOV).

\begin{figure}[tb]
	\centering
	\vspace{-1.3cm}
		\includegraphics[width=0.84\textwidth]{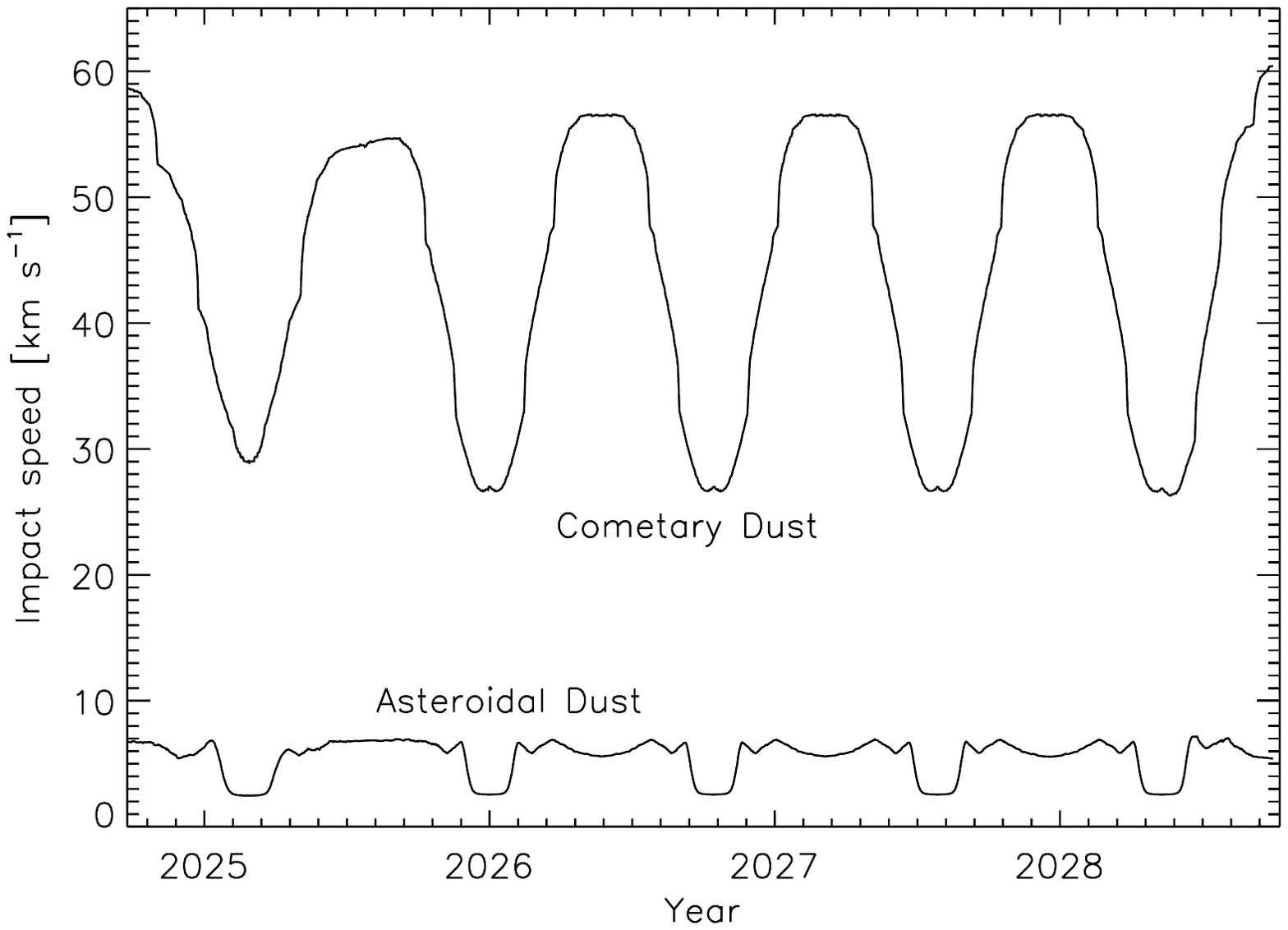}
		\vspace{-9cm}
		
		\includegraphics[width=0.84\textwidth]{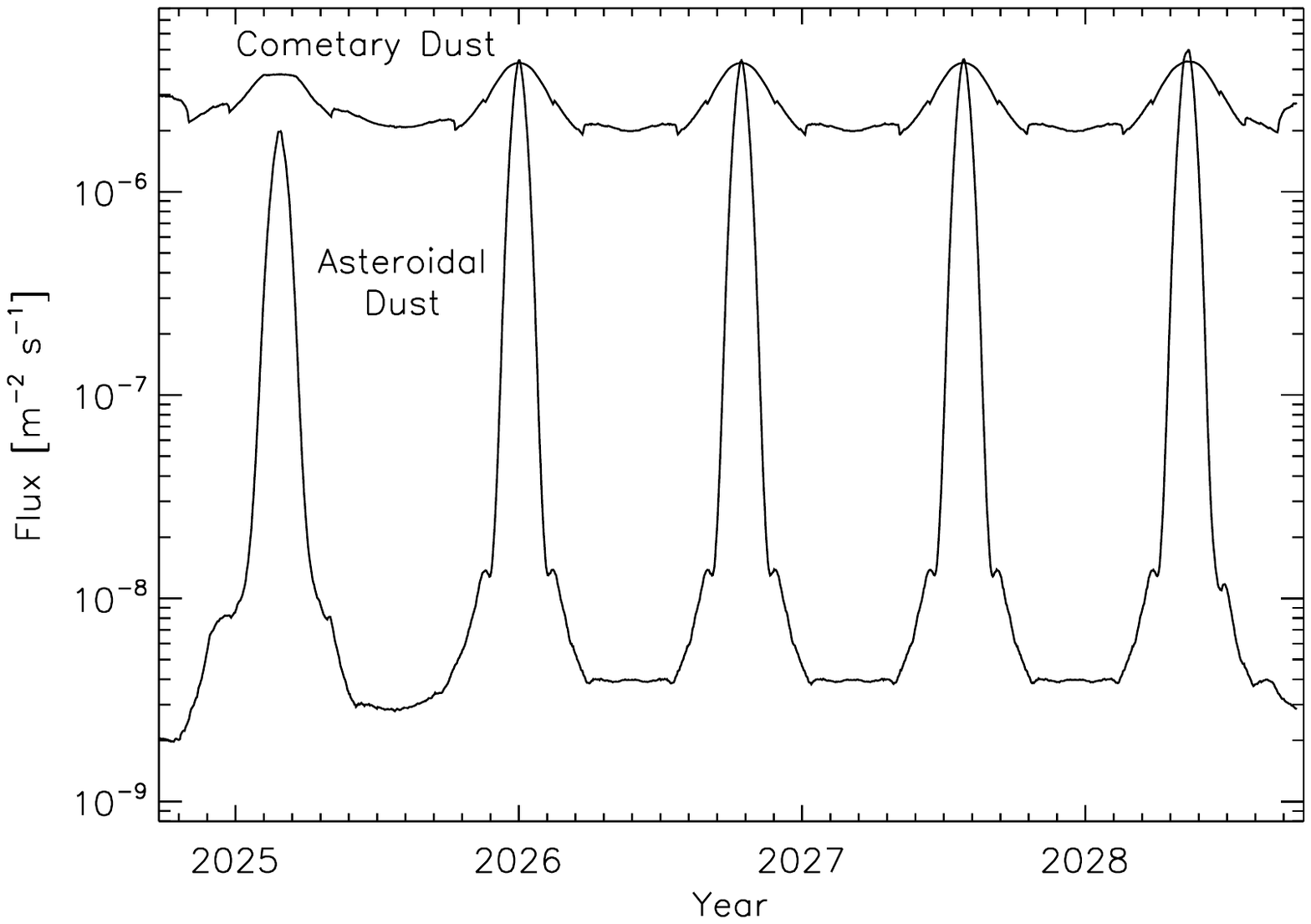}		
		\vspace{-7.5cm}
	\caption{Same as Figure~\ref{fig:imem_speed} but with the DDA sensor characteristics shown 
	in Figure~\ref{fig:dda_design}. 
	  }
	\label{fig:imem_dda}
\end{figure}

In Figure~\ref{fig:imem_speed} (top panel) we show the average impact speed  for  
interplanetary dust derived from IMEM. The speed modulation is due to the spacecraft motion 
around the Sun and the limited FOV of the dust sensor. 
For the asteroidal particles the average  speed varies 
between 4 and $12\,\mathrm{km\,s^{-1}}$, while that of the  
cometary particles ranges from 15 to $17\,\mathrm{km\,s^{-1}}$ (see also the sky maps in
Figures~\ref{fig:imem_com_ast_direction_1} to \ref{fig:imem_com_ast_direction_5}, right columns).  
The impact speed of cometary particles is larger on average than that of the asteroidal particles
due to their higher orbital eccentricities. 
It implies that, in 
particular during periods when the impact speed of asteroidal 
particles is low, the  speed measured by DDA can serve as a discriminator between these dust 
populations. 

Simulated dust fluxes of the interplanetary particles are displayed in the bottom panel of 
Figure~\ref{fig:imem_speed}. Their 
variability is similar to that of the impact speed, this is again due to the spacecraft motion. The flux of 
asteroidal particles varies by two orders of magnitude while that of the cometary particles 
varies only by about a factor of 5. 

\begin{figure}[tb]
	\centering
	\vspace{-0.3cm}
		\includegraphics[width=0.85\textwidth]{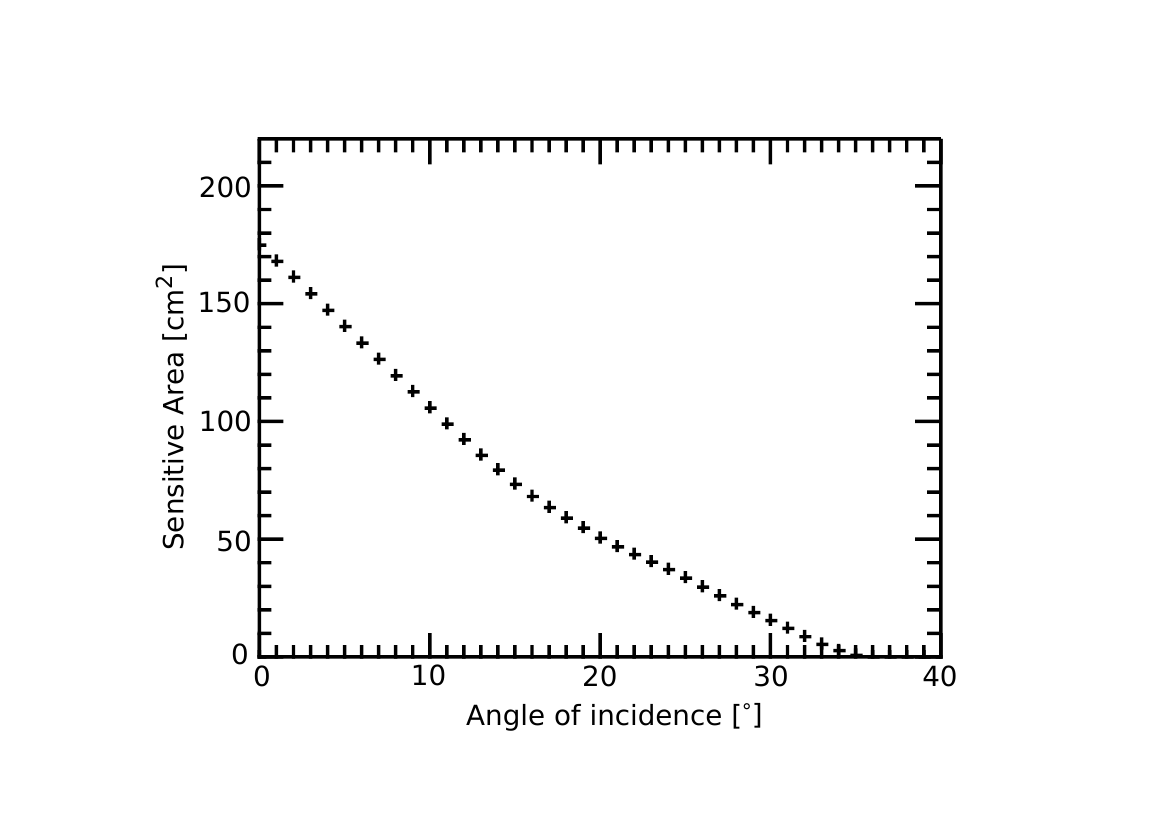}
		\vspace{-0.0cm}
			\caption{Field-of-view for  DDA  used for IMEM and IMEX simulations.  The sensitive area refers
			to a single DDA sensor head.
	  }
	\label{fig:dda_design}
\end{figure}

In Figure~\ref{fig:imem_dda} we show the same parameters -- impact speed and dust flux -- but 
for the DDA sensor FOV (Figure~\ref{fig:dda_design}). Again we assumed 
a sensor pointing parallel to the spacecraft velocity vector. Due to the much narrower FOV of DDA
as compared to the flat plate sensor, the dust
fluxes are reduced by more than three orders of magnitude. Furthermore, the difference in the average 
impact speeds between asteroidal and
cometary particles is much larger: for asteroidal particles the average speed is below $10\,\mathrm{km\,s^{-1}}$
while that of the cometary particles varies between approximately 30 and $\mathrm{55}\,\mathrm{km\,s^{-1}}$.
This supports the discrimination of asteroidal and cometary particles from
the impact speed measurement. 

Variations in the impact speeds and fluxes 
are anti-correlated in Figures~\ref{fig:imem_speed} and ~\ref{fig:imem_dda}, i.e. 
during periods when high impact speeds occur, the dust fluxes are low, and vice versa, 
which at first glance 
seems to be counter-intuitive. 
Figures~\ref{fig:imem_com_ast_direction_1} to \ref{fig:imem_com_ast_direction_4} in
Appendix~1, however, show that particles with different impact speeds approach the spacecraft from different 
directions
which is a consequence of the distributions of orbital elements for the asteroidal and cometary dust particles
used in the IMEM model  \citep{dikarev2004}: In the spacecraft frame of reference slow particles are much
more abundant than fast particles and one has to keep in mind that particles orbiting the Sun with higher 
eccentricity and
inclination have  higher impact speeds. Furthermore, due to the spacecraft motion around the Sun 
(Figure~\ref{fig:dest_traj}) 
the particle impact pattern significantly changes with time. Given 
that the flat plate sensor, and  much more so the narrow-angle DDA sensor, can detect only particles
from a very narrow range in impact directions, the sensor cuts out only a very limited fraction
of the total number of particles approaching the spacecraft, which are concentrated in the center
of the plots shown in Figures~\ref{fig:imem_com_ast_direction_1} to \ref{fig:imem_com_ast_direction_4}.

We also performed test runs with the IMEM model simulating dust fluxes on to the Earth. Due to the
rotational symmetry of the interplanetary dust cloud implemented in the model and the practically
circular Earth orbit with zero inclination, these simulations  did 
not show a temporal variation, neither for a spherical (4$\pi$) nor a flate plate (2$\pi$) 
sensor, as expected. The temporal modulation only appears in the simulation runs with the
\dpl\ trajectory.

\subsubsection{Dust Fluences}

\label{sec:IMEM_fluences}

In this subsection we give estimates for dust fluences detectable with 
 DDA during the \dpl\ mission derived from our IMEM simulations (Table~\ref{tab:imem_fluences}). A
flat plate sensor has been assumed for most of our simulations so far, i.e. the sensor has a FOV of
$\pm 90^{\circ}$,  and its sensitivity profile as a function of incidence angle is described by a cosine 
function. 
Note that DDA will have an angular sensitivity 
 considerably smaller than that of a 
flat plate (Figure~\ref{fig:dda_design}). 

\begin{table}[tb]
  \centering
      \caption{Particle detections predicted with IMEM for a mission duration of 1474~days and a sensor 
      normal vector pointing parallel to the spacecraft velocity vector. 
      Column~2 gives the average impact speed, 
      column~3 the average flux, 
      and columns~4 to 6 give the fluence of particles integrated over the entire mission. 
      Column~4 lists the fluence for a $\mathrm{1\,m^{2}}$ sensor, while columns~5 and 6 give 
      the fluence for a $\mathrm{0.035\,\mathrm{m^2}}$
      sensor, i.e two DDA sensor heads. Columns~4 and 5 give the fluence for $\pm 90^{\circ}$ FOV half-cone,  column~6  for the 
      DDA sensor  
      (Figure~\ref{fig:dda_design}).}
      \begin{tabular}{lccccc} 
          &&&&& \\
\hline
 \multicolumn{1}{c}{Population} & Average       & Average          &   \multicolumn{3}{c}{Fluence} \\  
                      &             speed       &  flux            &   \multicolumn{2}{c}{Flat plate ($\pm 90^{\circ}$)} & DDA \\ 
                      & [$\mathrm{km\,s^{-1}}$] & [$\mathrm{m^{-2}\,s^{-1}}$] & [$\mathrm{m^{-2}}$] & \multicolumn{2}{c}{$\mathrm{[0.035\,\mathrm{m^{-2}}}]\,\,\,\,\,\,\,\,\,\,$}\\
 \multicolumn{1}{c}{(1)} &          (2)         &    (3)           &     (4)         &    (5)       &     (6)           \\
       \hline
                                   &&&& \\[-2.2ex]
Asteroidal dust, pop.~2  &   7.8                & $5.4 \cdot 10^{-4}$ & $6.8 \cdot 10^4$    &  2380  &   75 \\
Cometary dust, pop.~4    & 15.7                 & $5.7 \cdot 10^{-3}$ & $7.3 \cdot 10^{5}$  &  25550 &   685 \\
Total interplanetary dust &  11.7               & $6.3 \cdot 10^{-3}$ & $8.0 \cdot 10^5$    &  27930 &   760 \\
         \hline
 \end{tabular}
   \label{tab:imem_fluences}
\end{table}

\begin{figure}[tb]
	\centering
	\vspace{-6.7cm}
		\includegraphics[width=0.85\textwidth]{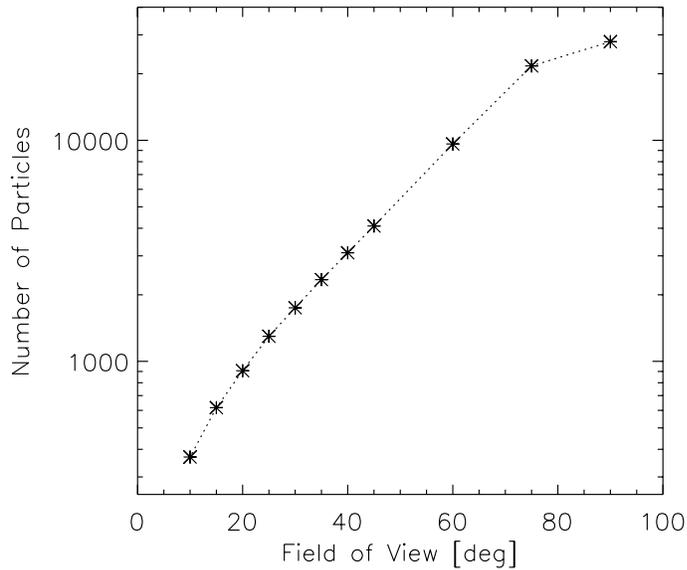}
		\vspace{-1.4cm}
			\caption{Fluence of interplanetary dust particles for 
			a flat plate sensor with $\mathrm{0.035\,\mathrm{m^2}}$ sensor area and varying sensor FOVs 
			from IMEM simulations. The DDA field-of-view corresponds to an angle of $\mathrm{17^{\circ}}$  
			in this diagram.
	  }
	\label{fig:imem_sensor}
\end{figure}

Next we study the effect of the  FOV on the dust fluences. To this end, we  
perform IMEM simulations assuming a flat plate sensor whose normal vector is 
oriented parallel to the spacecraft velocity vector as before. This time, however, we  
vary the  FOV. The result is shown in Figure~\ref{fig:imem_sensor}\footnote{In Figure~\ref{fig:imem_sensor}, to describe the sensitive area as a function of
incidence angle, we use a cosine function  which is cut off at the specific angle given on the x axis of that Figure. 
This leads to somewhat higher fluences than those derived with the DDA sensitivity profile for a 
comparable FOV}..
It is obvious 
that the dust fluences are significantly reduced for a narrower  FOV. 

Finally, we use the 
DDA sensitivity profile (Figure~\ref{fig:dda_design}) to simulate dust fluences,  
again with a sensor  pointing 
parallel to the spacecraft velocity vector.  The result is listed in Table~\ref{tab:imem_fluences},
column~6. 
The simulations predict a total number of approximately  760 interplanetary particles detectable with two DDA
sensor heads during the entire \dpl\ mission, about 90\%  of them being of cometary origin. Note that the sensor pointing was not optimised to maximise the dust fluences, see Section~\ref{sec_discussion}.

\subsection{Interstellar Dust}

\label{sec_imex}

\begin{figure}[tb]
	\centering
	\vspace{-2.8cm}
		\includegraphics[width=0.9\textwidth]{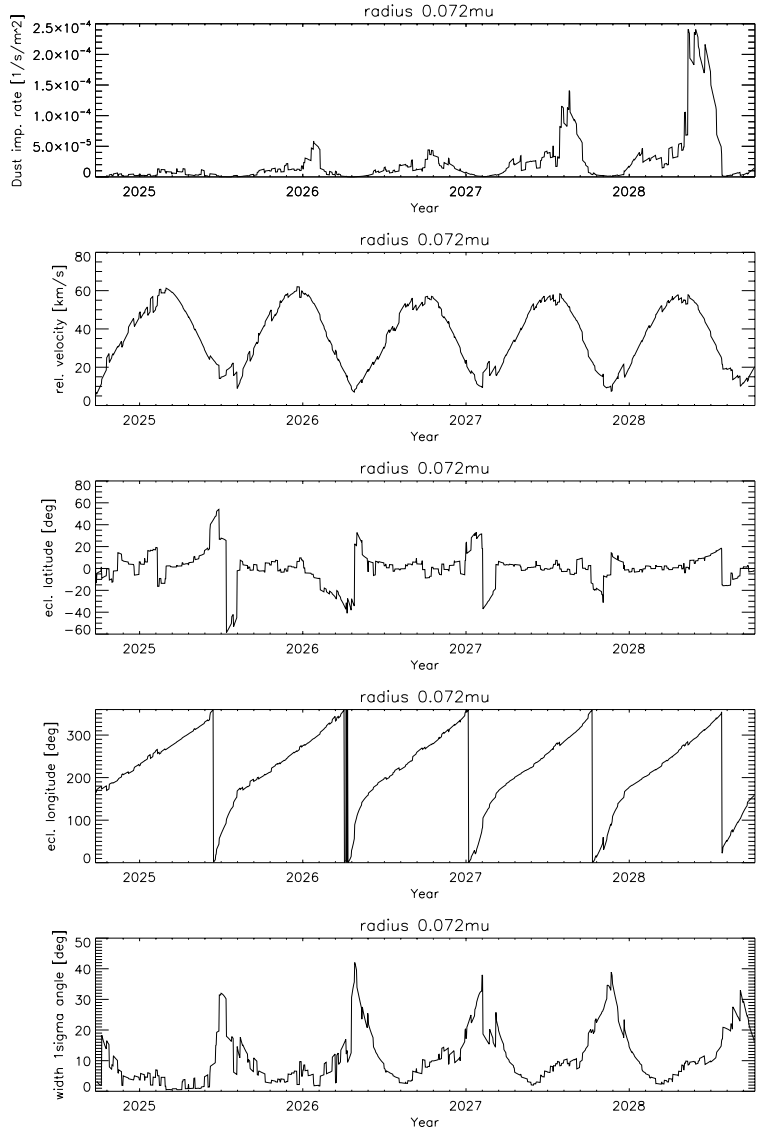}
		\vspace{-1.5cm}
	\caption{Simulated impact rate, and dynamical parameters for particles with 
	radius $r_d=\mathrm{0.072\,\mu m}$ in the spacecraft reference frame. From top 
	to bottom: impact rate, impact velocity, average approach direction of particles in ecliptic 
	latitude $\beta_{\mathrm{ecl}}$ and ecliptic longitude $\lambda_{\mathrm{ecl}}$,  and  the $1\,\sigma$ width of the 
	interstellar dust flow (note that for calculating impact rates 
	the model uses the DDA sensor profile oriented towards the effective interstellar 
	dust flow direction integrated over all particle size bins). 
	}
	\label{fig:imex_plots_2_1}
\end{figure}

\begin{figure}[tb]
	\centering
	\vspace{-1.8cm}
		\includegraphics[width=1.\textwidth]{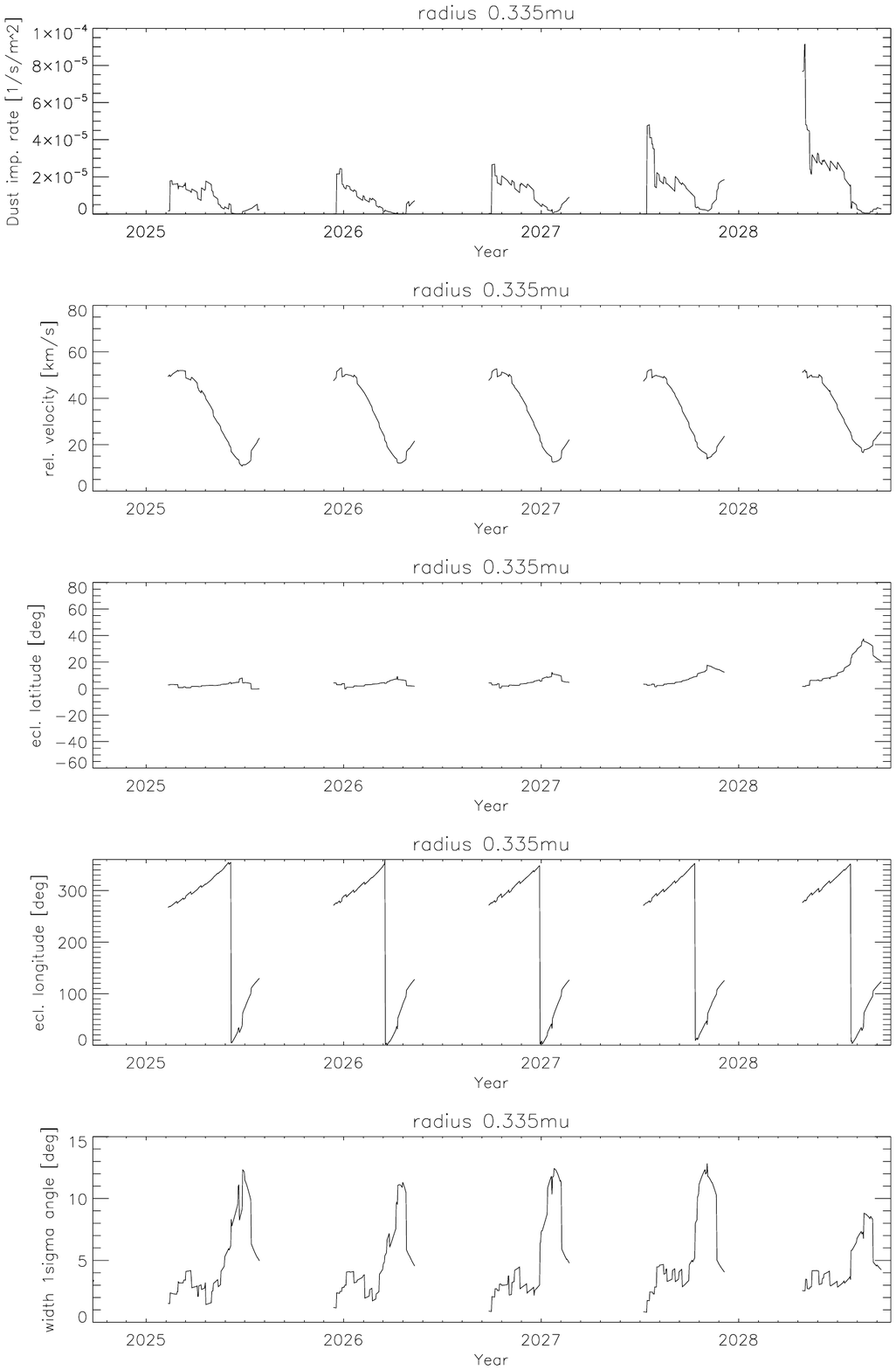}
		\vspace{-1.9cm}
	\caption{Same as Figure~\ref{fig:imex_plots_2_1} but for particle radius $r_d=\mathrm{0.335\,\mu m}$.
	 }
	\label{fig:imex_plots_2_2}
\end{figure}

\begin{figure}[tb]
	\centering
	\vspace{-1.8cm}
		\includegraphics[width=1.\textwidth]{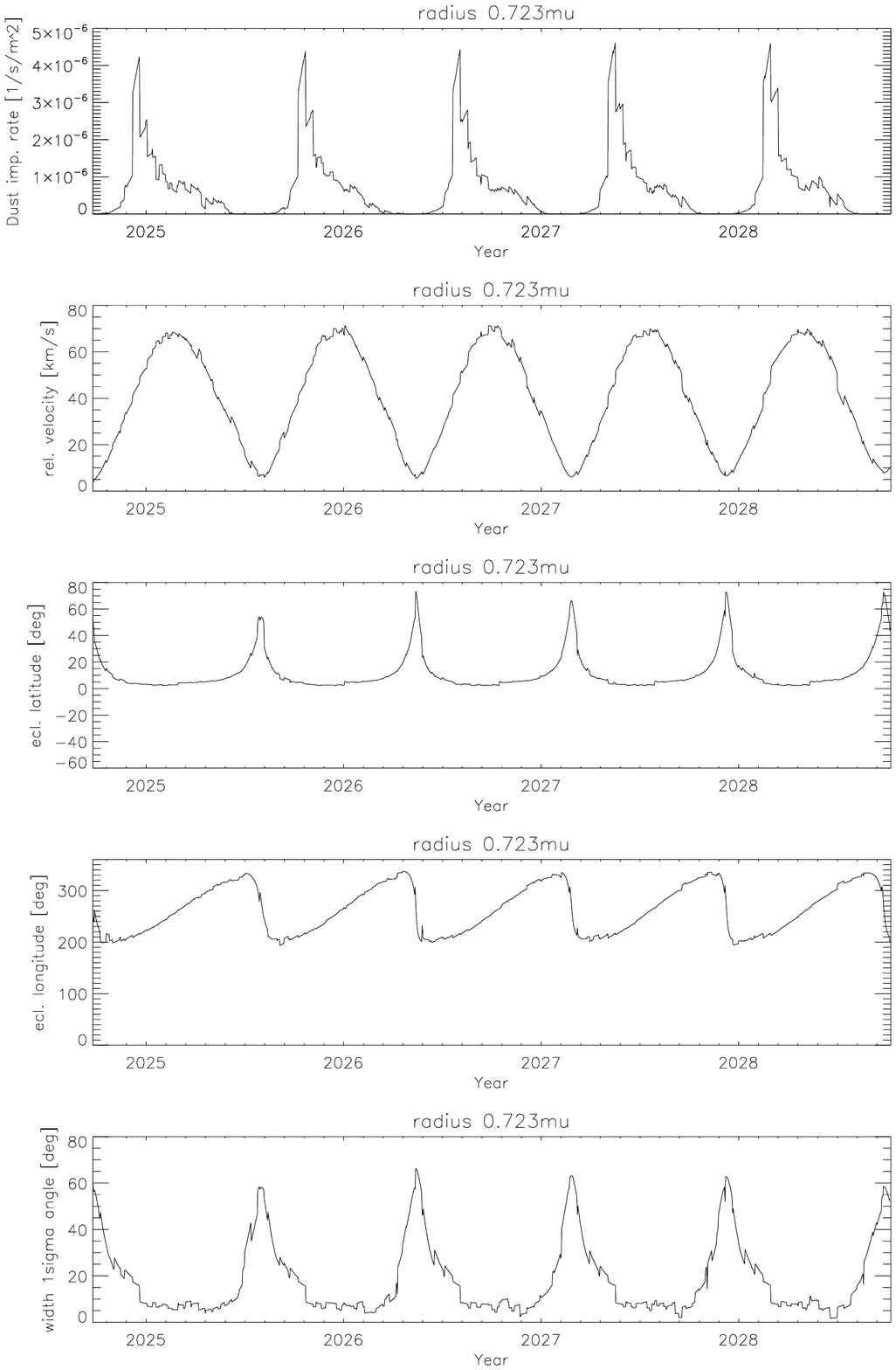}
		\vspace{-1.9cm}
	\caption{Same as Figure~\ref{fig:imex_plots_2_1} but for particle radius
	$r_d=\mathrm{0.723\,\mu m}$.
	}
	\label{fig:imex_plots_2_3}
\end{figure}

Previous simulations of interstellar dust in the solar system described  the interstellar dust flow
at larger heliocentric distances well, but they did not have the resolution to enable a good time-resolved 
description of the dust environment at Earth \citep{gruen1994a,landgraf2000b,sterken2012a}. 

Based on these earlier models and the dust measurements by the Ulysses spacecraft, 
\citet{strub2019}  executed high-resolution simulations in the context of the IMEX 
modelling effort (Interplanetary Meteoroid environment for EXploration) under ESA contract 
that included an 
interstellar dust module developed for this purpose. The authors simulated the dynamics of 
charged micrometer and sub-micrometer sized interstellar particles exposed to solar gravity, solar radiation pressure and a time-varying interplanetary magnetic field (IMF). 
The size distribution is 
represented by 12 particle radii 
between $\mathrm{0.049\,\mu m}$ and $\mathrm{4.9\,\mu m}$, and 
the dynamics of each of these sizes was simulated individually, assuming the adapted  $\beta$-curve 
for astronomical silicates \citep{sterken2012a}. In IMEX, the dust density in 
the solar system
is calibrated with the Ulysses interstellar dust measurements, again individually for each size bin 
\citep{strub2015}. Due to the variable IMF, the IMEX model
is time-dependent, contrary to  IMEM  (Section~\ref{sec_imem}). For details of  IMEX  the reader is
referred to \citet{strub2019}.
We use  IMEX  to simulate the time-resolved flux and dynamics of  interstellar dust particles 
in the inner solar system, assuming the DDA sensor profile shown in Figure~\ref{fig:dda_design}
with the sensor being oriented towards the effective approach direction of the interstellar particles
integrated over all dust size bins.

The IMEX model uses the same initial conditions as \citet{landgraf2000b} and \citet{sterken2012a,sterken2013a}: 
The simulated interstellar particles enter the solar system at a uniform direction and velocity, with an initial 
velocity of $\mathrm{v_{\infty} = 26\,km s^{-1}}$ and an inflow direction from an ecliptic longitude 
$\lambda_{\mathrm{ecl}} = 259^{\circ}$ and ecliptic latitude $\beta_{\mathrm{ecl}} = 8^{\circ}$ 
\footnote{Working values of a speed of $\mathrm{25.4 km\,s^{-1}}$ with directions from $255.7^{\circ}$ ecliptic 
longitude and +5.1$^\circ$ ecliptic latitude were recently suggested from Energetic Neutral Atom measurements by the Interstellar Boundary EXplorer mission (IBEX) by~\citet{mccomas2015b} and  confirmed by \citet{swaczyna2018}. 
These values agree with the speed 
of $24.5^{+1.1}_{-1.2}\,$\kms, the ecliptic longitude of $252 \pm5^{\circ}$ and the ecliptic latitude of 
$5\pm 5^{\circ}$ for the initial condition of the interstellar dust stream found by \citet{kimura2003a}.}. 
This is compatible with the inflow direction of 
the neutral gas into the solar system \citep{witte1996,lallement2014,wood2015}, and it is also compatible 
with the Ulysses measurements of the interstellar dust flow \citep{frisch1999a,strub2015,kimura2003b,kimura2003a}.
It is equivalent to the interstellar particles being at rest with respect to the local interstellar
cloud surrounding our solar system. In the following we will call this direction the nominal interstellar 
dust flow direction.

Measurements of interstellar dust inside the planetary system now provide a new
window for the study of solid interstellar matter at our doorstep \citep{frisch1999a}. 
The flow of the interstellar particles in the heliosphere is governed by two fundamental effects: 
(1) the combined gravitational  and  radiation pressure force of the Sun, and (2) the Lorentz force 
acting on
 a charged particle moving through the solar magnetic field ''frozen`` into the solar wind (the IMF). The former effect 
can be described as a multiplication of the gravitational force by a constant factor $(1 - \beta$), where the radiation pressure factor $\beta = |{\bf F}_{rad}|/|{\bf F}_{grav}|$ is a function of particle composition, size and morphology. Interstellar particles approach the Sun on hyperbolic trajectories, leading to either a radially symmetric focussing ($\beta  < 1$) or defocussing ($\beta > 1$) downstream of the Sun which is constant in time \citep{bertaux1976,landgraf2000b,sterken2012a}. Particle sizes observed by the Ulysses dust detector typically range from approximately $\mathrm{0.1\,\mu m}$ to several micrometers, corresponding to $\mathrm{0 \lesssim \beta \lesssim 1.9}$~\citep{kimura2003b,landgraf1999a}\footnote{\citet{landgraf1999a} found a range of  $1.4 < \beta < 1.8$  from Ulysses measurements, and~\citet{kimura2003b} found values for $\beta$ between 0 and 1.9.}.
A detailed description of the forces acting on the particles and the resulting general interstellar dust flow characteristics was given by \citet{sterken2012a}. 

The interplanetary magnetic field (IMF) shows systematic variations with time, including the 25-day 
solar rotation 
and the 22-year solar magnetic cycle, as well as local deviations due to disturbances in the 
interplanetary magnetic field, due to, e.g. coronal mass ejections (CMEs). The dust particles in interplanetary 
space are typically 
charged to an 
equilibrium potential of +5~V \citep{mukai1981,kimura1998,kempf2004}.  Small particles have a 
higher charge-to-mass ratio, hence their dynamics is more sensitive to the interplanetary magnetic 
field. The major effect of the magnetic field on the charged interstellar dust  is a 
focussing and defocussing relative to the solar equatorial plane  with the 22-year 
magnetic cycle of the Sun \citep{landgraf2000b,landgraf2003,sterken2012a,sterken2013a}. 
Modifications of the particle dynamics by solar radiation pressure and the Lorentz force acting on 
 charged dust particles
have to be taken into account for a proper interpolation of the interstellar dust properties
to the interstellar medium outside the heliosphere where these particles originate from  \citep{slavin2012}.

In Figures~\ref{fig:imex_plots_2_1} to \ref{fig:imex_plots_2_3} we show the temporal variations
of the dynamical parameters for  particles in three representative size bins. Along the \dpl\ trajectory the dust 
spatial density and dynamical parameters of  
the interstellar particles  depend on spacecraft position, time in the solar (Hale) cycle 
and particle size.  At this distance from the Sun, 
each size range is dominated by a different force   \citep{landgraf1998a,sterken2012a,strub2019}: 
Electromagnetic interaction ($r_d=\mathrm{0.072\,\mu m}$),
radiation pressure ($\mathrm{0.335\,\mu m}$), and solar
gravity ($\mathrm{0.492\,\mu m}$).  
Dust spatial densities in the inner solar system for these particle sizes during the \dpl\ mission 
are shown in Figures~\ref{fig:imex_3d_3} to \ref{fig:imex_3d_1} in Appendix~2.

Strong modulations of the dust impact rate over time are obvious in 
Figures~\ref{fig:imex_plots_2_1} to \ref{fig:imex_plots_2_3}. 
To first order,  
maxima and minima are caused by the approximately annual periodicity of the spacecraft motion 
around the Sun, and by the varying particle impact speed  
(see Figure~\ref{fig:dest_traj}): When the spacecraft moves against the 
dust flow (approximately in quadrants II and III) the impact 
speed reaches up to 
$60\,\mathrm{km\,s^{-1}}$, while at other times  it is close to zero 
(at $\mathrm{Y \approx 0}$ in quadrants I and IV). Fluxes and impact speeds are highly correlated, 
high fluxes coincide with high impact speeds. 

In addition to this modulation by the spacecraft motion, size-dependent forces acting
on the particles lead to  further alterations as described above. Particles with 
$r_d \lesssim \mathrm{0.1\,\mu m}$  strongly interact with the IMF 
(Figure~\ref{fig:imex_plots_2_1}, top panel). Therefore, the phase of the 22-year IMF cycle 
strongly affects their spatial density and flow:
In 2022 the overall configuration of the IMF will change from a defocussing to a focussing 
configuration, and it is expected to reach its maximum focussing condition approximately in 2031. 
It leads to an overall increase in the spatial density of these small particles 
in the inner solar system and, hence, to an increase in the impact rate during the \dpl\ mission, 
in addition to the approximately annual modulation caused by the spacecraft motion alone. 
This is evident by the increase in the dust density seen in the left column of 
Figure~\ref{fig:imex_3d_1} in the Appendix. Even under  optimal focussing conditions of the IMF, strong 
filtering of the heliosphere remains effective for these particles, and the flux at Earth's orbit is
reduced by orders of magnitude with respect to the unfiltered flux outside the heliosphere
\citep{landgraf2000a,krueger2015a}.

Interstellar particles with a ratio of solar radiation pressure over gravity  $\beta > 1.4$ 
cannot be observed at Earth orbit because the solar radiation 
pressure prevents them from entering the inner solar system (the avoidance cone due to 
radiation pressure filtering is seen in the left and middle columns of Figure~\ref{fig:imex_3d_2}
in  Appendix~2).  
We use the same $\beta$-curve as \citet{sterken2013a} which was adapted from the 
curve for astronomical silicates given in \citet{gustafson1994}, by scaling it to 
a maximum value $\beta_{\mathrm{max}} \simeq 1.6$, in agreement with the range 
$ 1.4 \lesssim \beta_{\mathrm{max}} \lesssim 1.8$  measured by Ulysses 
\citep{landgraf1999a}.
For this assumption, particles 
with sizes $\mathrm{0.1\,\mu m} \lesssim r_d \lesssim \mathrm{0.3\,\mu m}$ have $\beta > 1.4$ 
and are absent at Earth orbit. In the simulations, this applies to 
particles in the size bins $r_d=\mathrm{0.156\,\mu m}$ and $\mathrm{0.229\,\mu m}$, and partially to
particles with $r_d=\mathrm{0.106\,\mu m}$ which are detectable only during  short periods of time. 
On the other 
hand, particles with $r_d\gtrsim \mathrm{0.335\,\mu m}$ can enter the inner solar system and
are detectable by DDA. 

Finally, the dynamics of particles with $r_d \gtrsim \mathrm{0.5\,\mu m}$ are dominated by solar gravity. 
For these particles the modulation in
the  impact rate is due to the varying impact speed and gravitational focussing in  
the downstream region of the interstellar dust stream behind the Sun. Strong enhancements  
 in the impact rate in these regions are shown in Figure~\ref{fig:imex_plots_2_3}.
For a more detailed discussion  of the particle dynamics at 1~AU heliocentric distance
 see \citet{strub2019}.

\begin{figure}[tb]
	\centering
	\vspace{-9.3cm}
		\includegraphics[width=\textwidth]{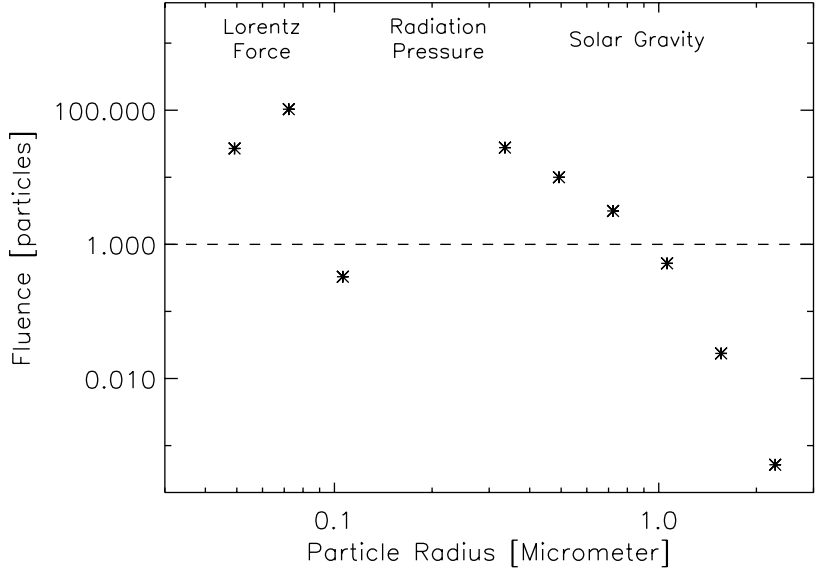}
		\vspace{-1.9cm}
	\caption{Fluence of interstellar particles during the 1474~days of the \dpl\ mission
	for a sensor area of $\mathrm{0.035\,\mathrm{m^{-2}}}$  with DDA pointing in the 
	effective interstellar dust flow direction in the spacecraft based reference frame. The horizontal dashed line indicates
	the limit of one particle impact detectable during the entire \dpl\ mission. The approximate size regimes 
	where the different forces dominate the dynamics and, thus, the spatial dust densities are indicated at the top.
	  }
	\label{fig:imex_fluence}
\end{figure}

In Figures~\ref{fig:imex_plots_2_1} to \ref{fig:imex_plots_2_3} the third and the forth panels 
show the deviation of the average particle impact direction from the nominal direction of the interstellar 
dust flow. Gradual shifts occur in ecliptic longitude for all three particle sizes. These shifts 
are more or less coincident in time so that, on average, all particles with all three sizes approach
from approximately the same direction. Thus, with a sufficiently large FOV, 
DDA may detect all particle sizes simultaneously.

The bottom panel in Figures~\ref{fig:imex_plots_2_1} to \ref{fig:imex_plots_2_3} 
shows the $1\sigma$ width of the interstellar dust stream. During periods of highest particle 
impact speeds the interstellar dust stream is narrowly collimated to within less then $10^{\circ}$, while
during periods of low speeds the stream width can reach $30^{\circ}$, even and up to $60^{\circ}$ 
for the $\mathrm{0.723\,\mu m}$ particles.

The fluence of  interstellar dust particles during the
entire \dpl\ mission is shown in  Figure~\ref{fig:imex_fluence}. The gap in the size range 
$ \mathrm{0.1\,\mu m} \lesssim r_d \lesssim \mathrm{0.3\,\mu m}$
is due to the radiation pressure filtering in the inner heliosphere, and the drop in the smallest  
size bin with $\mathrm{r_d = 0.049 \,\mathrm{\mu m}}$ is caused by electromagnetic filtering,  consistent 
with the Ulysses dust measurements between 3 and 5~AU \citep{landgraf2000a,krueger2015a}. 
Our simulations predict a total 
number of approximately 170 interstellar particles detectable with two DDA 
 sensor heads having a total sensitive area of 
$\mathrm{0.035\,\mathrm{m^{2}}}$ during the total measurement period of 1474~days. 
For particles with $r_d \gtrsim \mathrm{1\,\mu m}$ the 
predicted fluence is  below one particle impact during the entire \dpl\ mission,  and we
therefore  do not consider such relatively big particles here. 

\section{Discussion}

\label{sec_discussion}

A prerequisite for obtaining the dust fluences given in Section~\ref{sec_idp} is that DDA 
continuously measures these dust populations. In a real mission scenario, having a dust instrument with a 
restricted field-of-view, the instrument pointing  has to be optimised for each 
dust population individually, i.e. cometary, asteroidal, and interstellar dust. This implies
that lower dust fluences will likely be achieved in reality, unless more than one
population can be detected simultaneously with the same instrument pointing. This will be the 
case with DDA during some mission periods because the range in impact directions of the interplanetary 
impactors is much wider than that of interstellar particles. 
Given the variability of the expected dust fluxes, 
 the measurement periods and instrument pointing scenarios have to be optimised  in order to 
 maximise the overall number of measured dust particles for all populations. 
 
Our simulations  assumed a sensor pointing parallel to the spacecraft 
 speed vector in the case of interplanetary dust (IMEM) and towards the average interstellar
 dust inflow direction in the spacecraft reference frame (IMEX), respectively. 
An optimised pointing scenario, for example performing scans through the 
dust approach directions expected for different populations 
in order  to derive their relative abundances, will
 increase the  number of interplanetary particle  detections.
 More realistic predictions for dust fluences 
 measurable with DDA  require
 a detailed scenario for the  space\-craft orientation during the \dpl\ mission.

\subsection{Interplanetary Dust}

IMEM was designed as a tool to predict hazards imposed by dust particles on to Earth orbiting and 
interplanetary spacecraft. Sub-micrometer sized particles which are most susceptible to
radiation pressure usually impose a negligible threat to 
spacecraft structures. Therefore, the radiation pressure force was not included in IMEM when
the model was designed. This leads to  uncertainties in the impact directions
of dust particles with a high ratio of radiation pressure force over gravity, $\beta$. Their range 
in impact directions is likely much wider than predicted by our simulations so that the 
measurable fluxes of sub-micrometer sized particles could be lower than predicted 
by the IMEM model.  

The cometary populations implemented in IMEM are limited to Jupiter Family Comets (JFCs), thus our flux computations
are a conservative lower estimate. Dust particles released by Halley-type comets (HTCs) or Oort Cloud type
comets (OCCs) can produce particles on heliocentric retrograde orbits. The abundance of  retrograde
particles around 1~AU can be as high as 10\% of the abundance of prograde particles 
\citep{nesvorny2010,pokorny2014}
for an impact velocity of about 3 times the values derived for the JFCs. Hence, the flux of particles of cometary
origin along the trajectory of the spacecraft could be up to 30\% higher than computed with IMEM.

Figure~\ref{fig:imem_dda} (top panel) shows that a distinction between asteroidal and cometary dust can be 
accomplished from  the particle impact speed: asteroidal dust has average  speeds below
10\,\kms, while the average  speed of cometary particles exceeds 30\,\kms. Inspection of the bottom 
panel of Figure~\ref{fig:imem_dda} reveals that time intervals with the highest fluxes of asteroidal
dust approximately coincide with periods of increased interstellar dust flux (Figures~\ref{fig:imex_plots_2_1} 
to \ref{fig:imex_plots_2_3}). Given that the impact speeds of interstellar particles exceed 40\,\kms\ in these
time intervals,
a distinction between asteroidal and interstellar particles will be possible from the impact speed 
and  particle sizes 
during these periods. On the other hand, interstellar and cometary particles have comparable 
impact speeds and they have to be distinguished preferentially from the particle composition. 
Considering the  number of asteroidal particle detections expected from the 
IMEM simulations, DDA should be oriented towards the asteroidal particles during these time intervals 
(spikes in Figure~\ref{fig:imem_dda} bottom panel, $Y \approx 0$ in quadrant II in Figure~1).

On the other hand, cometary particles will be detectable with a relatively constant flux throughout the
mission. They should preferentially be measured during time intervals when the expected fluxes of 
interstellar particles and of asteroidal particles are low (quadrants I and IV in Figure~1). 
Based on the expected accuracy of the particle trajectory measurement with DDA, the measurements will 
allow us to constrain the source body from a backward tracing of the particle trajectory \citep{hillier2007}. 
 Together with improved modelling of the particle dynamics, we will
be able to derive the abundance of asteroidal and cometary particles and, hence, the contributions of each 
of these dust sources to  the zodiacal cloud.

\subsection{Interstellar Dust}

The Ulysses interstellar dust data set was chosen as the calibration dataset for IMEX because it contains the most 
comprehensive and homogeneous measurements by a single instrument over a period of 16 years, covering a 
large portion of the 22-year solar cycle. With a total of more than 900 identified interstellar particles 
it  has by far the largest dataset of all interstellar dust measurements performed to date
\citep{krueger2010b,krueger2015a,krueger2019b}. 

 Concerning the normalisation of the simulated fluxes, the temporal variability of the  
  flux  and of the flow  direction of the interstellar particles in the Ulysses  dataset are not entirely 
  reproduced by the model. Therefore, only the overall flux for each particle size bin was taken 
  into account for the normalisation, and each  bin was calibrated individually. For most of the 
  Ulysses measurement intervals the 
  model reproduces the dataset within a factor of 2 \citep{krueger2019b}, 
  only in 2005 is the discrepancy  more pronounced
when a rapid change in interstellar dust flow direction and density was observed 
\citep{krueger2007b}. The reason for this shortcoming remains an open question 
at the moment. It may be related to the material properties of the interstellar particles 
(e.g. composition and porosity), variable particle charging or the particle 
interaction with the heliospheric boundary \citep{sterken2015}, or it may be due to changes in the configuration 
of the heliospheric current sheet which are not taken into account 
in the present model. This likely marks the limits of our 
current understanding of the interstellar dust flow through the heliosphere.

Variations in the impact direction and the width of the interstellar  stream 
were measured with Ulysses between 3~AU  and 5~AU heliocentric distance \citep{strub2015}. 
The authors separated the
data set into two subsets, one with particles smaller than about $0.24\,\mathrm{\mu m}$, 
and the other one with larger particles. Their analysis showed that most of the time 
the average impact direction of the larger particles remained within approximately $\pm 20^{\circ}$
of the undisturbed interstellar dust flow direction, 
while the directions of the smaller particles frequently deviated by up to $60^{\circ}$, sometimes 
 exceeding $90^{\circ}$ \citep[][their Tables~4 and 5]{strub2015}. The stream widening for the
 large particles remained below $10^{\circ}$ most of the time, while that of the small particles
 usually stayed below $30^{\circ}$. It indicates that the  interstellar dust stream is rather  
 collimated, consistent with our modelling results for \dpl\ (Section~\ref{sec_imex}).
 
We also compared the IMEX model predictions to interstellar dust flux measurements from other missions, 
i.e. Helios, Cassini, and Galileo \citep{krueger2019b}. Despite different heliocentric distance ranges 
covered by these missions and different detection geometries of the instruments, the model predictions 
(based on a calibration using Ulysses data) agree with the measured fluxes to within about a factor 
of 2 to 3.  Typically, 
the model underestimates the measured dust fluxes.
Because of this, and the fact that the interstellar dust stream
is rather  collimated, the dust fluxes predicted for \dpl\ by the 
 IMEX model should also be realistic to within a factor of 2, with a tendency to underestimate the
 true fluxes. The largest uncertainties 
 arise for the small particles because 
 they are most strongly affected by the  heliospheric filtering \citep{landgraf2000a}. Our 
 present model assumes an undisturbed heliospheric current sheet which is a good approximation for the
  IMF during solar minimum conditions, while at solar maximum Coronal Mass Ejections (CMEs) can 
 significantly disturb the IMF,  preferentially affecting the dynamics of small particles. 

When Cassini was in orbit around Saturn, the CDA instrument also measured interstellar particles 
for limited periods of time \citep{altobelli2016}. Only relatively small  particles with masses below 
approximately $5\cdot10^{-16}\,\mathrm{kg}$ (corresponding to a particle radius  
 $r_d \approx 0.35\,\mathrm{\mu m}$) could be measured because of the limited instrument sensitivity
 (i.e. instrument saturation for bigger particles). 
The measured mass spectra show  a depletion of carbon,  indicating
that organic constituents may be rare or even absent in these particles. A carbon depletion
in   dust in the local interstellar cloud (LIC) was suggested 
from  derived gas-mass abundances  \citep{slavin2008}. Loss of carbon from
the dust  may  occur due to particle destruction by shock waves in the LIC
 \citep{kimura2015}.

The Stardust mission revealed  seven interstellar particles which are diverse in 
elemental composition, crystal structure, and size. The presence of 
crystalline grains and multiple iron-bearing phases, including sulfide, in some particles 
indicates that individual interstellar particles diverge from any one representative model
of interstellar dust inferred from astronomical observations and theory  \citep{westphal2014b}. 
The Stardust particles also showed 
that interstellar dust with mass $\mathrm{3 \cdot 10^{-15}\, kg}$ might 
be  porous and has higher $\beta$ and charge-to-mass ratios \citep{sterken2014}.
Interstellar dust with mass exceeding $\mathrm{5 \cdot 10^{-16}\, kg}$ might be  porous 
aggregates of submicron-sized silicate grains \citep{sterken2015,kimura2017}. 
Silicate grains do not stick to each other 
in the interstellar medium, but organic matter would assist them in sticking, 
if their surfaces are covered by organic matter. Therefore, submicrometer-sized grains
 in  porous aggregates might still retain organic matter.
 
 Micrometer-sized porous particles generally have higher charge-to-mass ratios \citep{ma2013} and
higher $\beta$ values \citep{kimura1999a} than 
compact particles of the same mass. 
With DDA we will be able to measure the electrical charge 
and mass of interstellar particles in much the same way as was successfully done for 
interplanetary dust particles with Cassini CDA \citep{kempf2004}. Hence, these 
parameters together with the measured dust spatial densities and  
dynamical modelling will better constrain the particle porosities in the future.

Gravitational focussing deflects and concentrates  particles whose dynamics are 
dominated by the gravitational field of a celestial body. In the case of the interstellar dust 
stream in the solar system, particles with  $r_d \gtrsim 0.5 \,\mathrm{\mu m}$ are concentrated 
in the downstream direction behind the Sun (Figure~\ref{fig:imex_3d_3};  $X \gtrapprox 0$ in quadrant I in 
Figure~\ref{fig:dest_traj}). The 
interstellar dust flow is inclined 
by $8^{\circ}$ with respect to the ecliptic plane, so that \dpl\ will not
traverse the region with the highest dust density (see the middle panel of 
Figure~\ref{fig:imex_3d_3}).

Even though an increased dust impact rate is expected in this region
(the spikes in Figure~\ref{fig:imex_plots_2_3}, top panel), the detection of only 
14 bigger interstellar particles is predicted in the size range 
$\mathrm{0.5\,\mu m}\lesssim r_d \lesssim \mathrm{1.0\,\mu m}$  during the entire 
\dpl\ mission. This number 
takes into account that the spatial density of such particles is enhanced in the region 
downstream of the Sun, due to  focussing by solar gravity 
(see Figure~\ref{fig:imex_3d_3} in  Appendix~2, and the spikes in the top panel of  
Figure~\ref{fig:imex_plots_2_3}). Hence, the DDA pointing should be 
optimised for the detection of such big particles preferentially during the time intervals
when DDA will traverse this region (at $X \gtrapprox 0$ in quadrant I in 
Figure~\ref{fig:dest_traj}). 
The gravitational focussing   increases the  impact speed to about
40 to $50\,\mathrm{km\,s^{-1}}$ in this  region (Figure~\ref{fig:imex_plots_2_3},
second panel from top),  however, these high speeds are expected to restrict the 
detectability of organic compounds in the DDA impact spectra because complex organic 
molecules are mostly destroyed \citep{khawaja2016}.  Another limitation is
imposed by the DDA sensor itself: In order to avoid abundant noise events in the data
set, the angle between the instrument boresight and the Sun direction has to exceed 
$90^{\circ}$ according to the present instrument design, 
 restricting the detectability of interstellar dust in this spatial region downstream of the
 Sun. 

A comparison of Figures~\ref{fig:imex_3d_3} and \ref{fig:imex_3d_2} in  
Appendix~2 shows that
the spatial distribution of the intermediate sized  particles with 
$r_d = 0.335\,\mathrm{\mu m}$ is completely different from  that of the larger 
$r_d = 0.723\,\mathrm{\mu m}$ particles: While the larger particles show a concentration
in the region downstream of the Sun due to gravitational focussing, there is a 
deficiency of particles in this spatial region in the intermediate size particles due
to the filtering by the radiation pressure. The size of this avoidance region 
depends on $\beta$ (and, hence, particle size and optical properties) and in three dimensions it has the 
approximate shape of a paraboloid. DDA can detect approximately  28
interstellar particles in this intermediate 
size range throughout quadrants III and IV in Figure~\ref{fig:dest_traj} during the
entire mission. In quadrants I and II these particles are undetectable. A temporal
variability is also evident in Figure~\ref{fig:imex_3d_2} with an increase in particle 
density at the boundary of the paraboloid between 2025 and 2029 due to
the heliospheric filtering of the IMF which also affects the intermediate sized particles. 

Finally, the spatial density of the smallest particles in our simulations having 
$r_d = 0.072\,\mathrm{\mu m}$ shows a strong temporal variation with an overall
increase  from 2025 to 2029. In this
time interval the  
heliosphere gradually switches from its defocussing to its focussing configuration, 
leading to  dust densities in the inner solar system increasing with time. Similar to the intermediate 
sized particles, these small
particles are preferentially detectable in quadrants III and IV. The maximum focussing 
configuration is
expected approximately in 2031, afterwards the IMF will become defocussing 
again \citep{strub2019}. Hence,   DDA measurements of such small 
interstellar particles should be concentrated  towards the second half
of the presently planned \dpl\ mission. The dust spatial density in the inner solar 
system still increases  after 
the presently planned end of the mission (cf. Table~\ref{tab:schedule}).

\section{Conclusions}

\label{sec_conclusions}

We used two up-to-date computer models which are readily available to investigate the dynamics of 
interplanetary and interstellar dust particles in the inner heliosphere, namely IMEM 
developed by \citet{dikarev2004,dikarev2005a,dikarev2005b}, and IMEX developed by 
\citet{sterken2012a,sterken2013a} and \citet{strub2019}, which is based on the work of 
\citet{landgraf2000b}. 
We studied the detection conditions for such particles with a 
Dust Analyser (DDA) on board the \dpl\  mission to the active asteroid (3200) Phaethon.
The mission is presently under development by the Japanese space agency JAXA/ISAS.  
The dust detection conditions during the Phaethon flyby were not the subject of this paper. 
Our results can be summarised as follows:

\begin{itemize}
\item The dust flux, average impact speed and impact direction of interplanetary and 
interstellar dust particles on to DDA are
strongly variable in time. The modulation is largely due to the 
spacecraft motion around the Sun, but also due to size-dependent forces acting on the particles,
leading to particle size-dependent variations in dust spatial density.
\item A statistically significant number of interplanetary and interstellar   dust particles 
can be  detected and analysed in-situ with DDA during the interplanetary voyage
of \dpl\, which is presently foreseen to last four years. 
\item During long mission periods the particle impact direction and speed can be used to discriminate
between interstellar and interplanetary particles and likely also to distinguish between
cometary and asteroidal  particles.  
\item The average approach direction of small interstellar particles ($\lesssim \mathrm{0.3\,\mu m}$)  
 is rather independent of particle size. 
\item Larger interstellar particles which are dominated by gravity can be preferentially detected in the
focussing region downstream of the Sun. 
\end{itemize}

\section*{Acknowledgements}

HK gratefully acknowledges support from the Japan Society for the Promotion of Science (JSPS grant S16740) 
for a research visit at the Planetary Exploration Research Center at Chiba Institute of Technology during
which a significant part of the work presented in this paper was performed.  
The IMEM and IMEX models were developed under ESA funding (contracts 21928/08/NL/AT and 4000106316/12/NL/AF - IMEX).
Support by Deutsches 
Zentrum f\"ur Luft- und Raumfahrt (DLR, grant 50OO1801), by JAXA (grant UA18-DU2-03-U1000), 
and by MPS is also gratefully acknowledged.
We thank
Valery Dikarev, Alexey Mints and Rachel Soja for valuable discussions about the IMEM model. 
 We are grateful to two
anonymous referees whose comments substantially improved the presentation of our results.


\clearpage

\section*{Appendix 1}

\begin{figure}[th]
	\centering
		\includegraphics[width=0.49\textwidth]{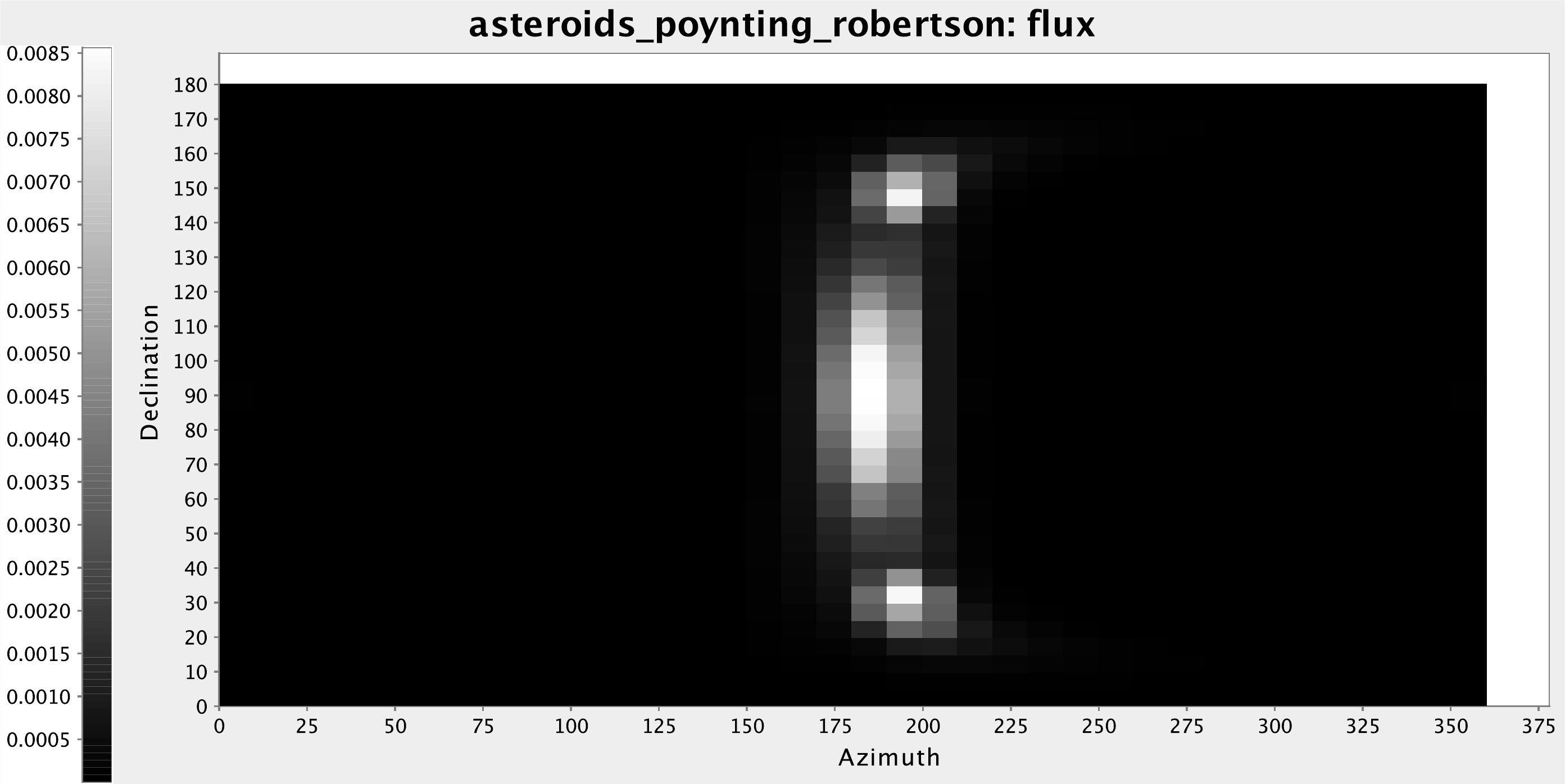}
		\includegraphics[width=0.49\textwidth]{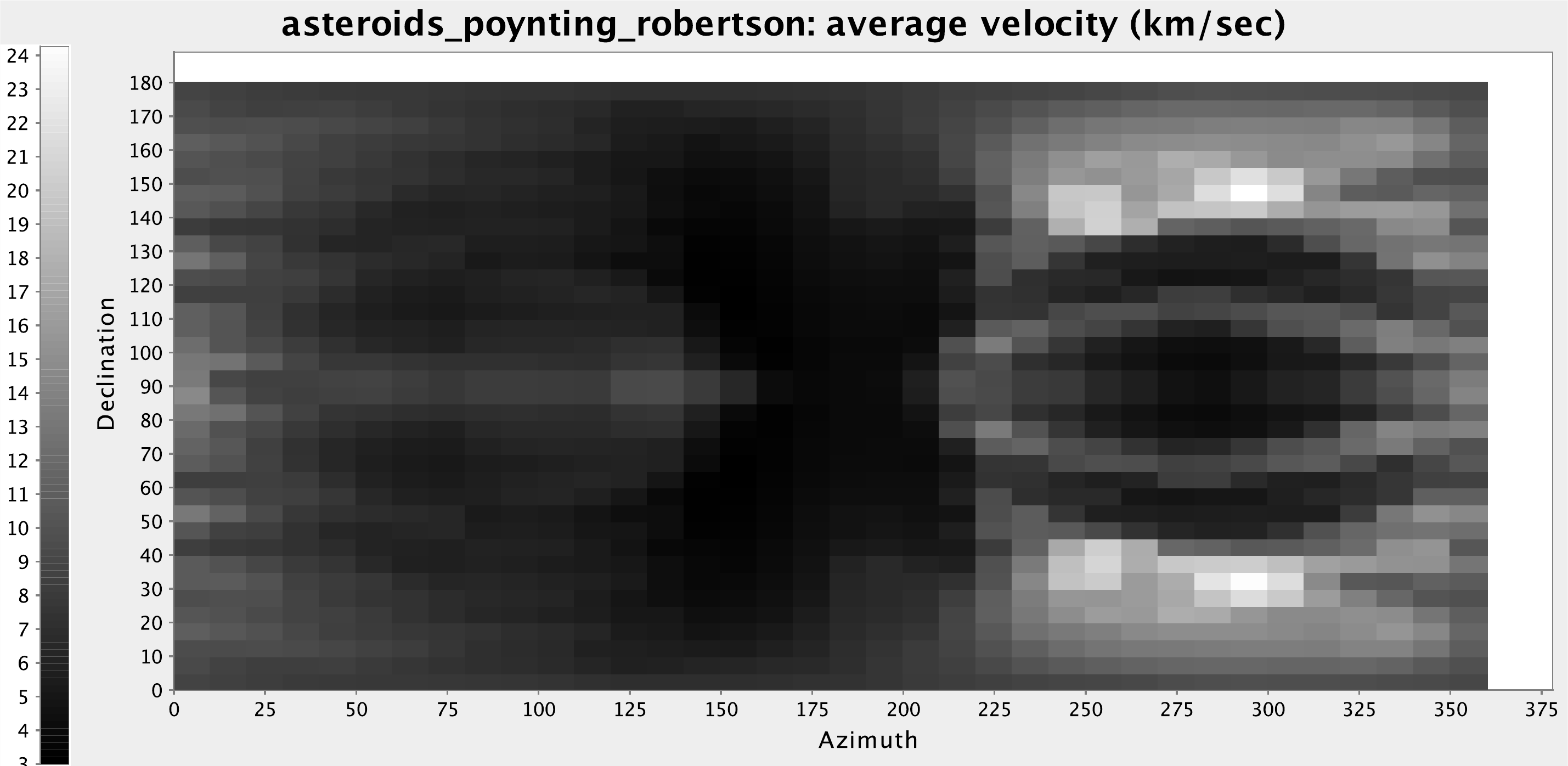}
		\includegraphics[width=0.49\textwidth]{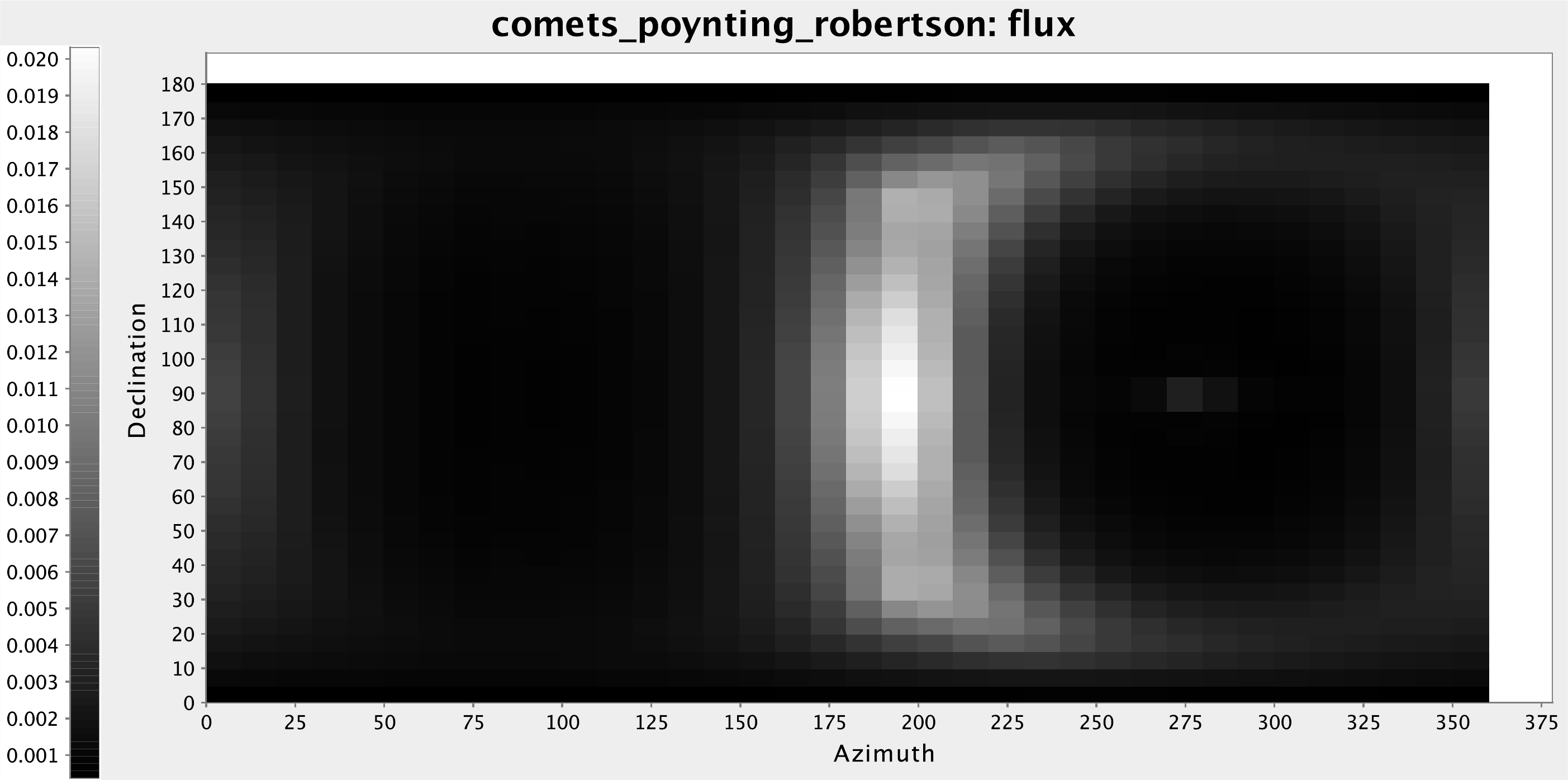}
		\includegraphics[width=0.49\textwidth]{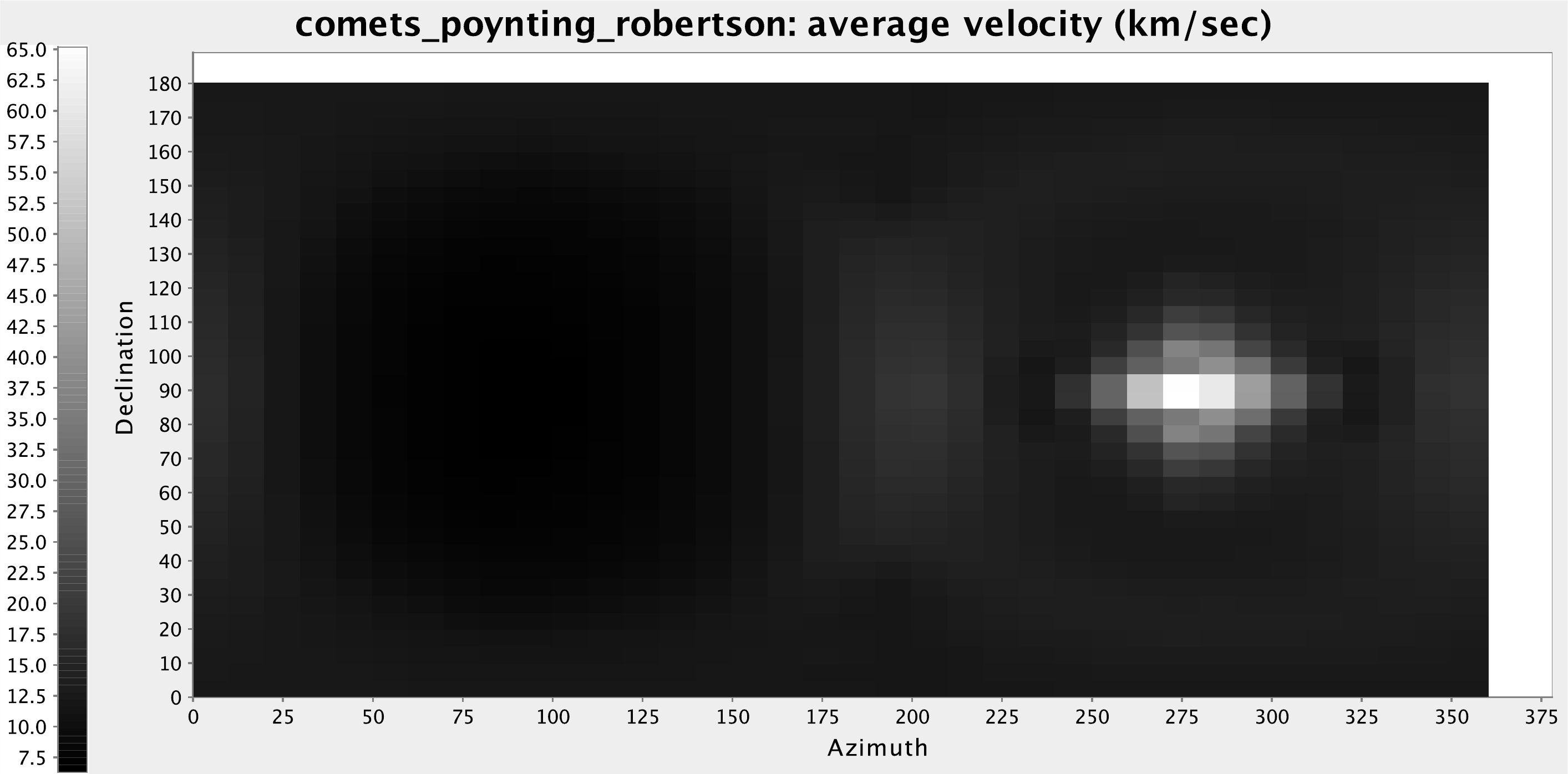}	
		\includegraphics[width=0.49\textwidth]{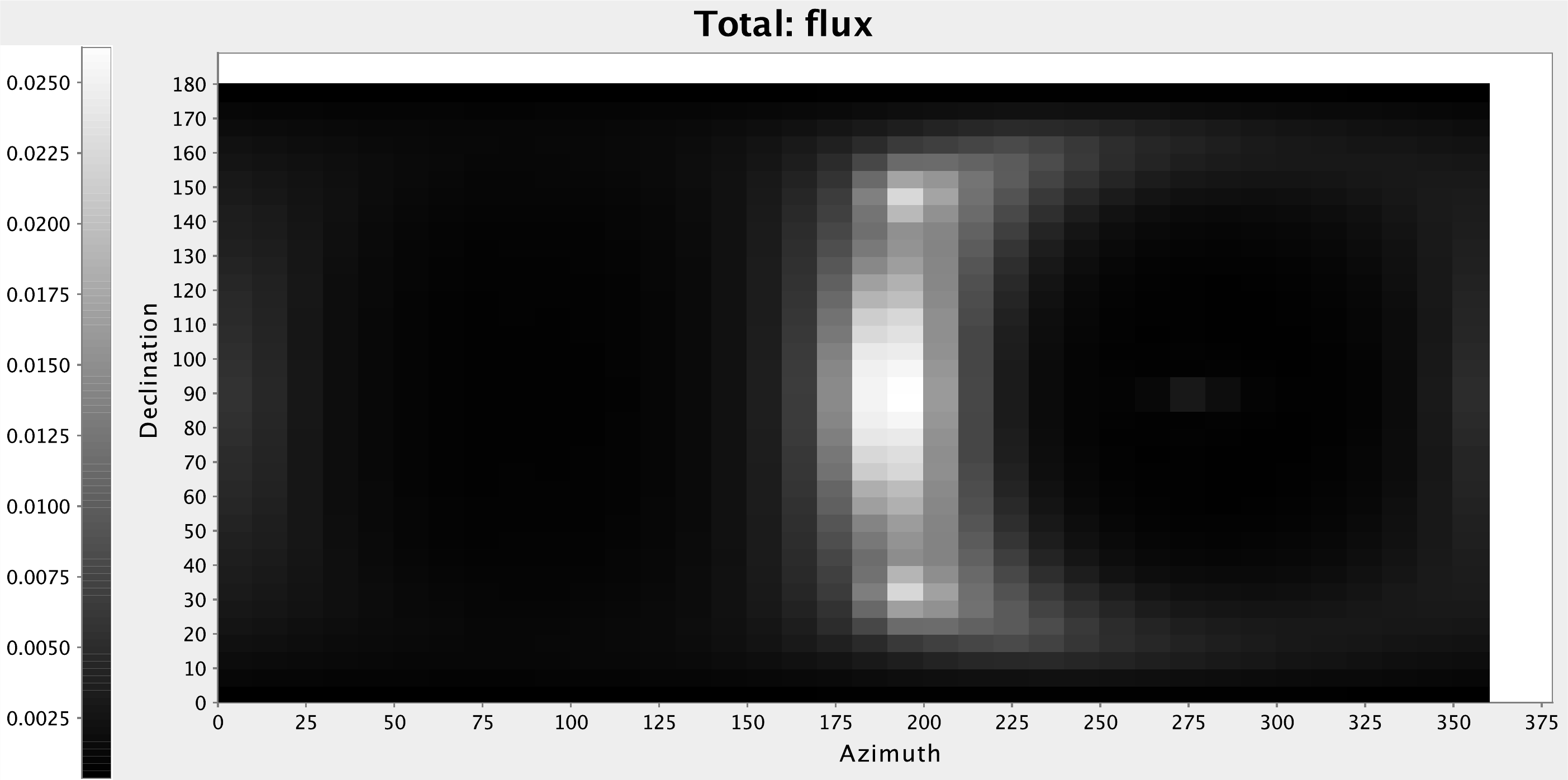}
		\includegraphics[width=0.49\textwidth]{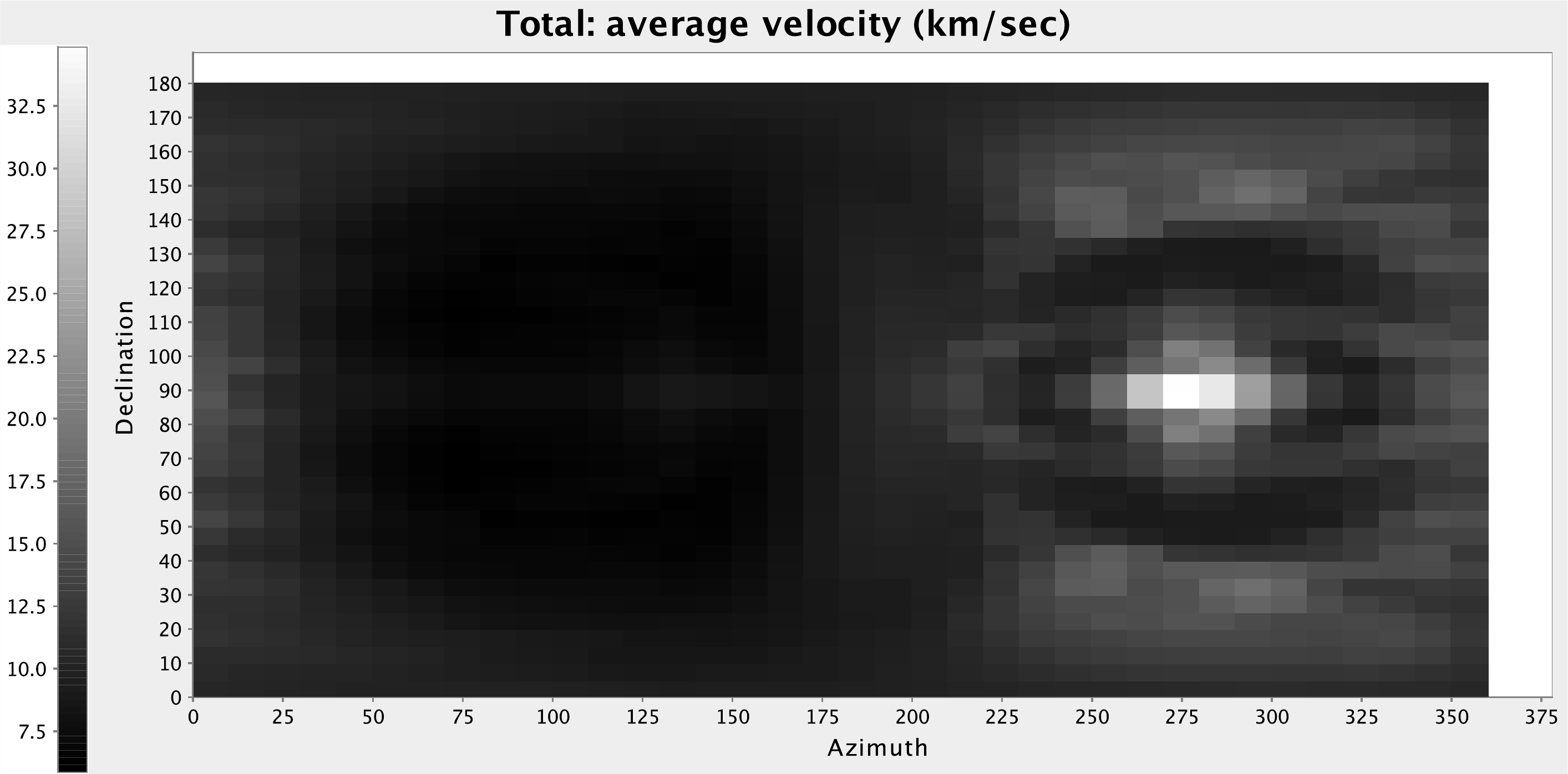}	
	\caption{Sky maps showing the distribution of dust fluxes ({\em left column}) and 
	impact speeds ({\em right column})   for the time interval 01 to 15 December 2024.
  {\em Top row:} asteroidal particles (population~2), 
	{\em Middle row}: cometary particles (population~4),  {\em bottom row}:
	total fluxes (populations~2 and 4 together). The x axis shows the azimuth angle 
	(ranging from 0 to $360^{\circ}$) and the y axis the declination (ranging from 
	0 to $180^{\circ}$). An azimuth angle  $\mathrm{90^{\circ}}$ and declination  $90^{\circ}$ corresponds to the direction of 
	 vernal equinox.  Due to the orientation of
	the \dpl\ trajectory, a declination of $90^{\circ}$ 
	is close to the ecliptic plane. }
\label{fig:imem_com_ast_direction_1}
\end{figure}

\begin{figure}[t]
	\centering
		\includegraphics[width=0.49\textwidth]{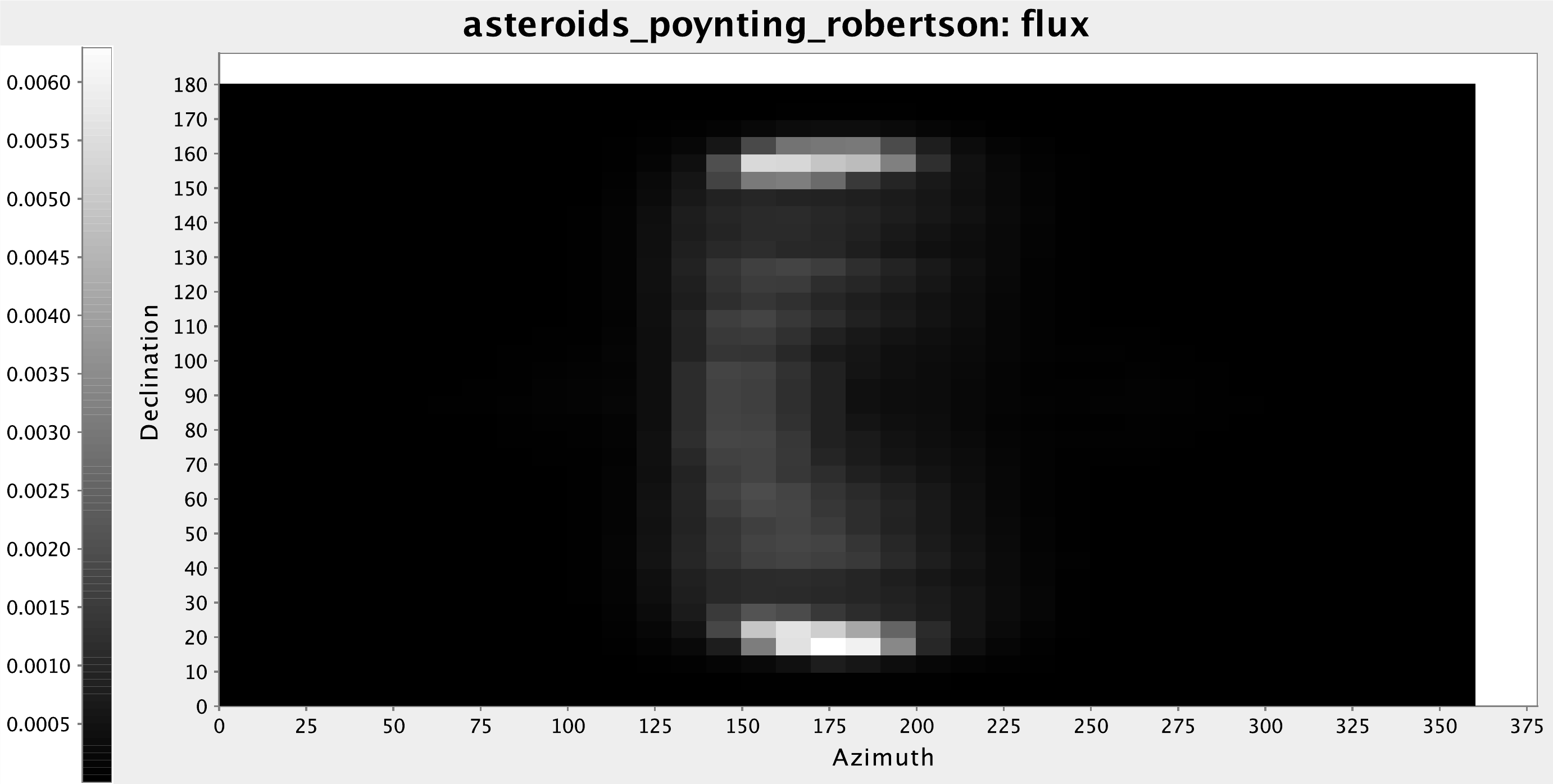}
		\includegraphics[width=0.49\textwidth]{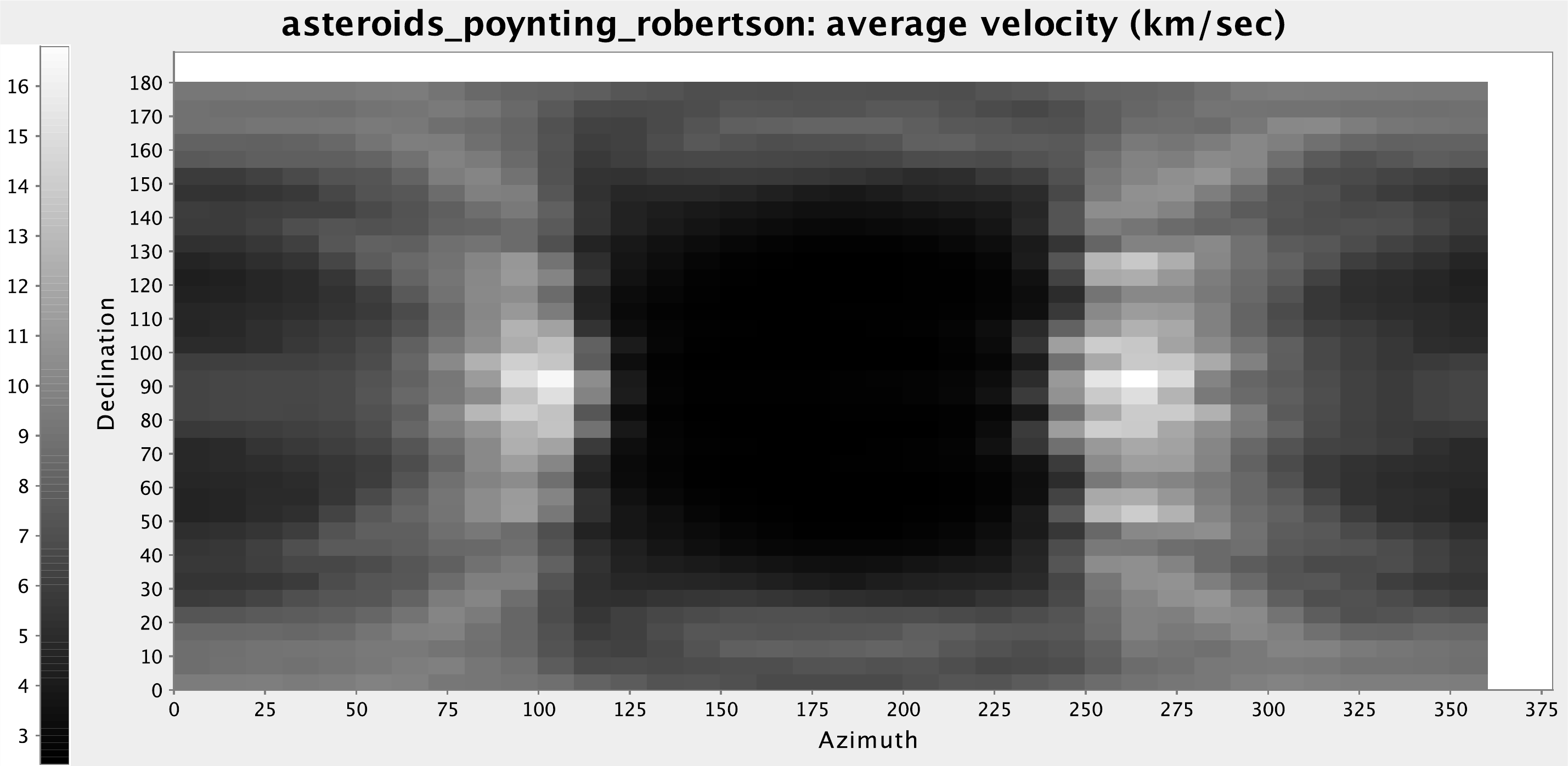}
		\includegraphics[width=0.49\textwidth]{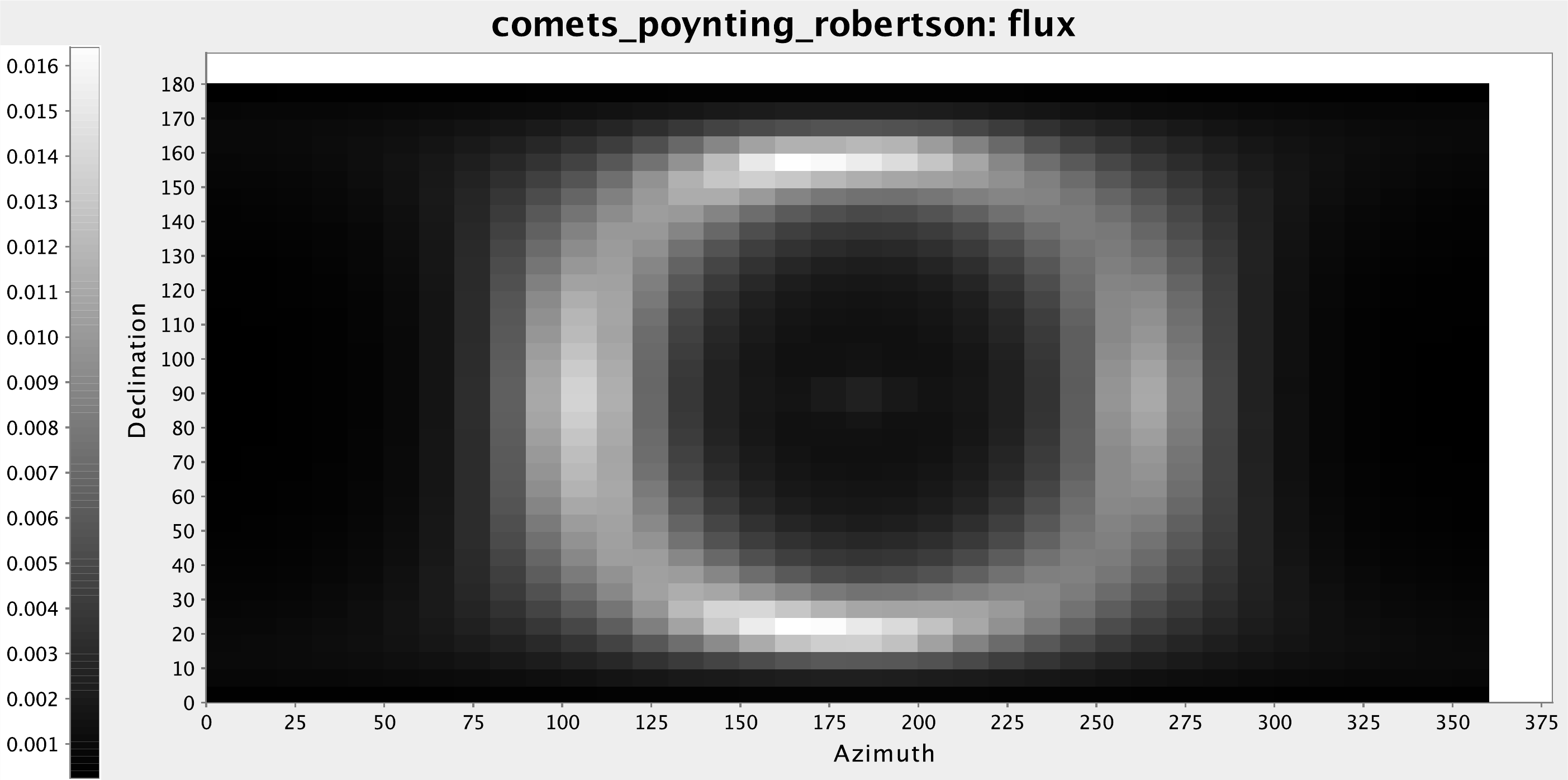}
		\includegraphics[width=0.49\textwidth]{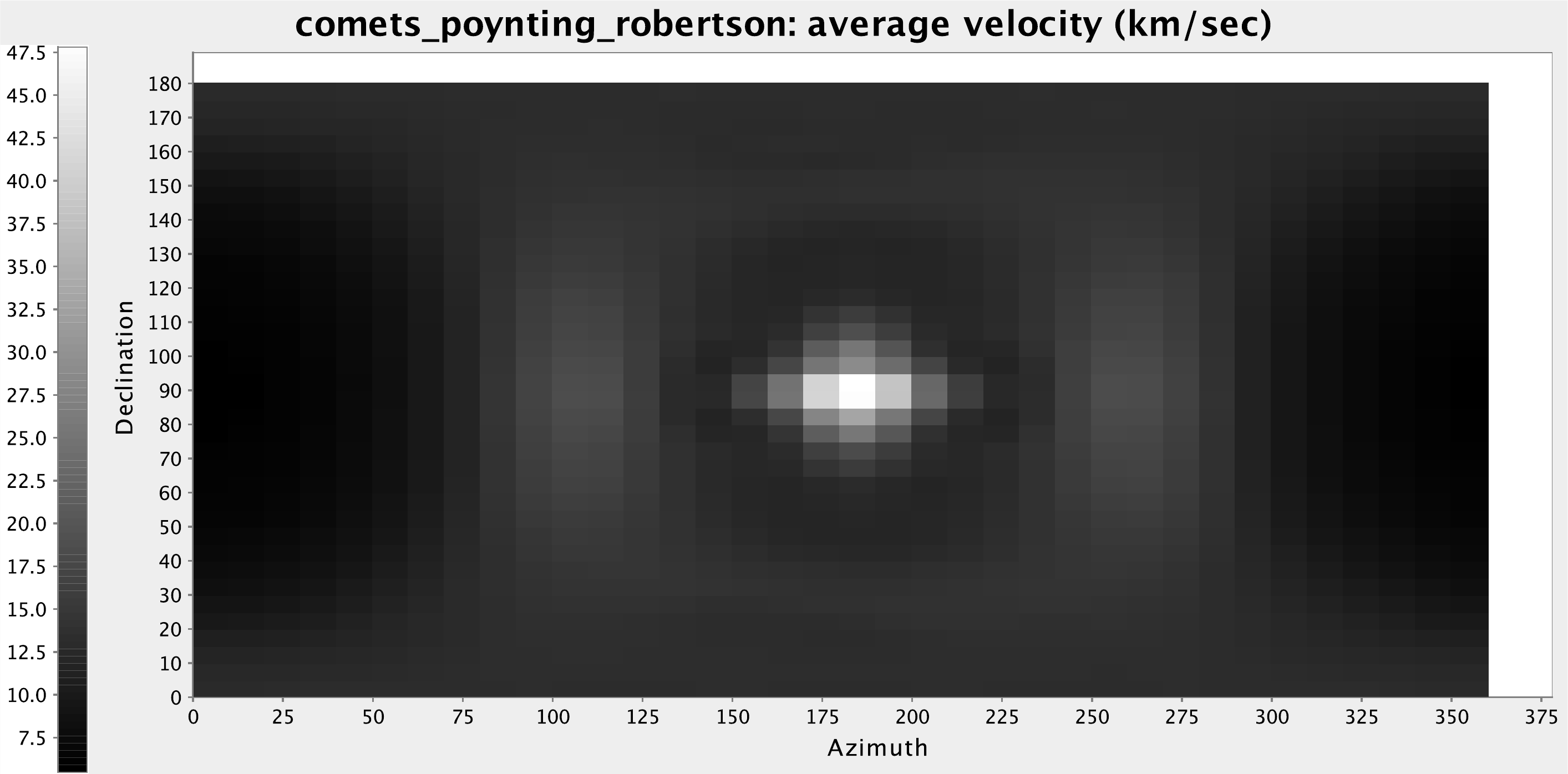}	
		\includegraphics[width=0.49\textwidth]{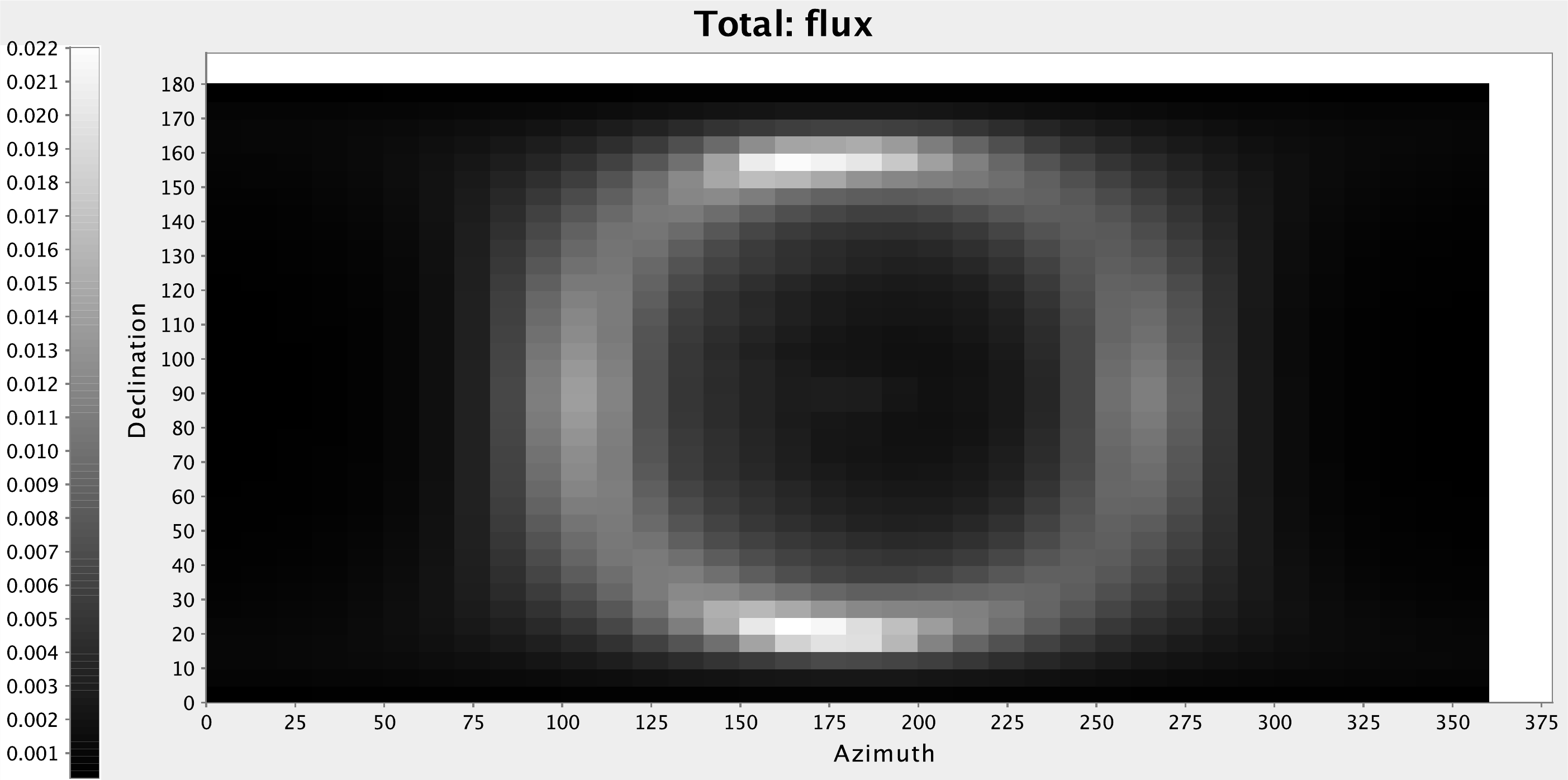}
		\includegraphics[width=0.49\textwidth]{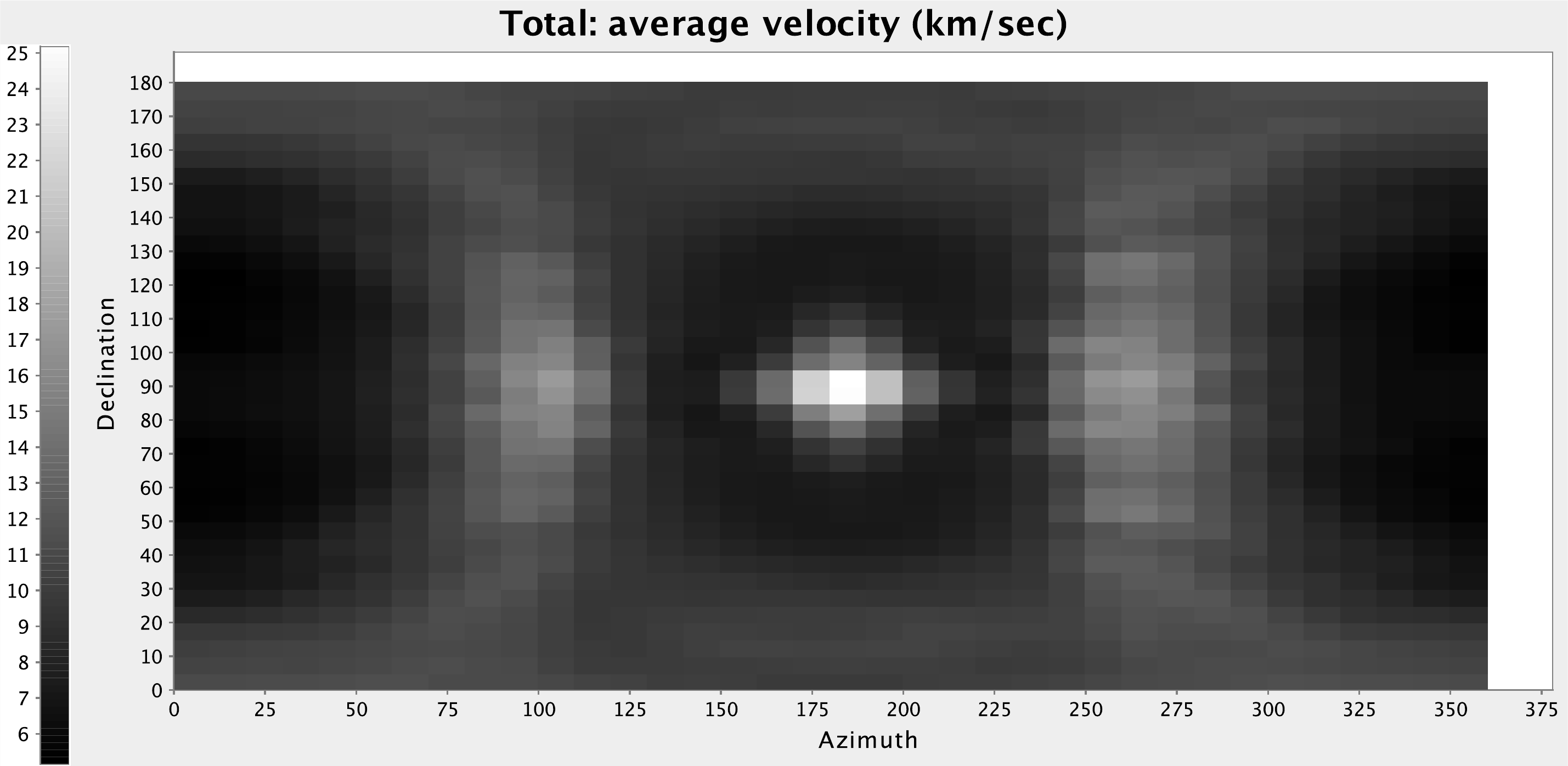}	
	\caption{Same as Figure~\ref{fig:imem_com_ast_direction_1} but for the time interval 14 to 28 February 2025.}
	\label{fig:imem_com_ast_direction_2}
\end{figure}

\begin{figure}[t]
        \centering
                \includegraphics[width=0.49\textwidth]{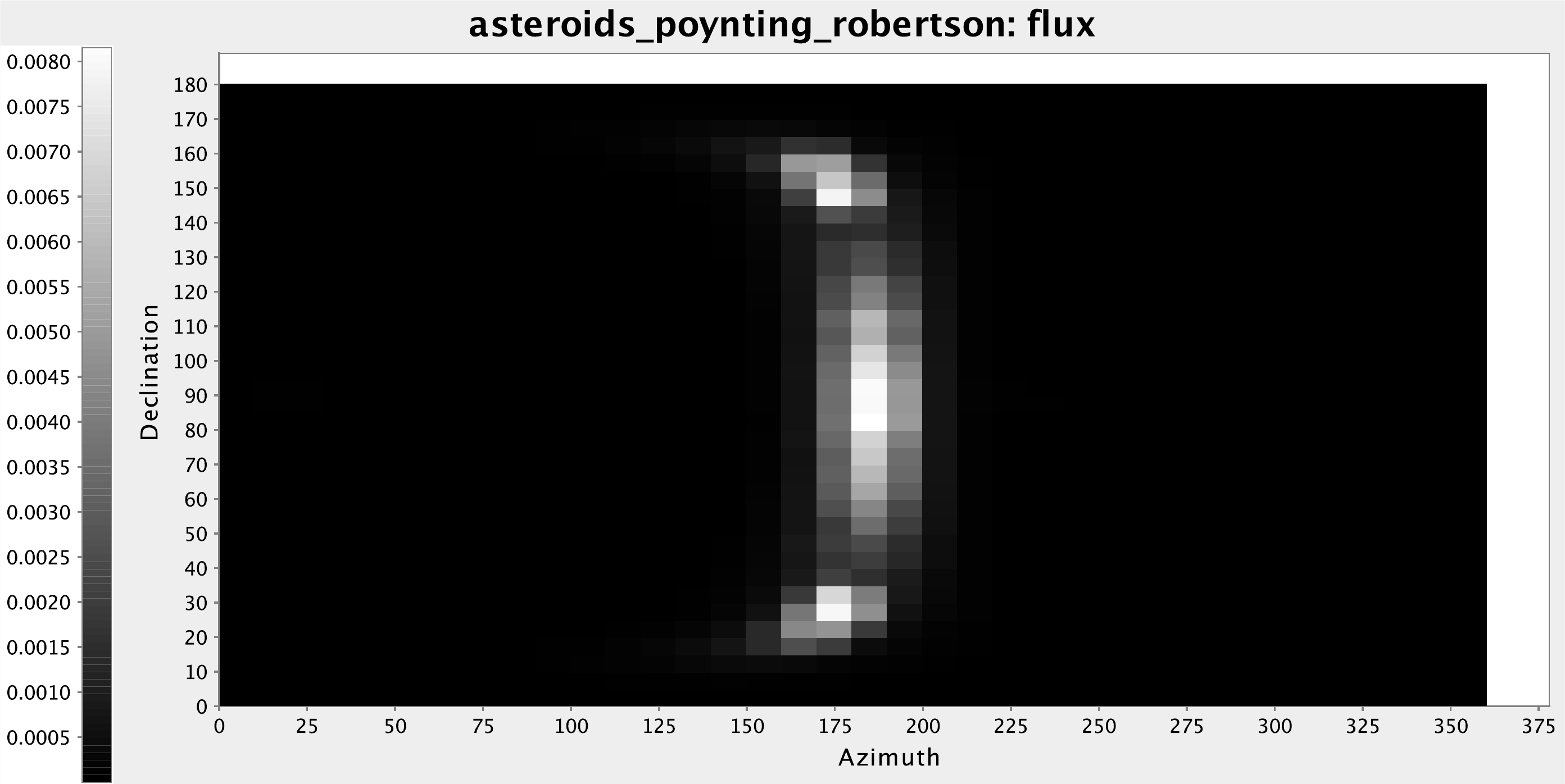}
                \includegraphics[width=0.49\textwidth]{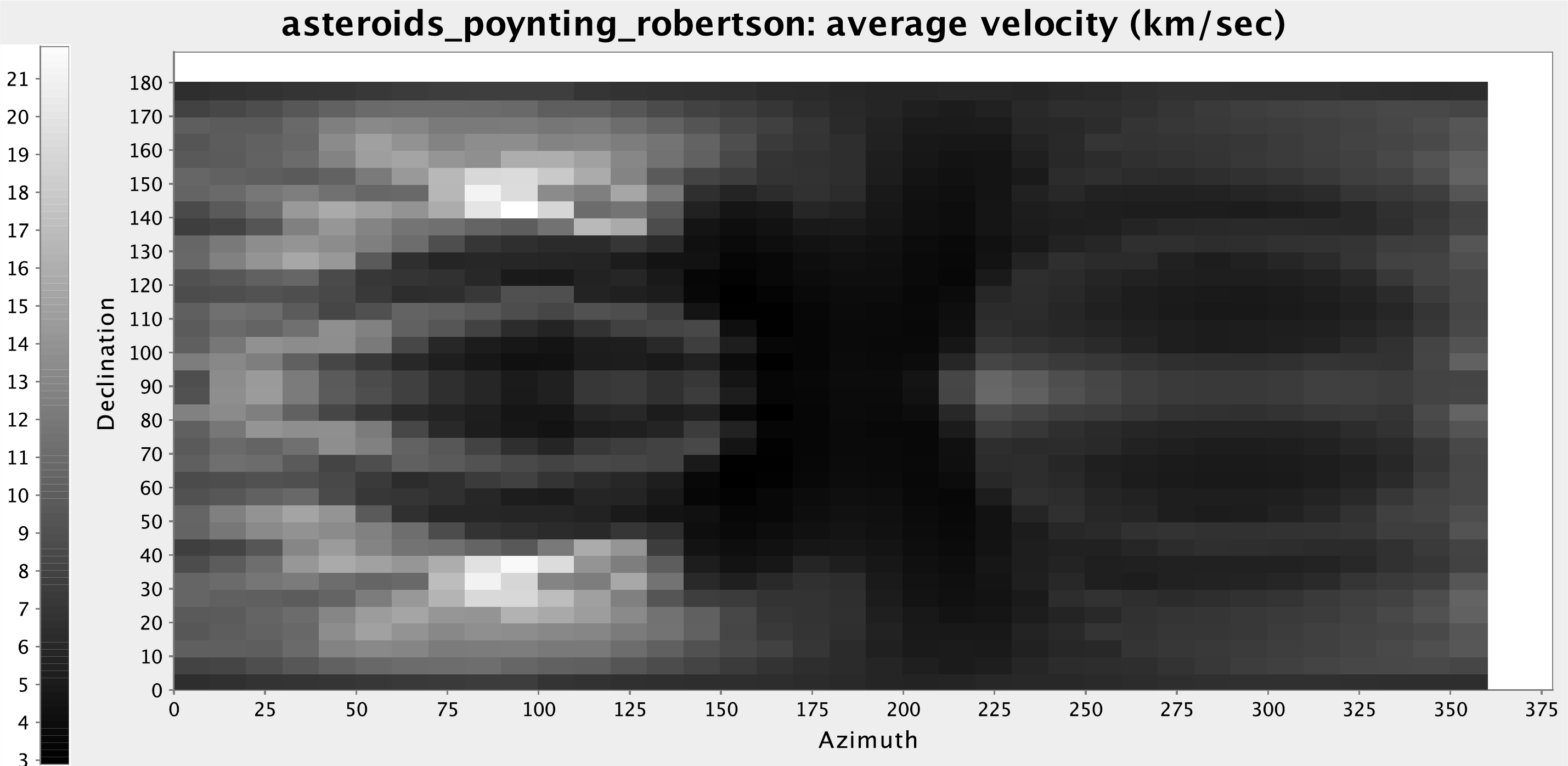}
                \includegraphics[width=0.49\textwidth]{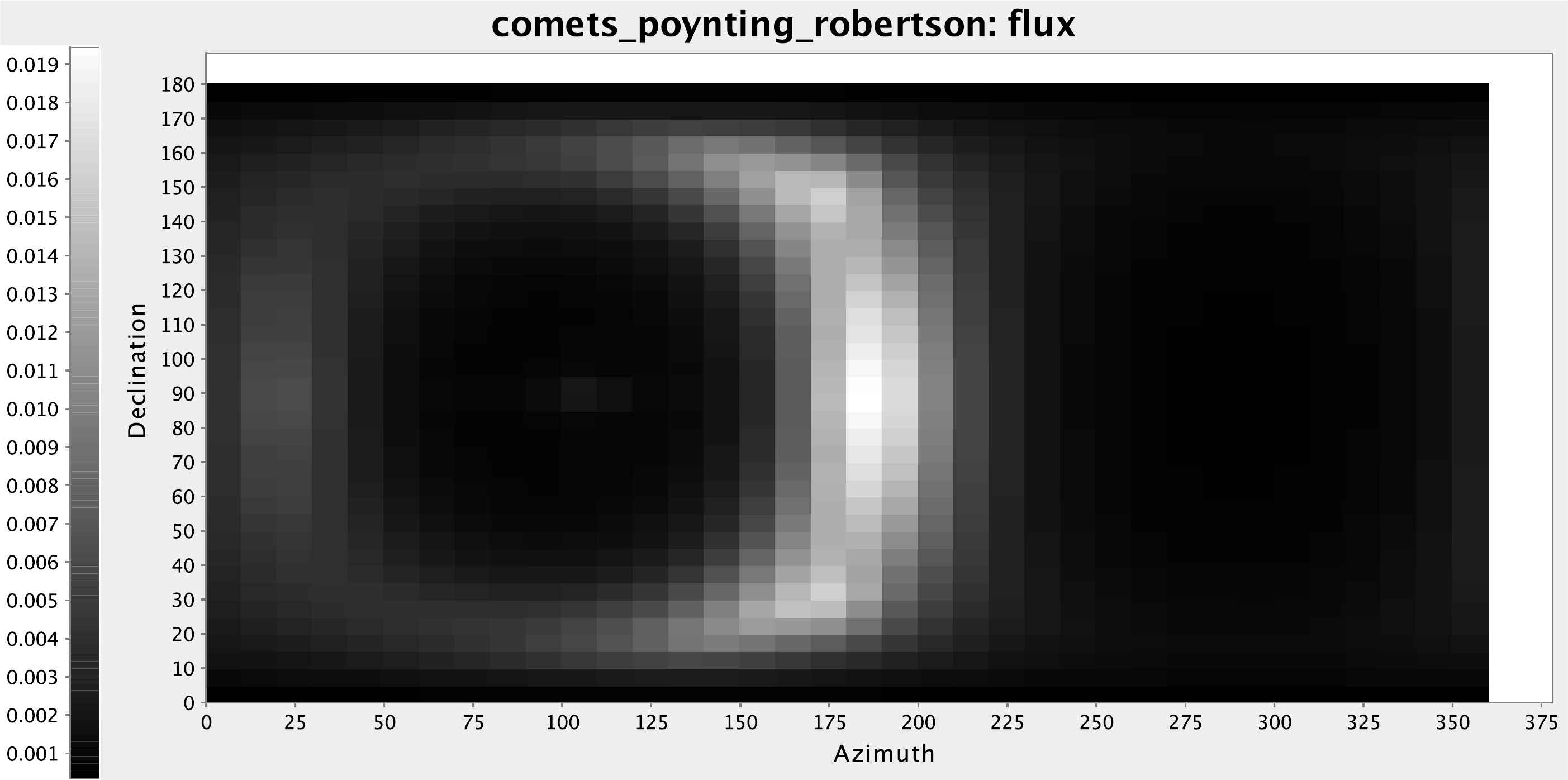}
                \includegraphics[width=0.49\textwidth]{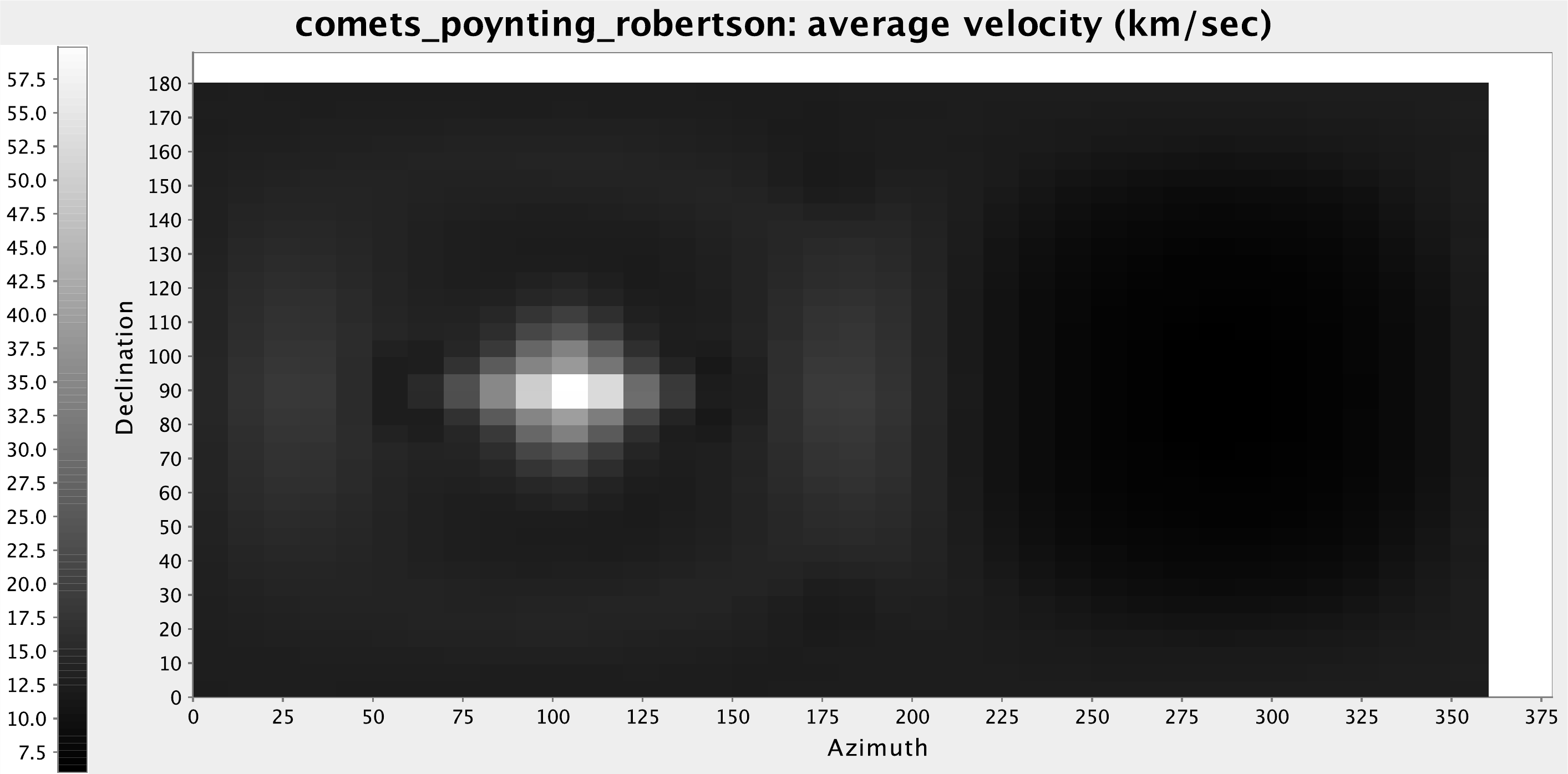}  
                \includegraphics[width=0.49\textwidth]{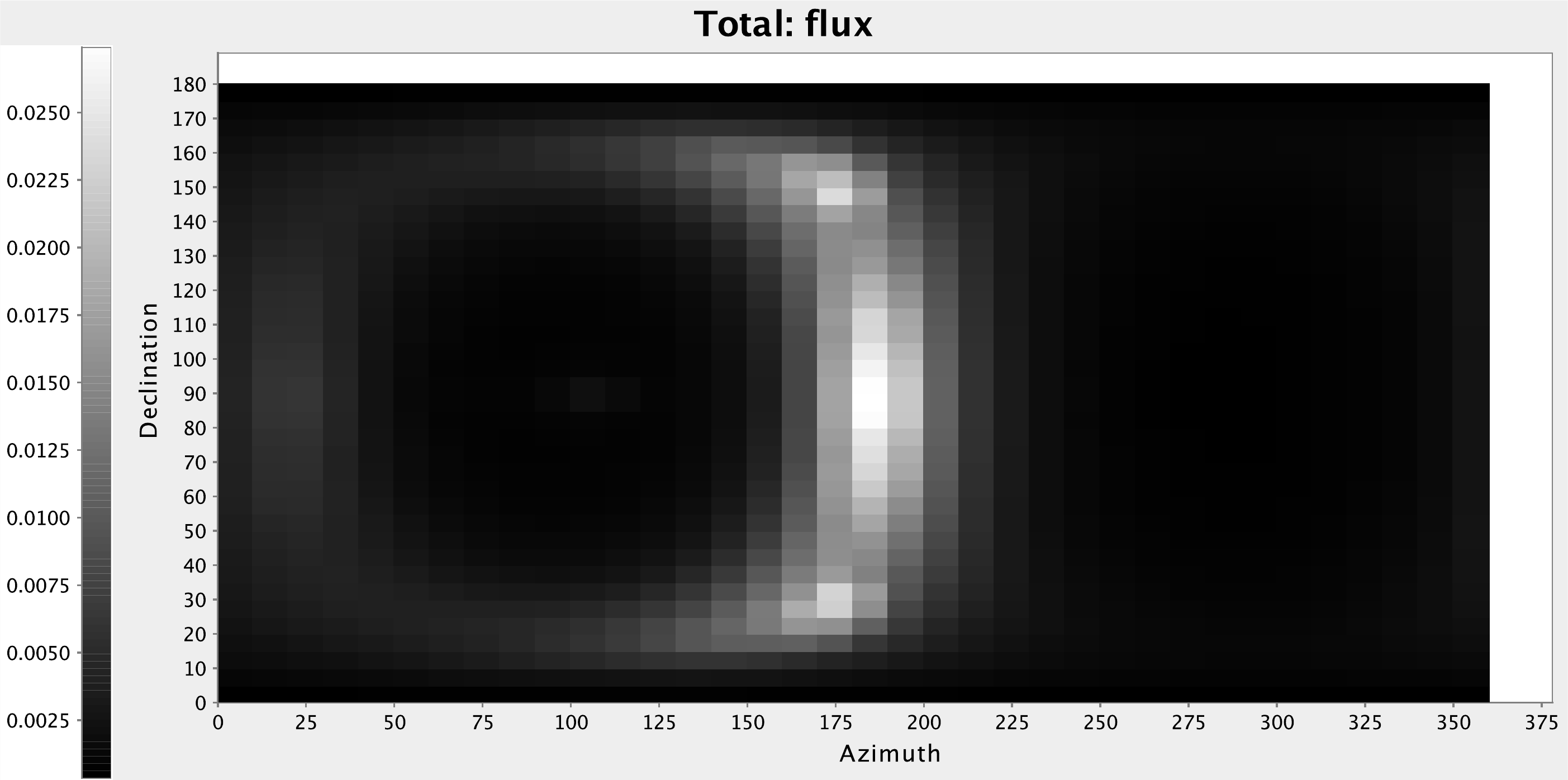}
                \includegraphics[width=0.49\textwidth]{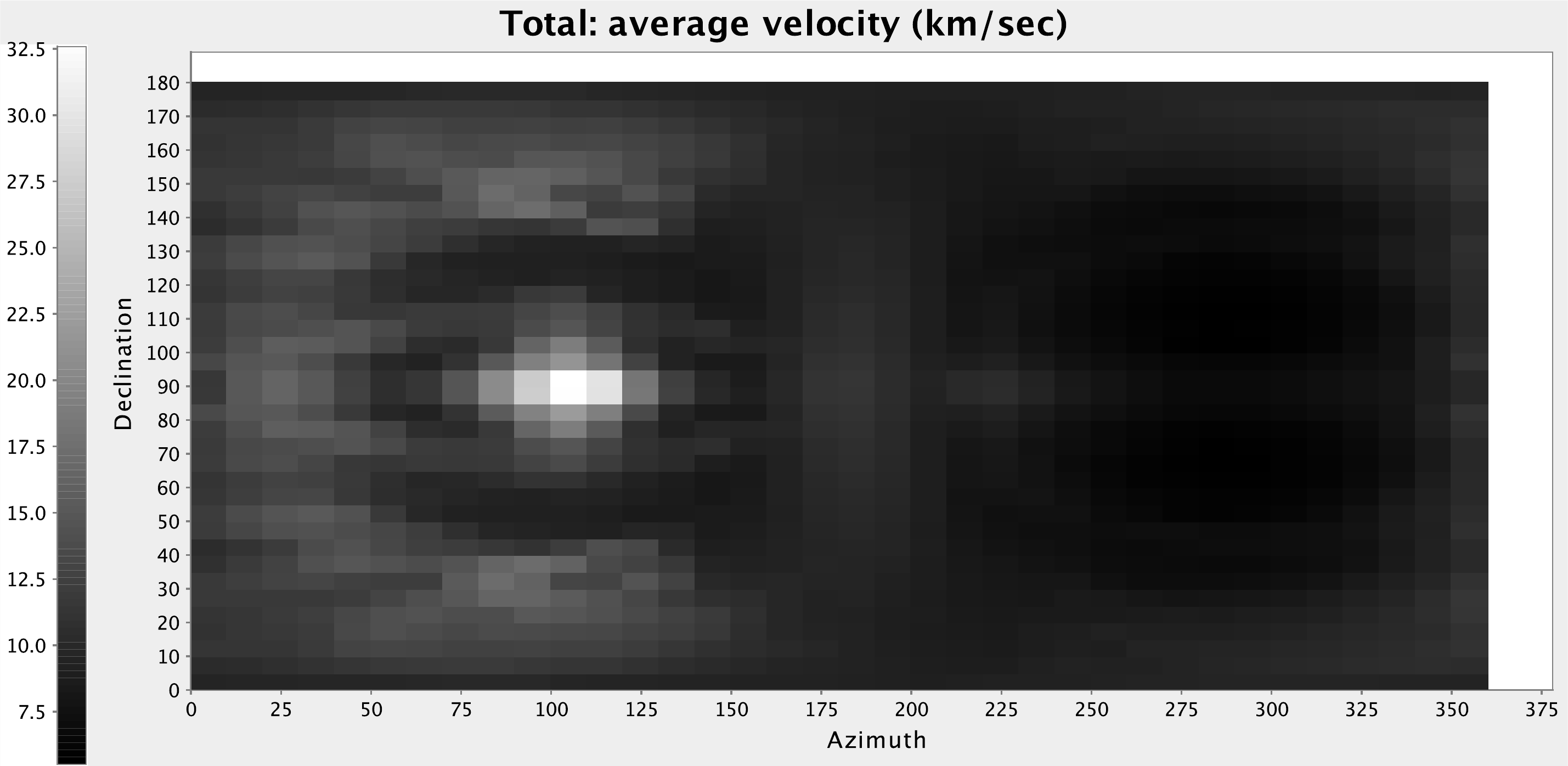}  
        \caption{Same as Figure~\ref{fig:imem_com_ast_direction_1} but for the time interval 16 to 30 April 2025.}
        \label{fig:imem_com_ast_direction_3}
\end{figure}

\begin{figure}[t]
        \centering
                \includegraphics[width=0.49\textwidth]{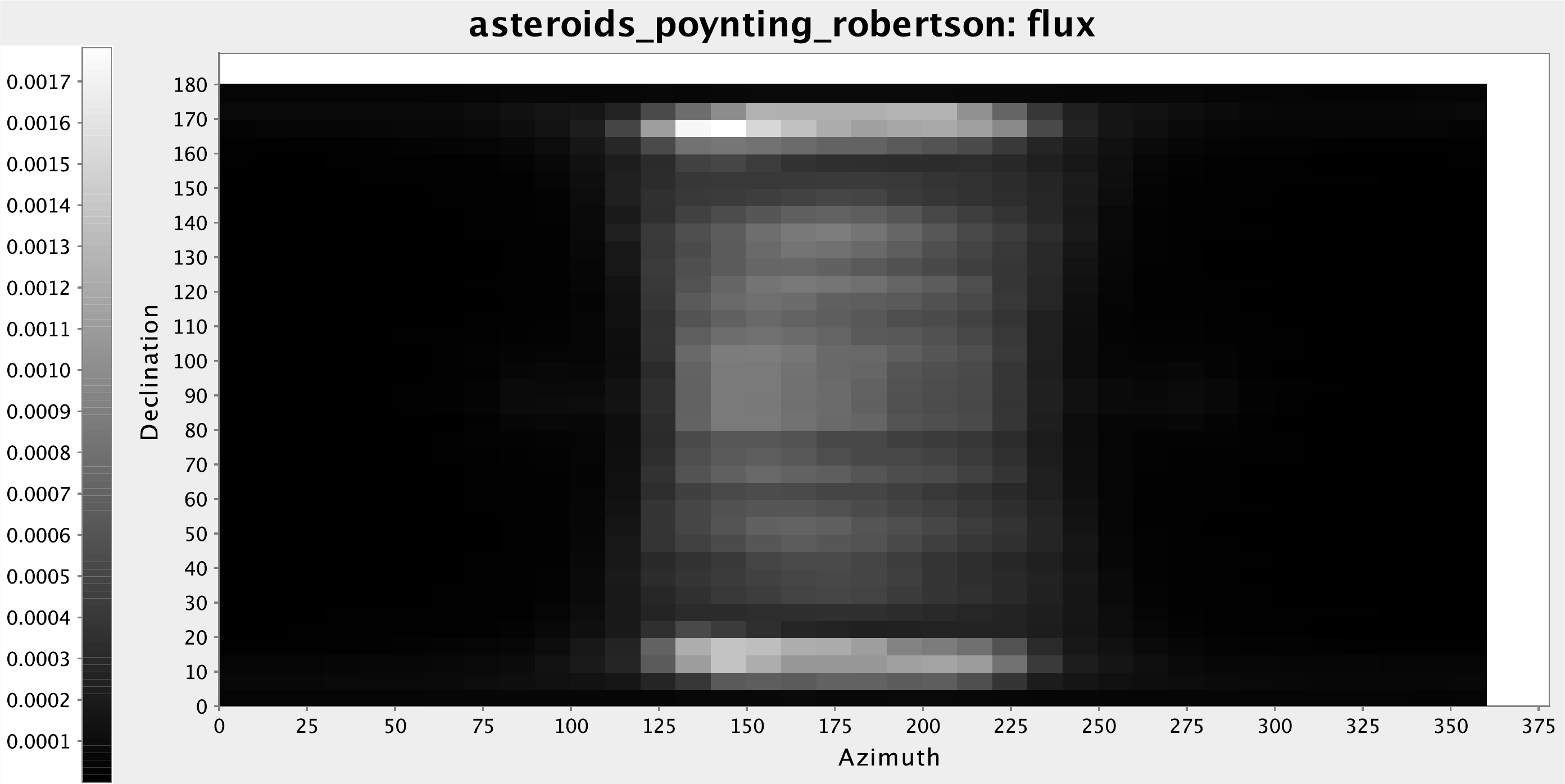}
                \includegraphics[width=0.49\textwidth]{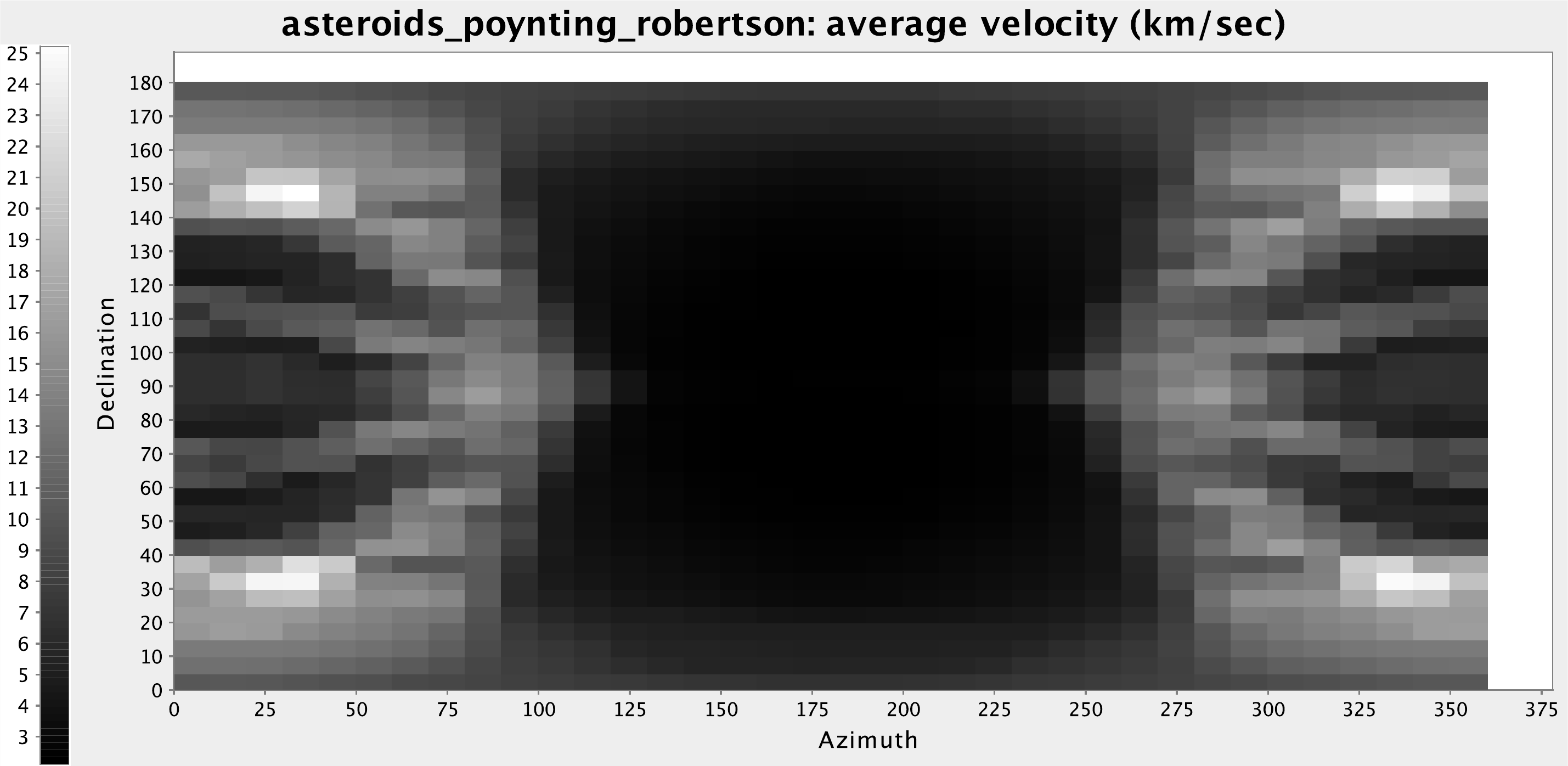}
                \includegraphics[width=0.49\textwidth]{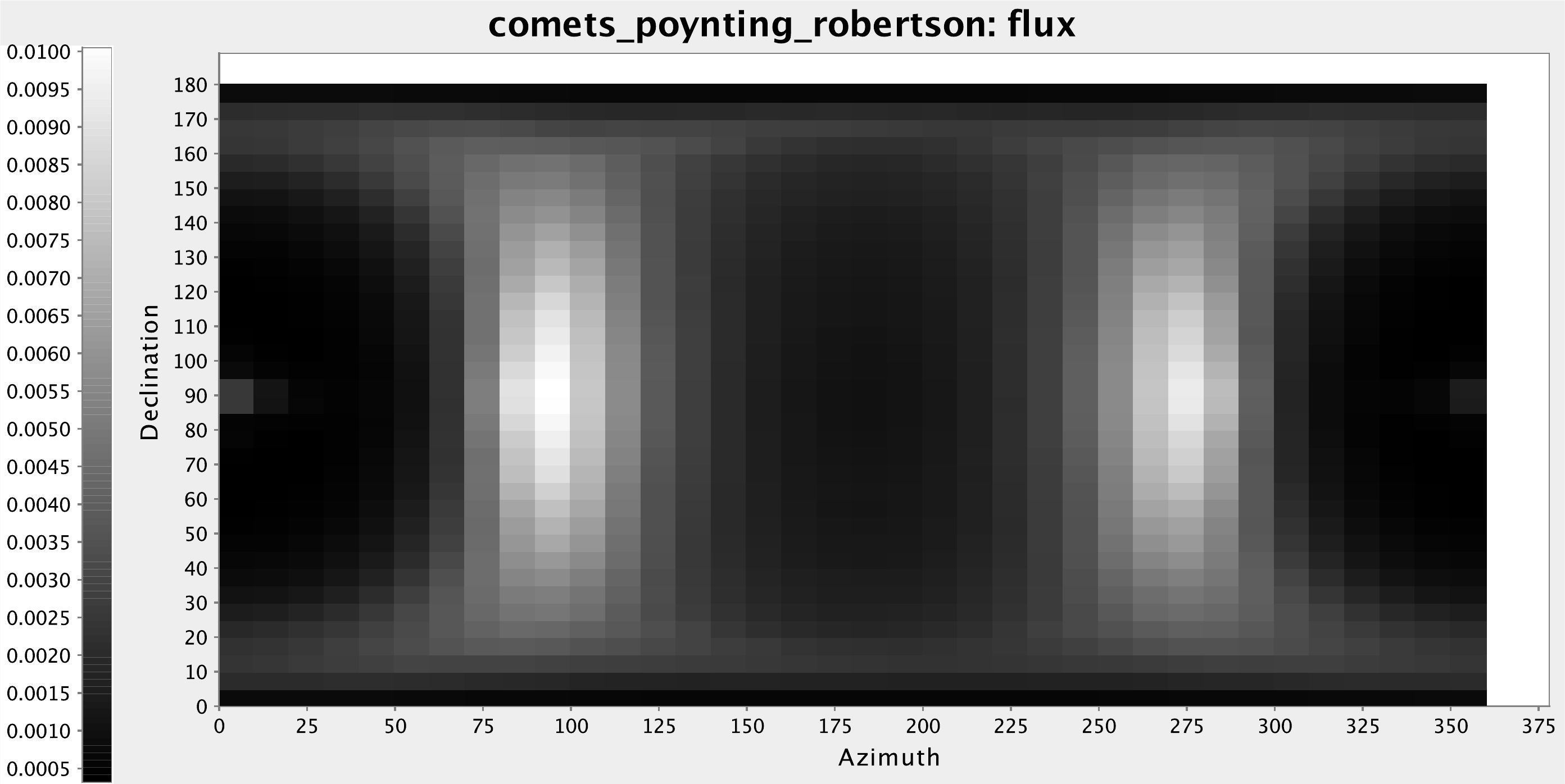}
                \includegraphics[width=0.49\textwidth]{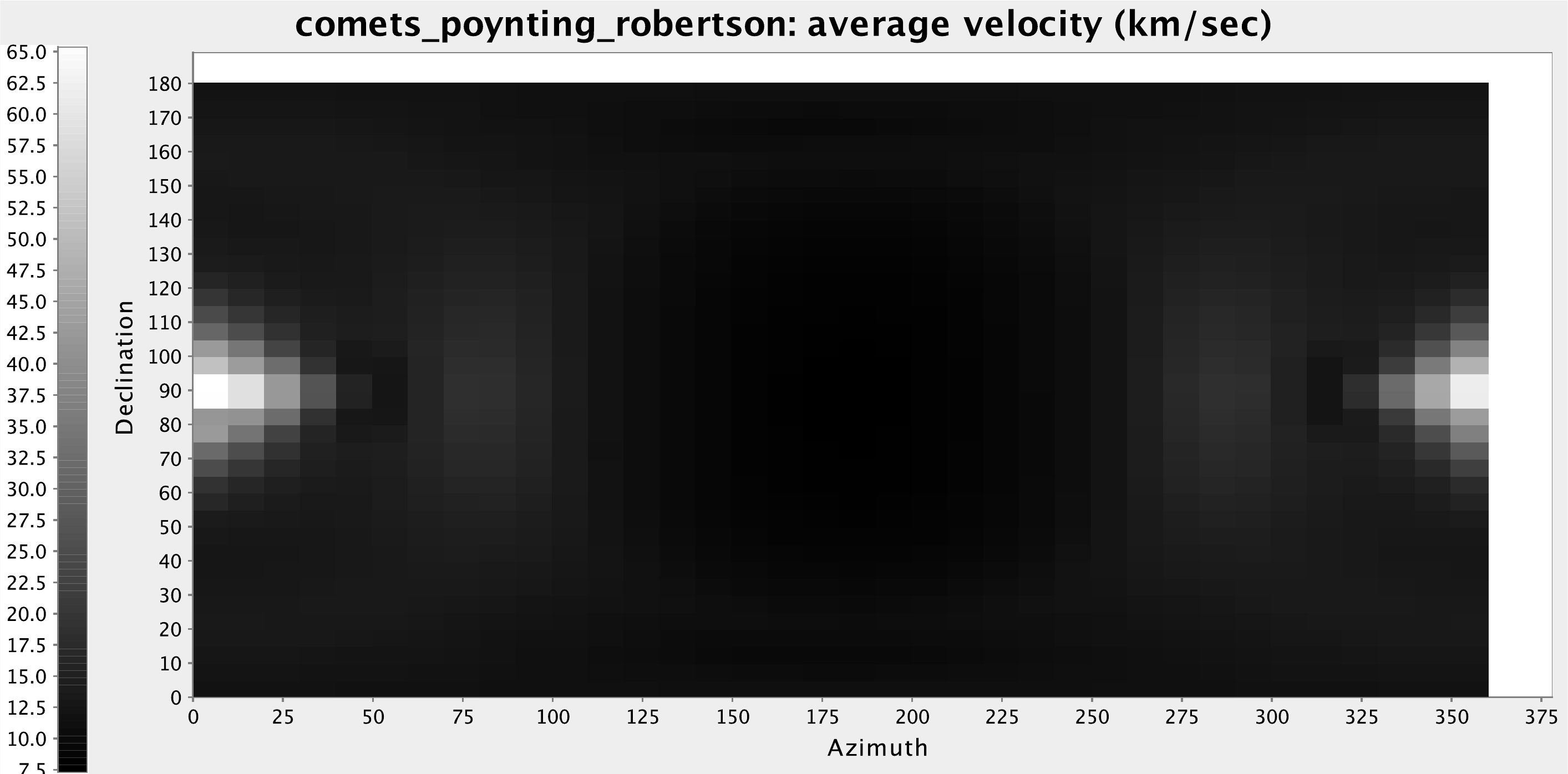}  
                \includegraphics[width=0.49\textwidth]{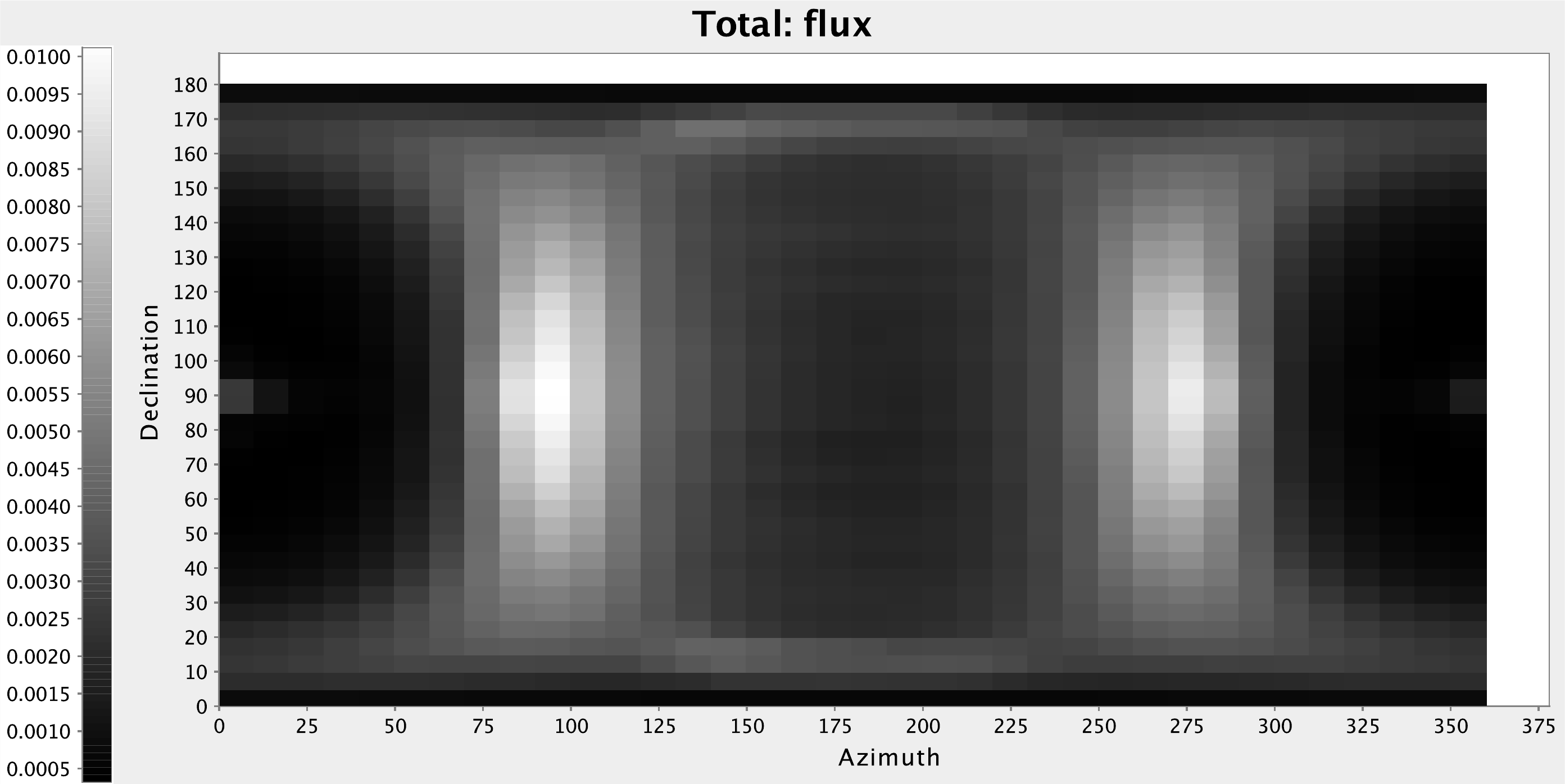}
                \includegraphics[width=0.49\textwidth]{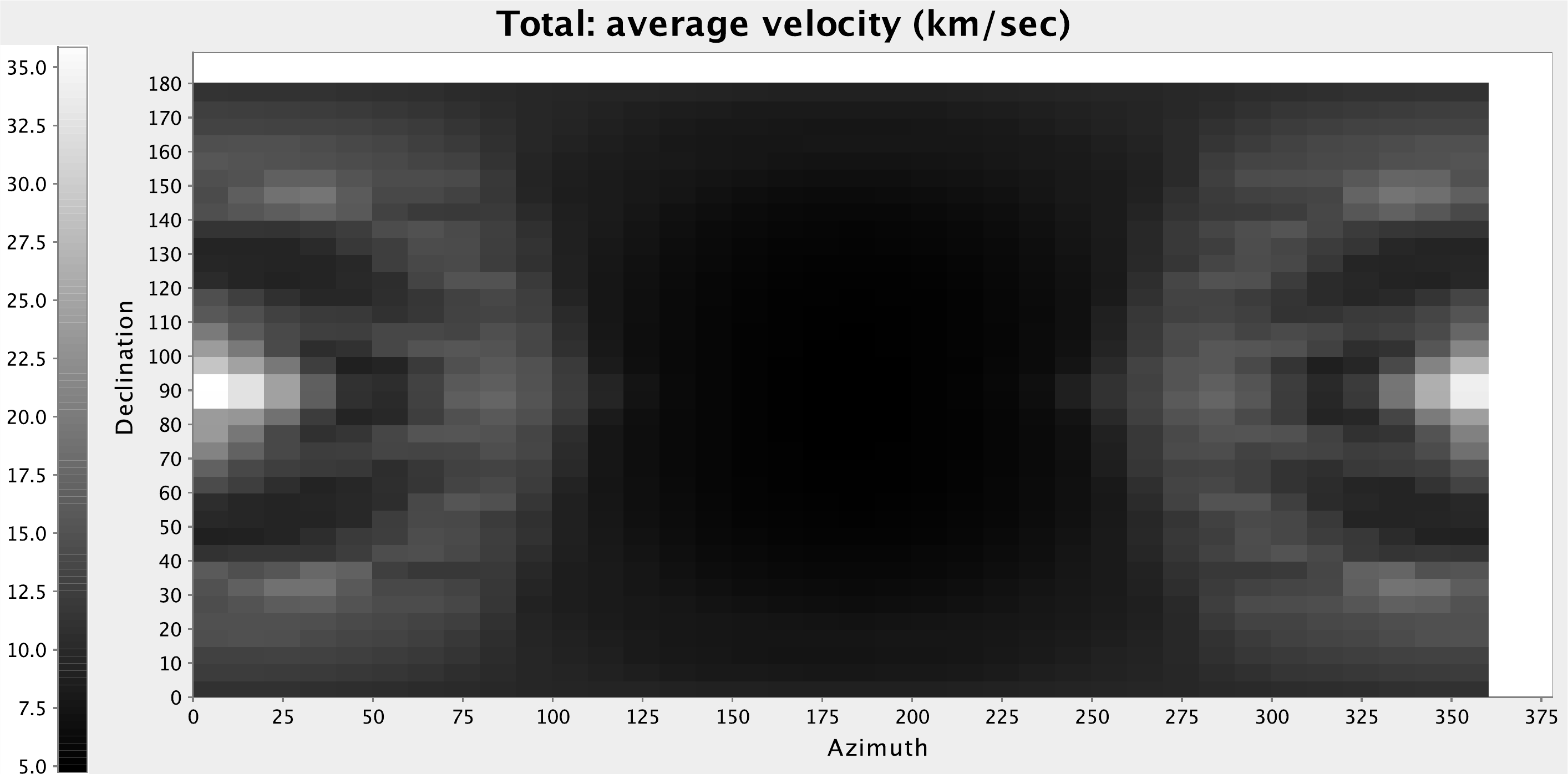}  
        \caption{Same as Figure~\ref{fig:imem_com_ast_direction_1} but for the time interval 16 to 30 July 2025.}
        \label{fig:imem_com_ast_direction_4}
\end{figure}

\begin{figure}[t]
        \centering
                \includegraphics[width=0.49\textwidth]{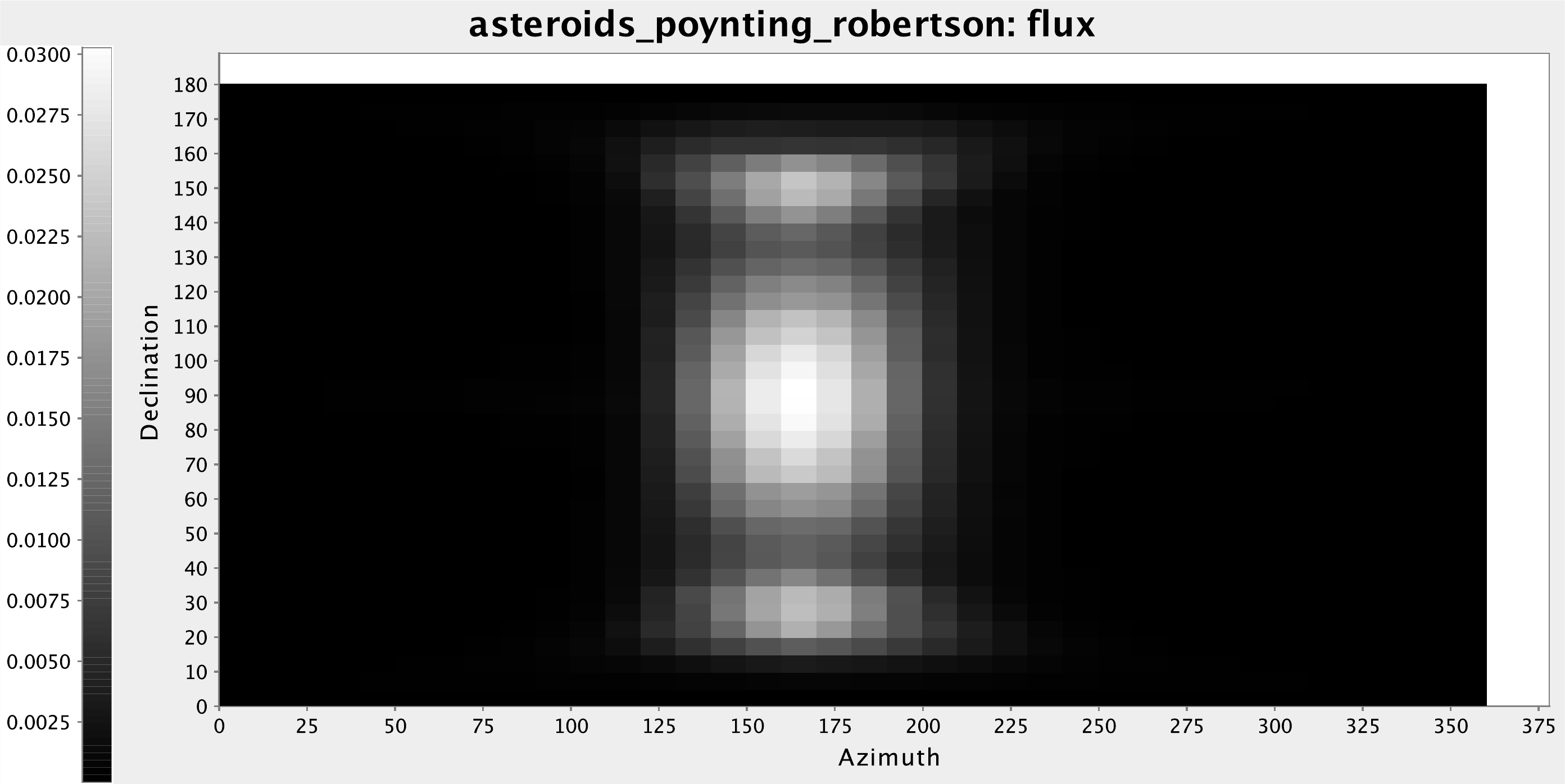}
                \includegraphics[width=0.49\textwidth]{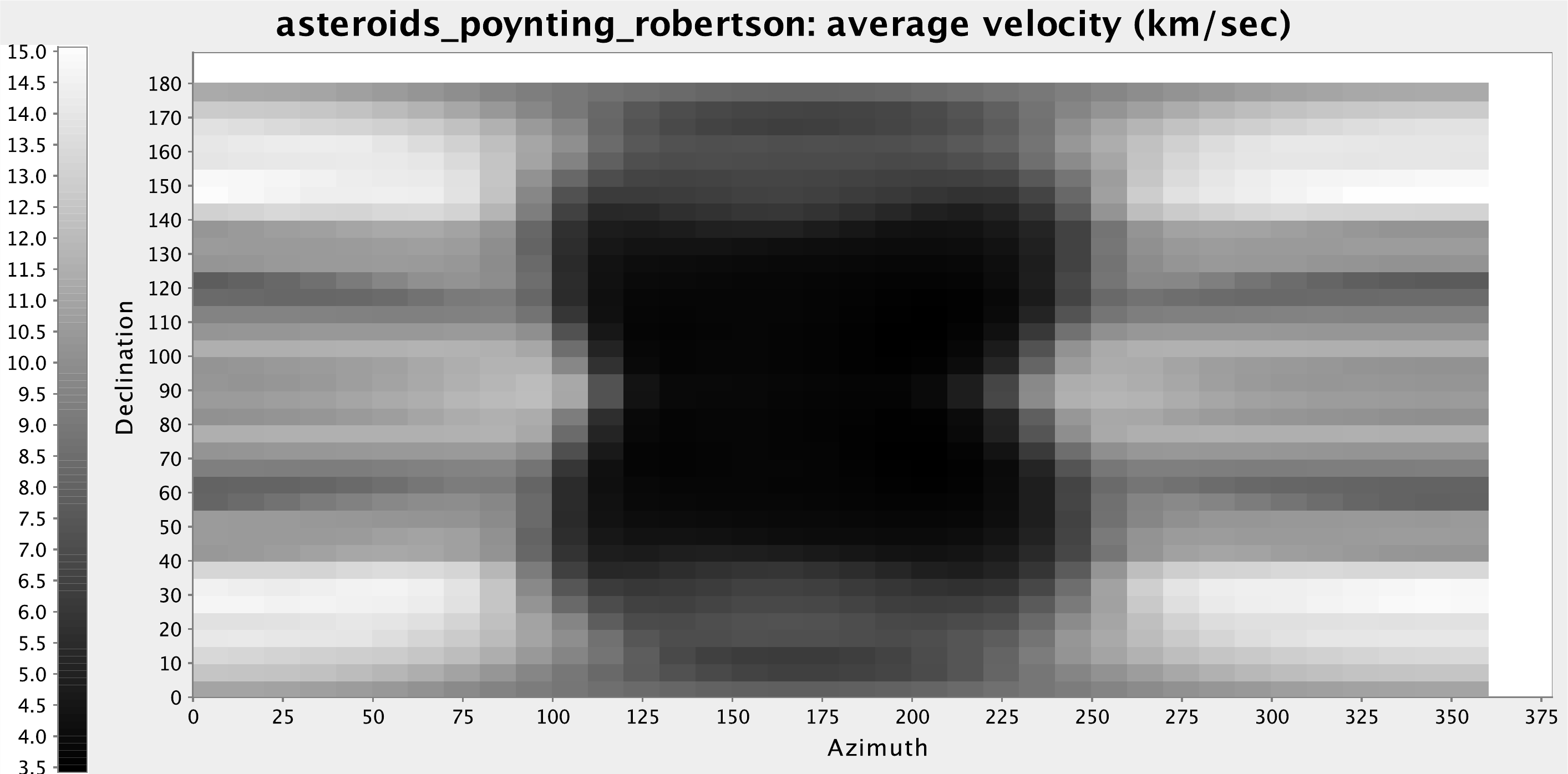}
                \includegraphics[width=0.49\textwidth]{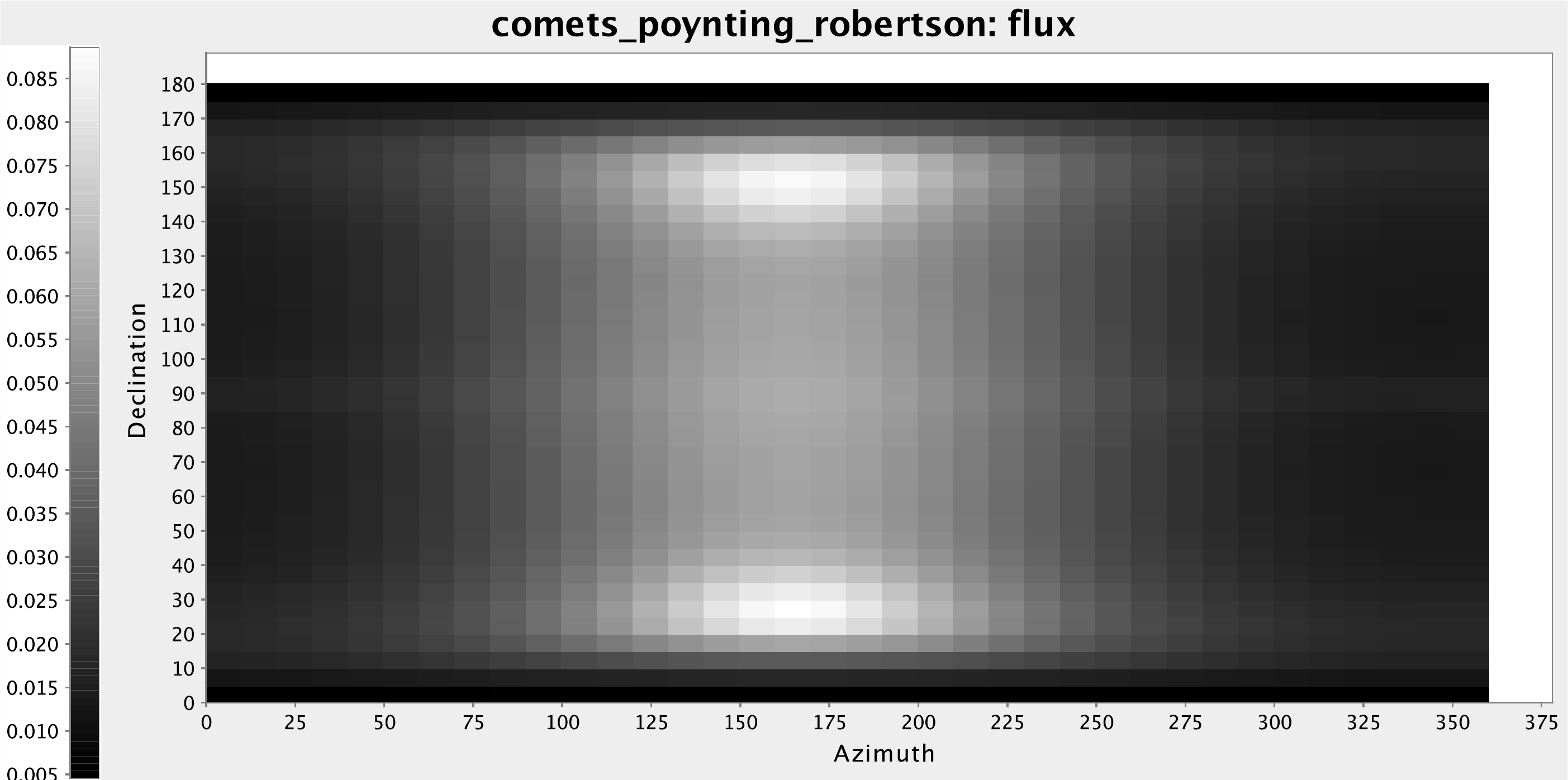}
                \includegraphics[width=0.49\textwidth]{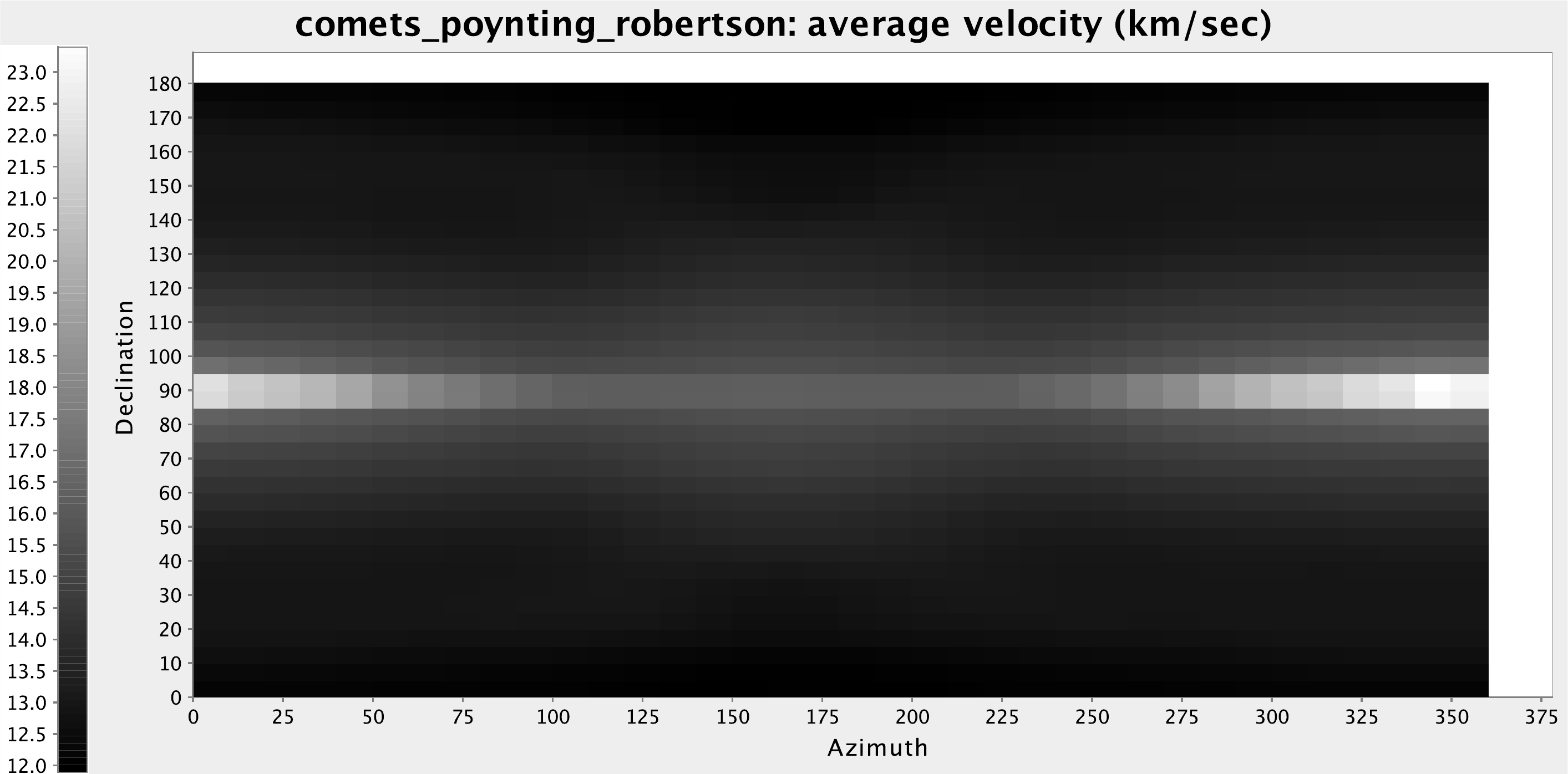}  
                \includegraphics[width=0.49\textwidth]{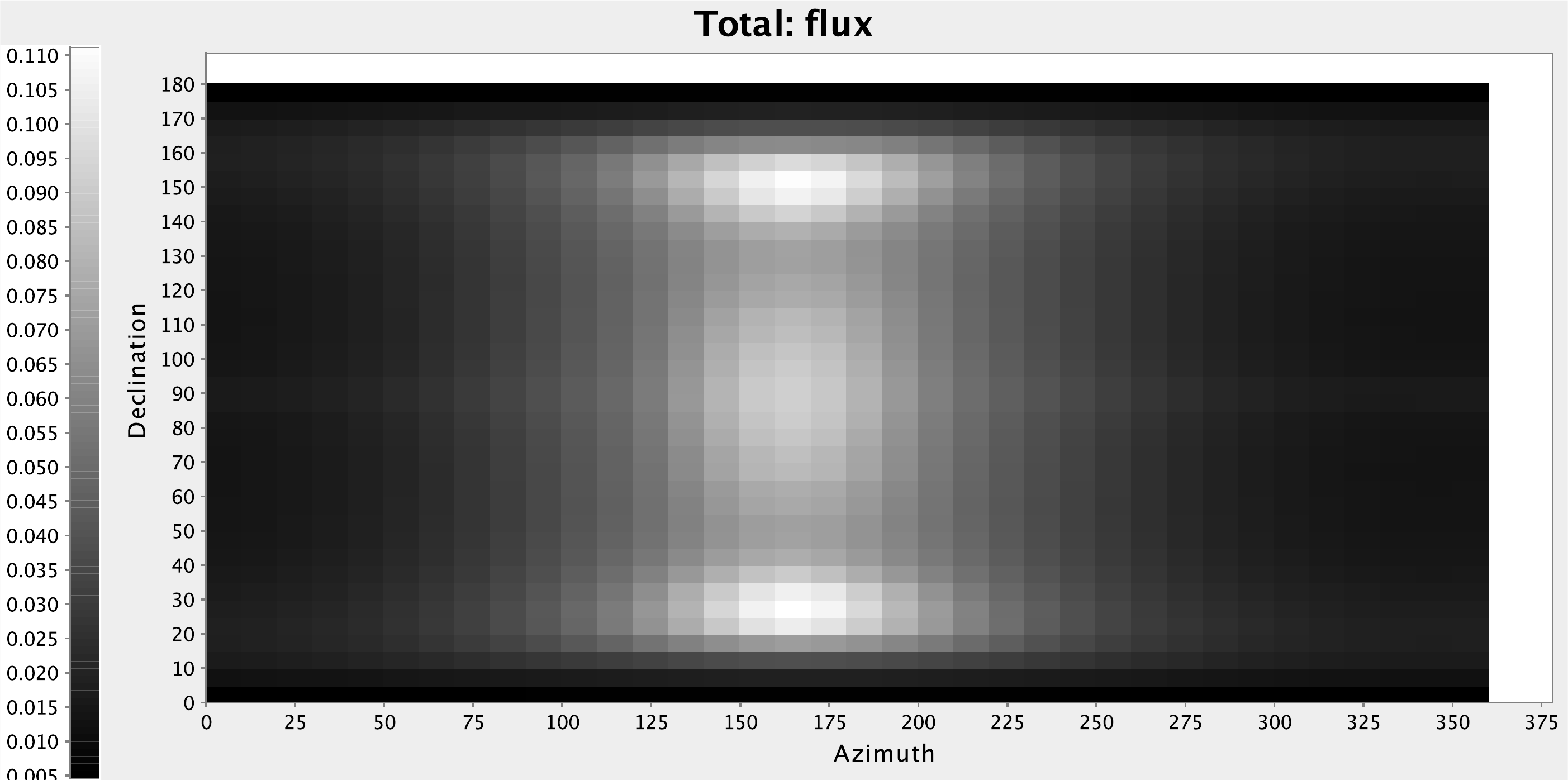}
                \includegraphics[width=0.49\textwidth]{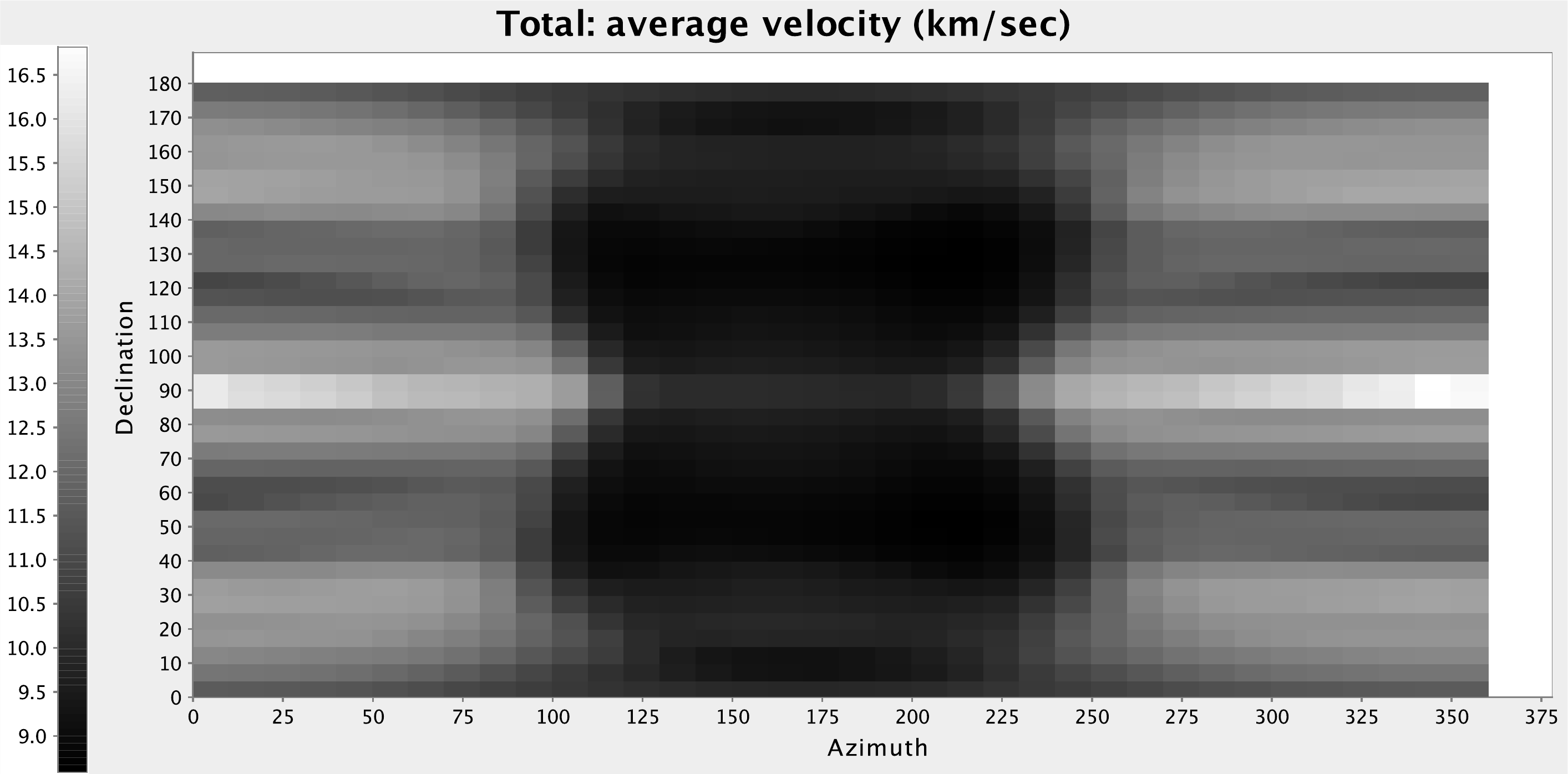}  
        \caption{Same as Figure~\ref{fig:imem_com_ast_direction_1} but for the entire \dpl\ mission.
        }
        \label{fig:imem_com_ast_direction_5}
\end{figure}

\begin{figure}[tbh]
	\centering
	\vspace{-2.9cm}
		\includegraphics[width=1.1\textwidth]{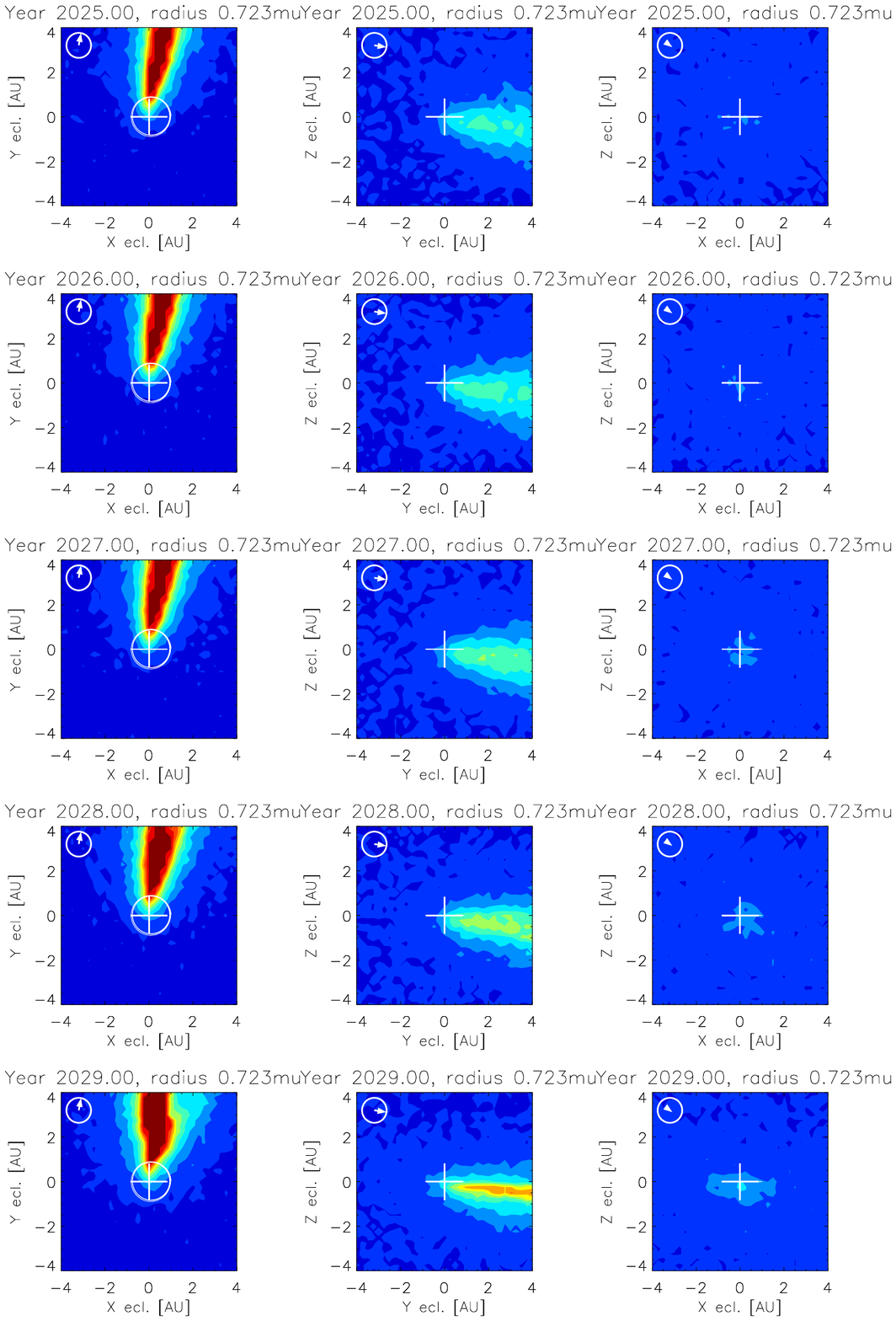}
	    		\vspace{-3.2cm}		
	\caption{Cross sections along the ecliptic coordinate planes through the simulated  spatial
	density cubes for particles 
	with radius $r_d=\mathrm{0.723\,\mu m}$ during the \dpl\ mission, at the beginning of each indicated year. 
	The Sun 
	is at the center,  and the almost circular \dpl\ trajectory is shown in white in the left column. The dust density is color coded: dark blue: low dust density; green, yellow, 
	and red represent density enhancements with respect to the initial 
	density at 50~AU. The projection of the original  interstellar dust flow direction (at 50~AU) 
	is shown as an arrow in
	the top left corner of each plot.}
	\label{fig:imex_3d_3}
\end{figure}

Figures~\ref{fig:imem_com_ast_direction_1} to \ref{fig:imem_com_ast_direction_5} show sky maps with 
fluxes and impact speeds for the asteroidal and cometary dust particles. To illustrate the variations 
during the \dpl\ mission we
show sky maps for four different time intervals lasting 15 days each (Figures~\ref{fig:imem_com_ast_direction_1} 
to \ref{fig:imem_com_ast_direction_4}), and for the total  1474 
mission  days (Figure~\ref{fig:imem_com_ast_direction_5}).

\clearpage

\section*{Appendix 2}

Spatial distribution of interstellar dust in the solar system from IMEX simulations for three particle 
sizes. 

\begin{figure}[tbh]
	\centering
	\vspace{-2.9cm}
		\includegraphics[width=1.25\textwidth]{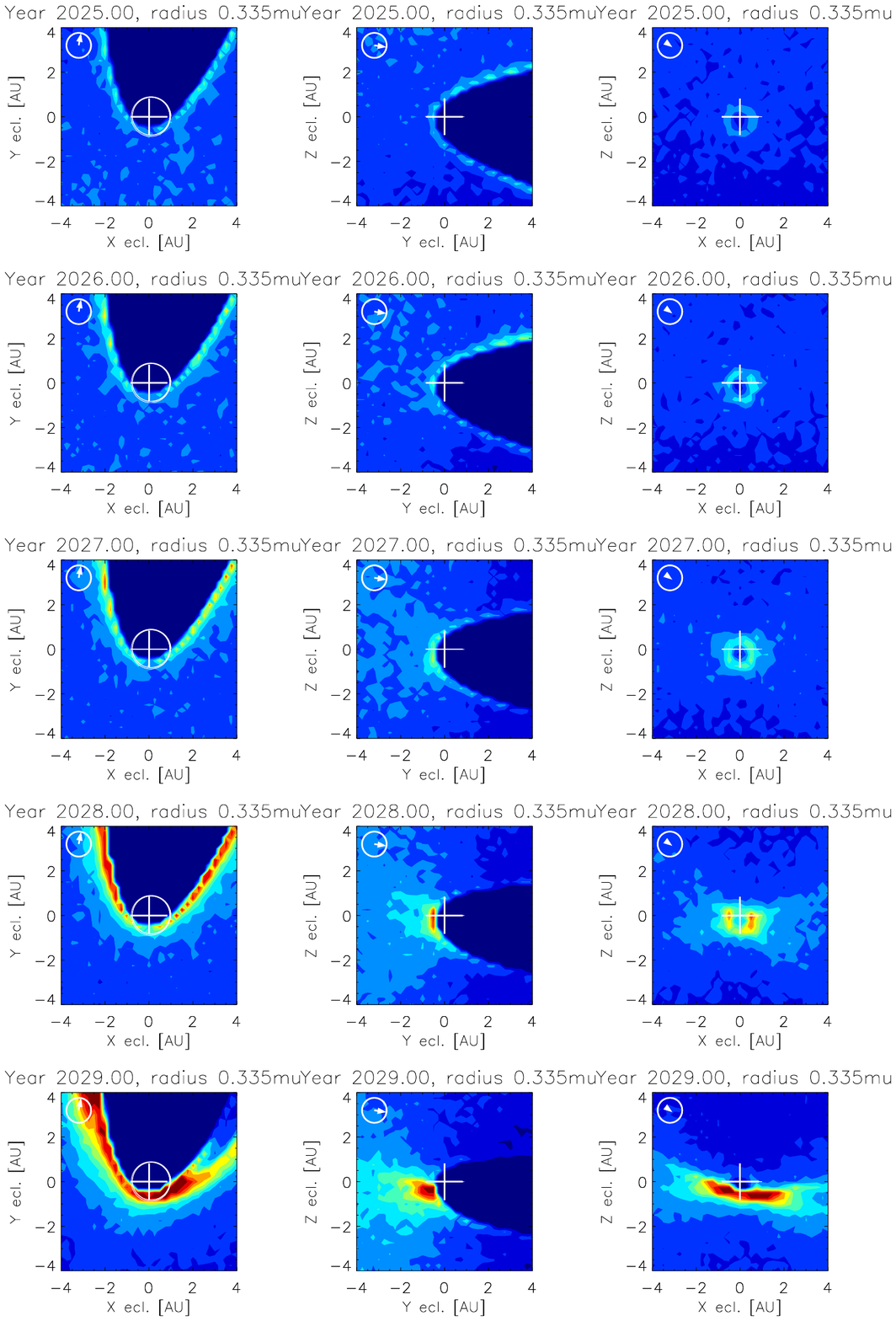}
	    		\vspace{-3.2cm}		
	\caption{Same as Figure~\ref{fig:imex_3d_1} but for particle radius $r_d=\mathrm{0.335\,\mu m}$.}
	\label{fig:imex_3d_2}
\end{figure}

\begin{figure}[tbh]
	\centering
	\vspace{-2.9cm}
		\includegraphics[width=1.25\textwidth]{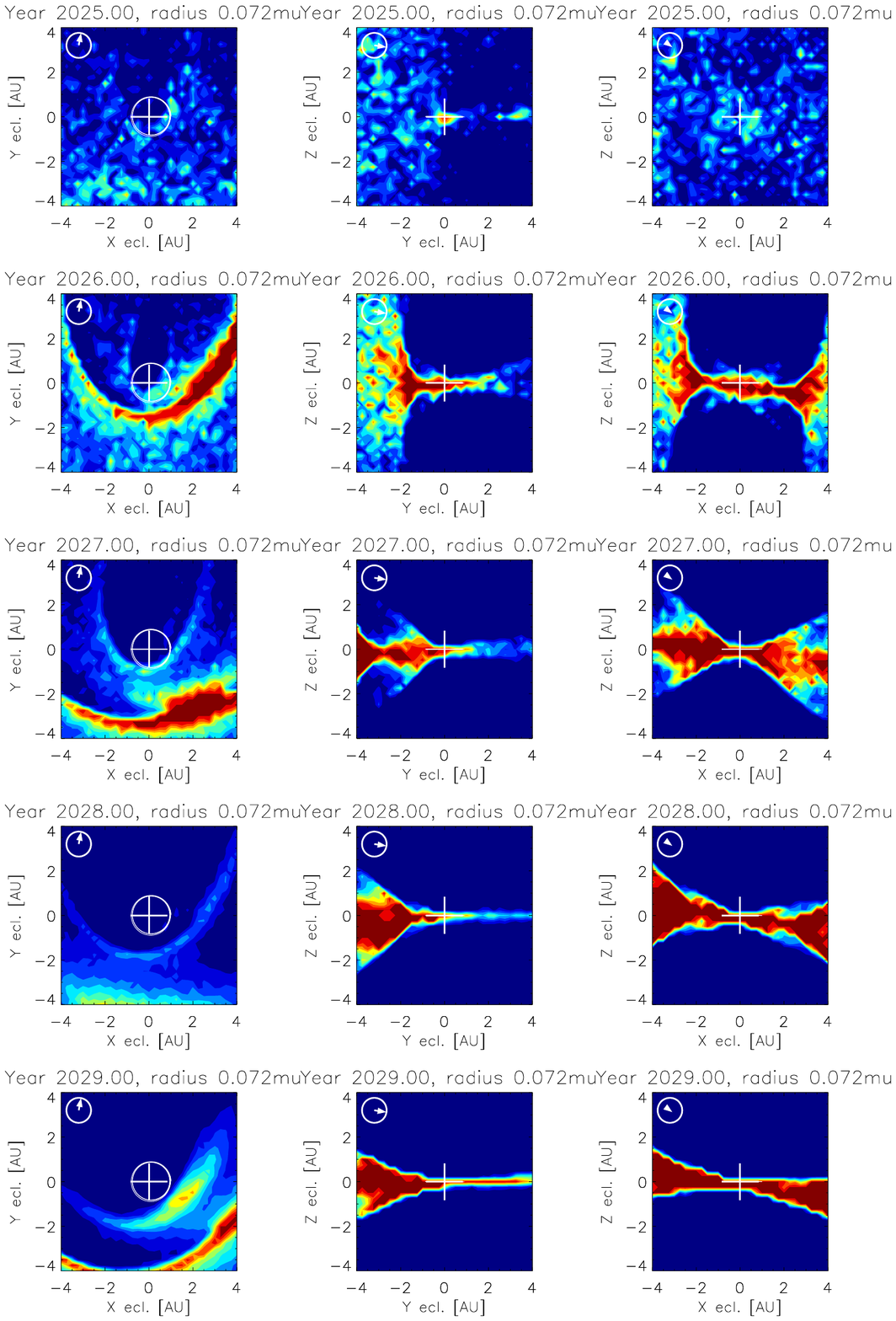}
	    		\vspace{-3.2cm}		
	\caption{Same as Figure~\ref{fig:imex_3d_1} but for particle radius $r_d=\mathrm{0.072\,\mu m}$.}
	\label{fig:imex_3d_1}
\end{figure}

\end{document}